\documentclass[10pt,a4paper,twocolumn]{article}

\usepackage[backend=bibtex,style=nature]{biblatex}
\usepackage[utf8]{inputenc} 
\usepackage[english]{babel} 
\usepackage[bf]{caption}
\usepackage[T1]{fontenc}
\usepackage{amsmath}
\usepackage{amssymb}
\usepackage{graphicx}
\usepackage{float}
\usepackage{stfloats}
\usepackage{tikz}
\usepackage{xcolor}
\usetikzlibrary{decorations.pathreplacing}
\usetikzlibrary{arrows.meta}
\usepackage{empheq}
\usepackage{geometry}
\usepackage[linktoc=all]{hyperref}
\hypersetup{
    colorlinks,
    citecolor=blue,
    filecolor=blue,
    linkcolor=blue,
    urlcolor=blue
}

\definecolor{colorT1}{rgb}{0.122,0.467,0.706}
\definecolor{colorT2}{rgb}{0.839,0.153,0.157}

\newcommand{\ndfrac}[2]{\mbox{\normalsize$\displaystyle\frac{#1}{#2}$}}

\def \step {0.8}
\def \stepFPSCartoon {1.5}
\def \sideFPSCartoon {5.0}
\definecolor{refcolor}{rgb}{0.173,0.627,0.173}

\newcommand{\bbD}{\mathbb{D}}
\newcommand{\bbI}{\mathbb{I}}
\newcommand{\bbK}{\mathbb{K}}
\newcommand{\bbL}{\mathbb{L}}
\newcommand{\bbM}{\mathbb{M}}

\newcommand{\bsb}{\boldsymbol{b}}

\newcommand{\bsu}{\boldsymbol{u}}

\DeclareMathOperator{\Spec}{Spec}
\DeclareMathOperator{\Tr}{Tr}

\title{\textbf{Selective and Collective Actuation in Active Solids}}
\author{P. Baconnier, D. Shohat, C. Hernández López, C. Coulais, V. Démery, G. Düring, O. Dauchot}

\date{}

\usepackage[switch, columnwise]{lineno}

\begin{document}


\twocolumn[
\begin{@twocolumnfalse}
\begin{center}
\LARGE \textbf{Selective and Collective Actuation in Active Solids}
\end{center}
\vspace{0.3cm}
\normalsize P. Baconnier$^{1}$, D. Shohat$^{1,2}$, C. Hernández López$^{3,4}$, C. Coulais$^{5}$, V. Démery$^{1,6}$, G. Düring$^{3,4}$, O. Dauchot$^{1,*}$ \\


\normalsize
\textbf{Active solids consist of elastically coupled out-of-equilibrium units performing work~\supercite{koenderink2009active, henkes2011active, menzel2013traveling, berthier2013non, ferrante2013elasticity, Prost:2015ev, ronceray2016fiber, woodhouse2018autonomous, briand2018spontaneously, giavazzi2018flocking, klongvessa2019active, maitra2019oriented, scheibner2020odd}. 
They are central to autonomous processes, such as locomotion, self-oscillations and rectification, in biological systems~\supercite{abercrombie1954observations, vilfan2005oscillations, szabo2006phase, mizuno2007nonequilibrium, banerjee2015propagating, smeets2016emergent, bi2016motility, chen2017weak,needleman2017active, holmes2018synchronized, peyret2019sustained, henkes2020dense}, 
designer materials~\supercite{brandenbourger2019non} 
and robotics~\supercite{brambilla2013swarm, pratissoli2019soft, li2019particle, dorigo2020reflections, oliveri2021continuous}. 
Yet, the feedback mechanism between elastic and active forces, and the possible emergence of collective behaviours in a mechanically stable elastic solid remains elusive. 
Here we introduce a minimal realization of an active elastic solid, in which we characterize the emergence of selective and collective
actuation resulting from the interplay between activity and elasticity. Polar active agents exert forces on the nodes of a two dimensional elastic lattice. The resulting displacement field nonlinearly reorients the active agents. For large enough coupling, a collective oscillation of the lattice nodes around their equilibrium position emerges. Only a few elastic modes are actuated and, crucially, they are not necessarily the lowest energy ones. Combining experiments with the numerical and theoretical analysis of an agents model, we unveil the bifurcation scenario and the selection mechanism by which the collective actuation takes place. Our findings may provide a new mechanism for oscillatory dynamics in biological systems~\supercite{abercrombie1954observations, smeets2016emergent, chen2017weak, peyret2019sustained} and the opportunity for bona-fide autonomy in meta-materials~\supercite{bertoldi2017flexible, pishvar2020foundations}}.

\vspace{0.3cm}


\end{@twocolumnfalse}
]

\begin{figure}[b!]
\centering
\begin{tikzpicture}

\node[anchor=south west,inner sep=0] at (-7.4,6.7)
{\includegraphics[width=.48\textwidth]{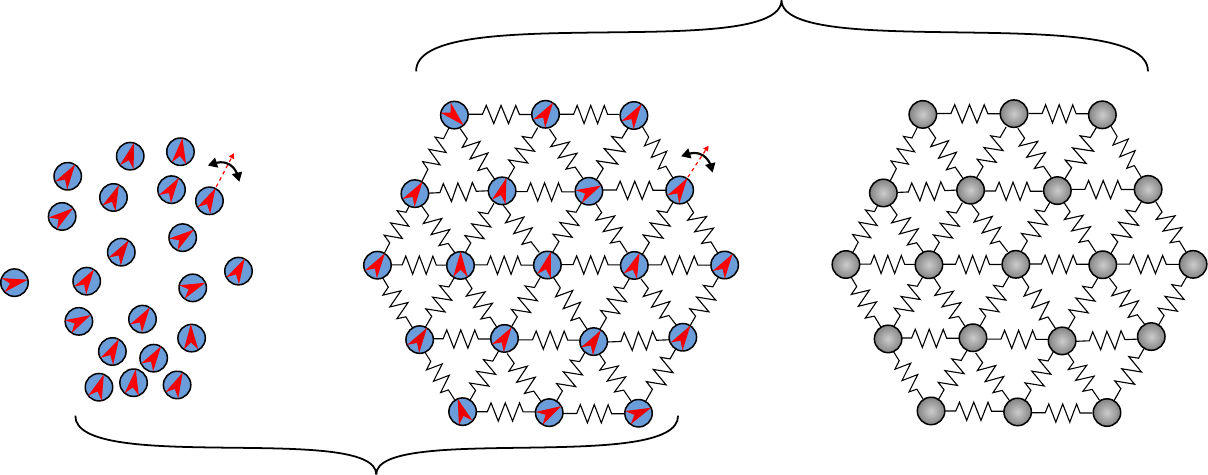}};

\node[] at (-4.8,6.6) {\scriptsize Polarities are free to rotate};
\node[] at (-2.0,10.1) {\scriptsize Reference configuration for the position};

\node[] at (-6.4,9.48) {\scriptsize \textbf{Active liquids}};
\node[] at (-3.65,9.48) {\scriptsize \textbf{Active solid}};
\node[] at (-0.4,9.48) {\scriptsize \textbf{Elastic solid}};

\node[anchor=west] at (-2.55,9.15) {\footnotesize \color{red} $\boldsymbol{\hat{n}}$};
\node[anchor=west] at (-5.85,9.15) {\footnotesize \color{red} $\boldsymbol{\hat{n}}$};


\node[anchor=south west,inner sep=0, rotate=-90] at (-4.1,5.45)
{\includegraphics[height=2.57cm]{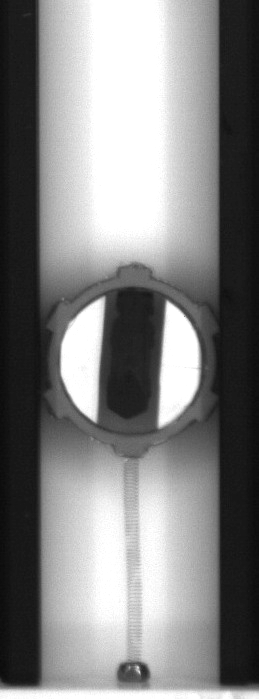}};
\node[anchor=south west,inner sep=0, rotate=-90] at (-4.1,4.05)
{\includegraphics[height=2.57cm]{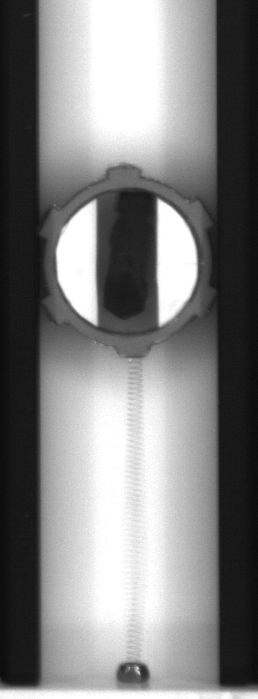}};

\draw[dashed] (-2.5,5.7) -- (-2.5,4.6);
\draw[dashed] (-2.12,5.7) -- (-2.12,3.1);

\draw[->] (-2.9,5.7) -- (-2.5,5.7);
\draw[->] (-1.72,5.7) -- (-2.12,5.7);

\node[anchor=south west] at (-2.1,5.75) {\small $l_{e}$};

\node[] at (-1.8,3.58) {\small \color{red} $\boldsymbol{F}\!_{a}$};
\draw[-{Latex[red]}, color=red, thick] (-2.63,3.58) -- (-2.15,3.58);

\node[anchor=south west] at (-4.1,5.4) {\small \textbf{OFF}};
\node[anchor=south west] at (-4.1,4.0) {\small \textbf{ON}};

\node[anchor=south west,inner sep=0] at (-0.85,3.1)
{\includegraphics[height=0.52cm]{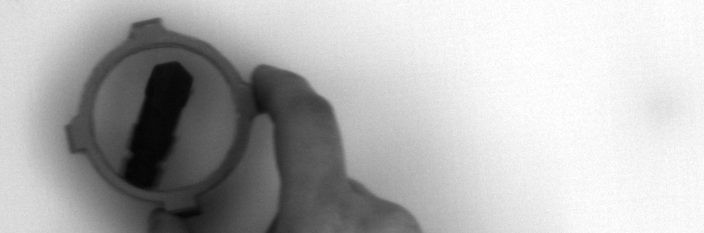}};
\node[anchor=south west,inner sep=0] at (-0.85,3.62)
{\includegraphics[height=0.52cm]{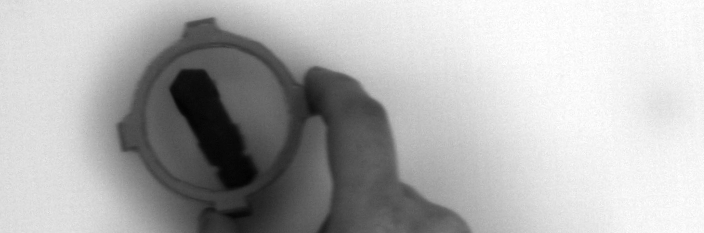}};
\node[anchor=south west,inner sep=0] at (-0.85,4.14)
{\includegraphics[height=0.52cm]{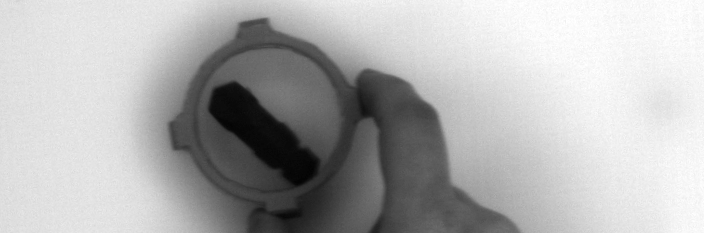}};
\node[anchor=south west,inner sep=0] at (-0.85,4.66)
{\includegraphics[height=0.52cm]{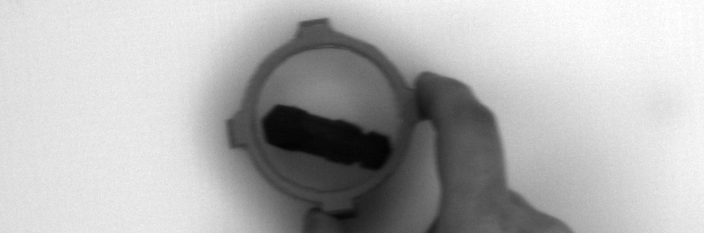}};
\node[anchor=south west,inner sep=0] at (-0.85,5.18)
{\includegraphics[height=0.52cm]{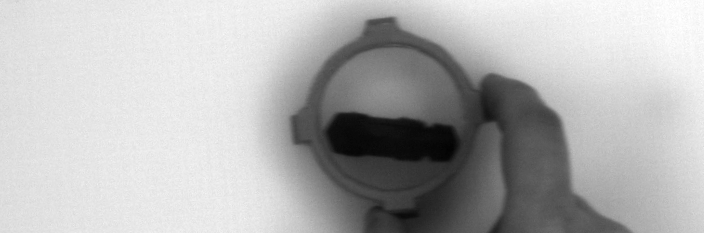}};

\draw[->] (-0.85,3.1) -- (-0.85,5.7);
\draw[->] (-0.85,3.1) -- (0.73,3.1);

\node[] at (0.9,3.1) {\small $x$};
\node[] at (-1.1,5.7) {\small $t$};

\draw[dashed] (-0.69,3.1) -- (-0.69,5.7);
\draw[dashed] (-0.18,5.18) -- (-0.18,5.7);
\draw[<->] (-0.69,5.7) -- (-0.18,5.7);

\node[anchor=south west] at (-0.63,5.75) {\small $l_{a}$};

\node[anchor=south west,inner sep=0] at (-7.5,3.1)
{\includegraphics[height=2.57cm]{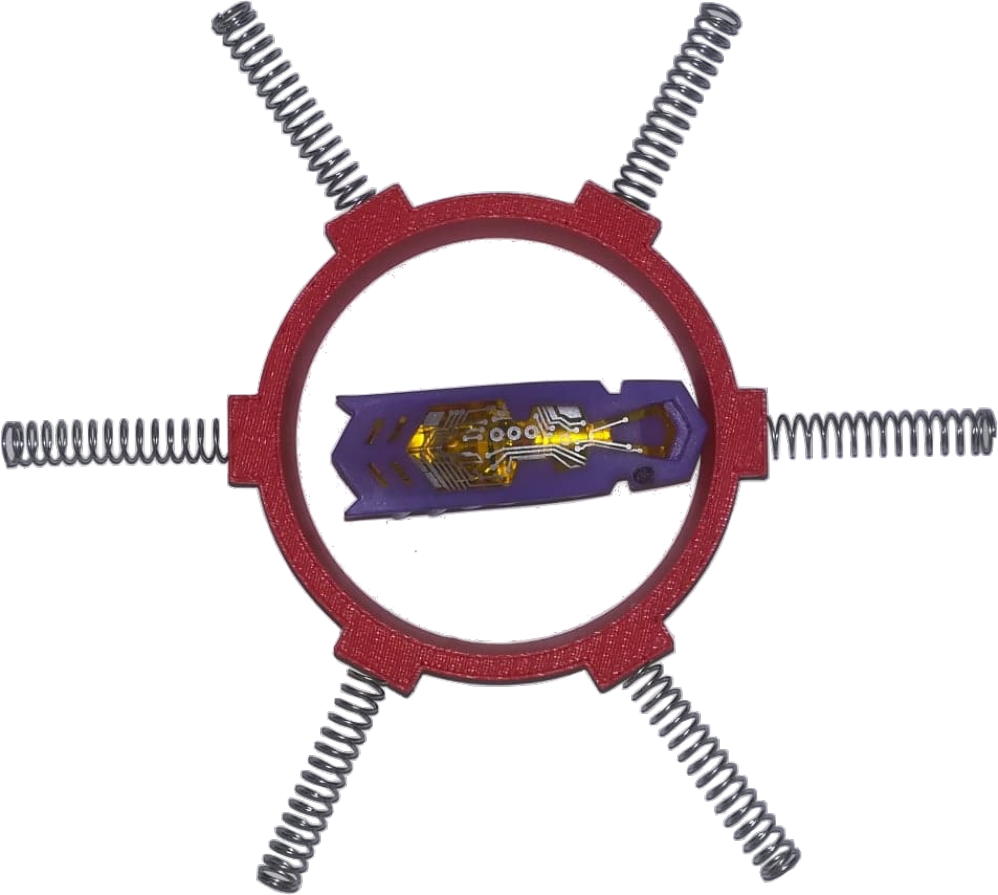}};
\node[anchor=south west,inner sep=0] at (-5.4,4.7)
{\includegraphics[height=0.55cm]{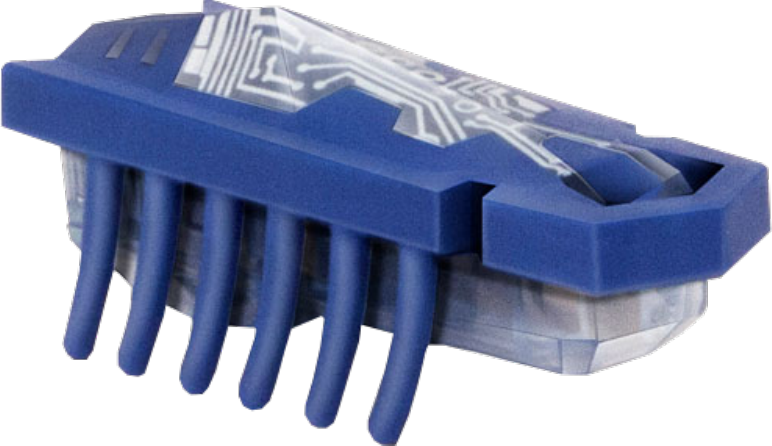}};

\node[anchor=west] at (-4.8,5.4) {\small \color{red} $\boldsymbol{\hat{n}}$};
\draw[-{Latex[red]}, color=red, ultra thick] (-5.05,5.2) -- (-4.35,4.95);

\node[anchor=north] at (-5.25,3.9) {\small $k$};

\node[anchor=south west,inner sep=0] at (-7.4,10.0) {\small \textbf{a}};
\node[anchor=south west,inner sep=0] at (-7.4,6.0) {\small \textbf{b}};
\node[anchor=south west,inner sep=0] at (-4.0,6.0) {\small \textbf{c}};
\node[anchor=south west,inner sep=0] at (-1.05,6.0) {\small \textbf{d}};

\end{tikzpicture}
\vspace*{-0.4cm}
\caption{\small{\textbf{Active solids design principle.} \textbf{a}, Active solids have positional degrees of freedom with a reference state, and a free to rotate polarity vector in the direction of the active force. \textbf{b}, Active unit : a hexbug is trapped in a 3$d$ printed cylinder. \textbf{c}, The active component, here confined in a linear track and attached to a spring of stiffness $k$, produces an active force of amplitude $F_{0}$ in the direction of the polarity $\boldsymbol{\hat{n}}$ and elongates the spring by a length $l_e = F_0/k$. \textbf{d}, The mechanical design of the hexbug -- mass distribution and shape of the legs -- is responsible for its alignement toward its displacement, here imposed manually, of the cylinder (see Supplementary Information for a quantitative measure of the self-alignment length $l_a$).}}
\label{fig:principle}
\end{figure}

Active solids~\supercite{koenderink2009active, henkes2011active, menzel2013traveling, berthier2013non, ferrante2013elasticity, Prost:2015ev, ronceray2016fiber, woodhouse2018autonomous, briand2018spontaneously, giavazzi2018flocking, klongvessa2019active, maitra2019oriented, scheibner2020odd} combine the central properties of simple elastic solids and active liquids~\supercite{toner1995long,Vicsek:2012ty, Bricard:2013jq, marchetti2013hydrodynamics, wittkowski2014scalar, peshkov2014boltzmann} (Fig.\ref{fig:principle}-a). On one hand the positional degrees of freedom of their constituting units have a well-defined reference state. On the other hand activity endows these units with an additional free degree of freedom in the form of polar, or dipolar, active forces. In active liquids, aligning interactions between these forces lead to collective motion. In active solids these active forces deform the elastic matrix, and induce a strain field, which depends on the forces configuration. This strain tensor will in turn reorient the forces. This generic nonlinear elasto-active feedback, a typical realization of which is the contact inhibition locomotion (CIL) of cells~\supercite{abercrombie1954observations, smeets2016emergent},  opens the path towards spontaneous collective excitations of the solid, which we shall call collective actuation. In this work we propose a minimal experimental setting and numerical model, in which we unveil the modal selectivity of collective actuation and its underlying principles.  

We consider crystalline lattices with, at the center of each node, an active particle with a fluctuating orientation (Fig.\ref{fig:principle}-b and Methods). Each node has a well defined reference position, but will be displaced by the active particles (Fig.\ref{fig:principle}-c). In contrast, the polarization of each particle is free to rotate and reorients towards its displacement (Fig.\ref{fig:principle}-d, Supplementary Information section 2.2.2 and Movie 1). This nonlinear feedback between deformations and polarizations is characterized by two length-scales: (i) the typical elastic deformation caused by active forces $l_e$ (Fig.\ref{fig:principle}-c) and (ii) the self-alignment length $l_a$ (Fig.\ref{fig:principle}-d). 
We complement the experiments with numerical simulations of elastically coupled self-aligning active particles~\supercite{dauchot2019dynamics} (Methods). In the over-damped, harmonic and noiseless limit, the model reads:
\begin{subequations} 
\vspace{-0.0cm}
\label{eq_dimensionless_noiseless_braket}
\begin{align}
 \dot{\boldsymbol{u}}_{i} &= \pi \boldsymbol{\hat{n}}_{i} - \mathbb{M}_{ij} \boldsymbol{u}_{j} \label{eq1:dimensionless_noiseless_braket} \\
 \dot{\boldsymbol{n}}_{i} &= (\boldsymbol{\hat{n}}_{i} \times \dot{\boldsymbol{u}}_{i} ) \times \boldsymbol{\hat{n}}_{i},
 \label{eq2:dimensionless_noiseless_braket} 
\end{align}
\vspace{-0.0cm}
\end{subequations}
where the ratio of the elasto-active and self-alignment lengths, $\pi = l_e/l_a$, which we refer to as the elasto-active feedback, is the unique control parameter. $\boldsymbol{\hat{n}}_i$'s are the polarization unit vectors, $\boldsymbol{u}_{i}$ is the displacement field with respect to the reference configuration and $\mathbb{M}$ is the dynamical matrix (Supplementary Information section 3.1).  

\begin{figure*}[t!]
\centering
\begin{tikzpicture}


\node[anchor=south west,inner sep=0] at (8.2,-3.95)
{\includegraphics[width=2.6cm]{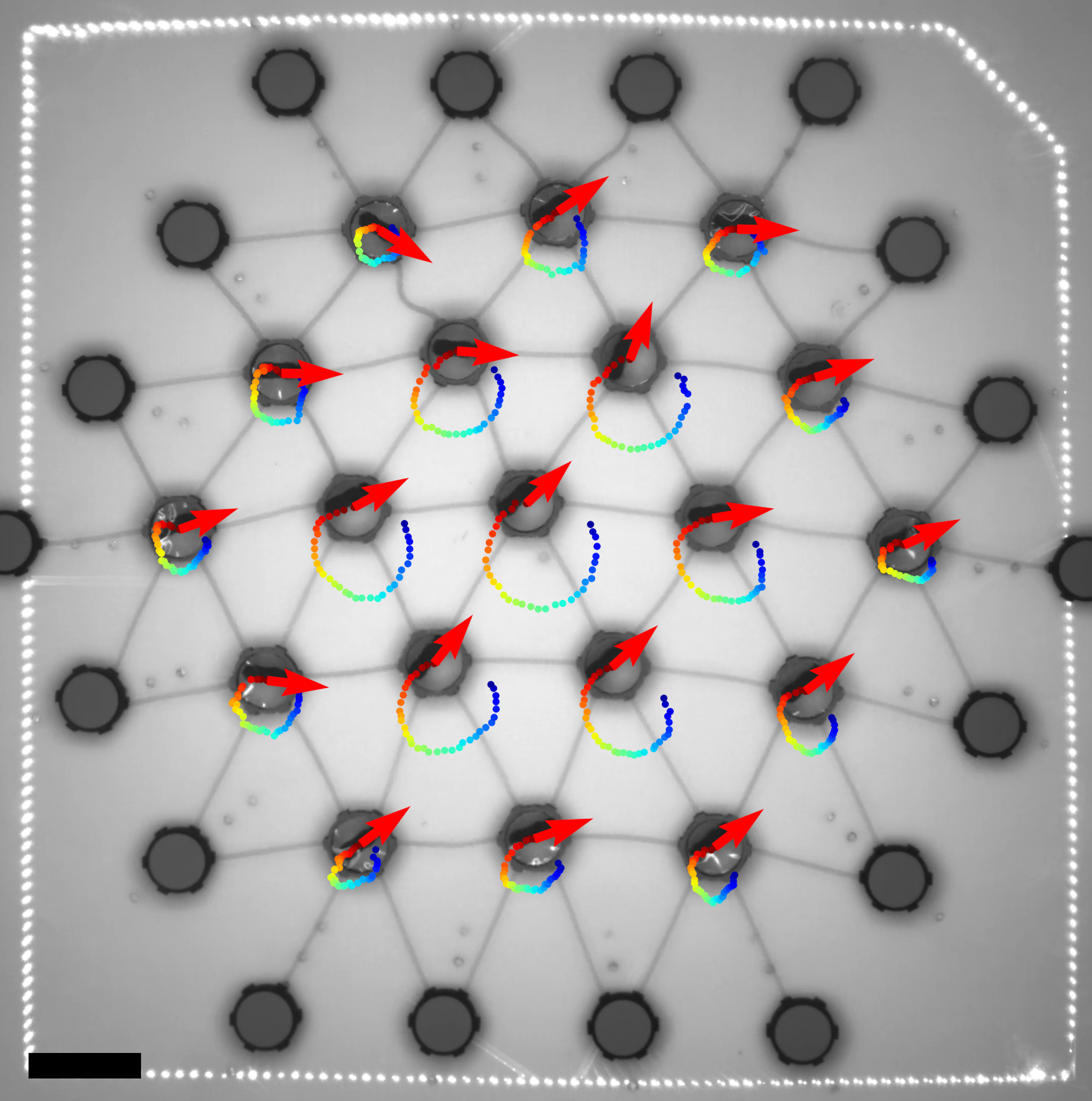}};
\node[anchor=south west,inner sep=0] at (8.2,-8.2)
{\includegraphics[width=2.6cm]{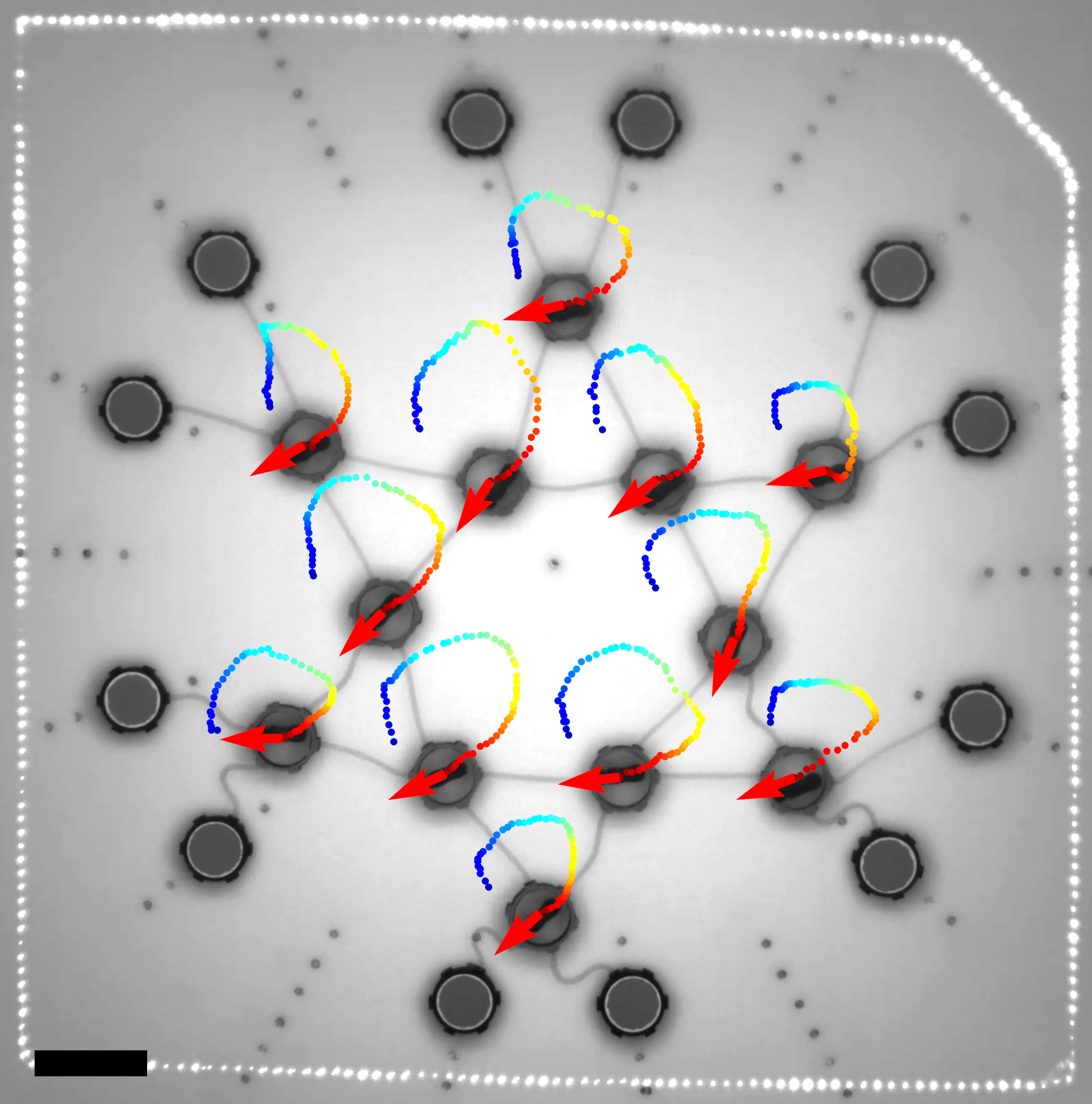}};

\node[anchor=south west,inner sep=0] at (16.4,-4.4)
{\includegraphics[height=3.5cm]{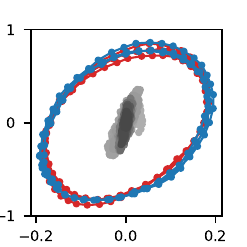}};
\node[anchor=south west,inner sep=0] at (16.4,-8.65)
{\includegraphics[height=3.5cm]{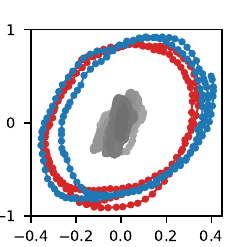}};

\node[anchor=south west,inner sep=0] at (11.45,-4.4) 
{\includegraphics[height=3.5cm]{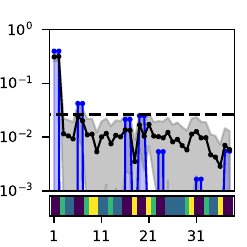}};
\node[anchor=south west,inner sep=0] at (11.45,-8.65) 
{\includegraphics[height=3.5cm]{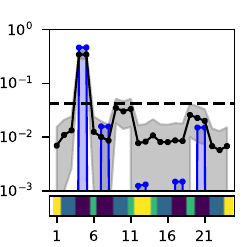}};

\draw[->] (17.1,-2.1) arc (142:117:2.0);
\draw[->] (19.3,-3.2) arc (-35:-60:2.0);

\draw[->] (17.05,-6.25) arc (145:125:2.3);
\draw[->] (19.35,-7.45) arc (-35:-56:2.0);

\node[anchor=north, rotate=90] at (15.95,-2.6) {\small $\langle \boldsymbol{\varphi}_{k} | \boldsymbol{\hat{n}} \rangle / \sqrt{N}$};
\node[anchor=north, rotate=90] at (10.9,-2.4) {\small $\langle \langle \boldsymbol{\varphi}_{k} | \boldsymbol{\hat{n}} \rangle^{2}  / N \rangle_{t}$};

\node[anchor=north, rotate=90] at (15.95,-6.9) {\small $\langle \boldsymbol{\varphi}_{k} | \boldsymbol{\hat{n}} \rangle / \sqrt{N}$};
\node[anchor=north, rotate=90] at (10.9,-6.8) {\small $\langle \langle \boldsymbol{\varphi}_{k} | \boldsymbol{\hat{n}} \rangle^{2} / N \rangle_{t}$};

\node[anchor=north] at (18.15,-4.4) {\small $\langle \boldsymbol{\varphi}_{k} | \boldsymbol{u} \rangle / \sqrt{N}$};
\node[anchor=north] at (13.65,-4.4) {\small $k$ index};

\node[anchor=north] at (18.15,-8.65) {\small $\langle \boldsymbol{\varphi}_{k} | \boldsymbol{u} \rangle / \sqrt{N}$};
\node[anchor=north] at (13.65,-8.65) {\small $k$ index};

\node[] at (15.5,-1.4) {\small $| \boldsymbol{\varphi}_{1} \rangle$};
\node[anchor=south west,inner sep=0] at (14.85,-2.7)
{\includegraphics[height=1.2cm]{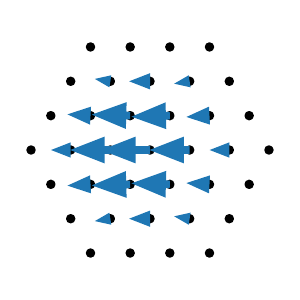}};
\node[] at (15.5,-2.9) {\small $| \boldsymbol{\varphi}_{2} \rangle$};
\node[anchor=south west,inner sep=0] at (14.85,-4.2)
{\includegraphics[height=1.2cm]{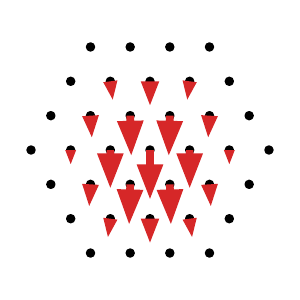}};

\node[] at (15.5,-5.7) {\small $| \boldsymbol{\varphi}_{4} \rangle$};
\node[anchor=south west,inner sep=0] at (14.85,-7.0)
{\includegraphics[height=1.2cm]{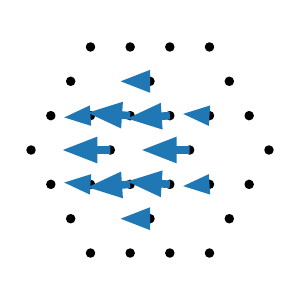}};
\node[] at (15.5,-7.2) {\small $| \boldsymbol{\varphi}_{5} \rangle$};
\node[anchor=south west,inner sep=0] at (14.85,-8.5)
{\includegraphics[height=1.2cm]{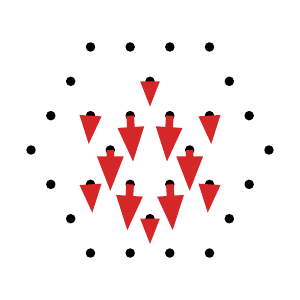}};

\node[anchor=south west,inner sep=0] at (20.1,-4.475)
{\includegraphics[height=3.5cm]{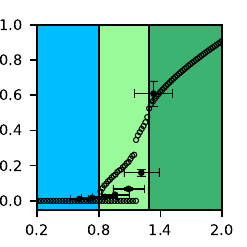}};
\node[anchor=south west,inner sep=0] at (20.1,-8.725)
{\includegraphics[height=3.5cm]{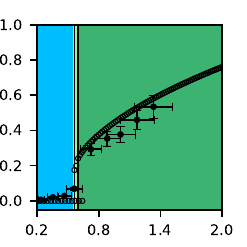}};

\node[anchor=north] at (21.9,-4.4) {\small $\pi$};
\node[anchor=north] at (21.9,-8.765) {\small $\pi$};

\node[anchor=north, rotate=90] at (19.6,-2.6) {\small $\Omega$};
\node[anchor=north, rotate=90] at (19.6,-6.9) {\small $\Omega$};

\node[anchor=south west,inner sep=0] at (23.1,-4.275)
{\includegraphics[height=3.5cm]{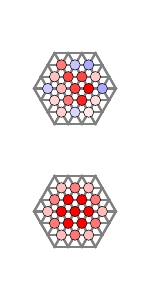}};

\node[] at (24.0,-2.5) {\scriptsize Exp.};
\node[] at (24.0,-3.9) {\scriptsize Num.};

\node[anchor=south west,inner sep=0] at (23.1,-8.525)
{\includegraphics[height=3.5cm]{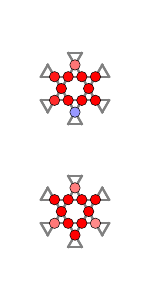}};

\node[] at (24.0,-6.75) {\scriptsize Exp.};
\node[] at (24.0,-8.12) {\scriptsize Num.};

\node[rotate=0] at (23.8,-4.8)
{\includegraphics[height=1.0cm]{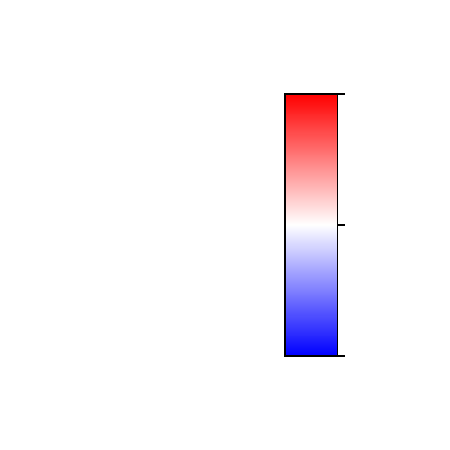}};
\node[] at (24.0,-4.35) {\scriptsize $\omega_{\text{max}}$};
\node[] at (24.0,-5.25) {\scriptsize $-\omega_{\text{max}}$};

\node[anchor=south west,inner sep=0] at (8.2,-1.1) {\small \textbf{a}};
\node[anchor=south west,inner sep=0] at (12.1,-1.1) {\small \textbf{b}};
\node[anchor=south west,inner sep=0] at (15.0,-1.1) {\small \textbf{c}};
\node[anchor=south west,inner sep=0] at (16.7,-1.1) {\small \textbf{d}};
\node[anchor=south west,inner sep=0] at (20.6,-1.1) {\small \textbf{e}};
\node[anchor=south west,inner sep=0] at (23.6,-1.1) {\small \textbf{f}};

\node[rotate=90] at (22.8,-3.25) {\scriptsize Collective};
\node[rotate=90] at (23.05,-3.25) {\scriptsize Actuation};
\node[rotate=90] at (21.7,-2.3) {\scriptsize Heterogeneous};
\node[rotate=90] at (20.8,-2.5) {\scriptsize Frozen-Disordered};

\node[rotate=90] at (22.8,-7.5) {\scriptsize Collective};
\node[rotate=90] at (23.05,-7.5) {\scriptsize Actuation};
\node[rotate=90] at (20.8,-6.75) {\scriptsize Frozen-Disordered};

\end{tikzpicture}
\vspace*{-0.3cm}
\caption{\small{\textbf{Selective and collective actuation in 2$d$ elastic lattices, pinned at their edges} (top) triangular lattice, $N=19$; (bottom) kagome lattice, $N=12$: \textbf{a}, When embedded in an elastic lattice, a large enough elasto-active feedback $\pi$ drives the system towards collective actuation dynamics (red arrows: polarities $\boldsymbol{\hat{n}}_{i}$; trajectories color coded from blue to red by increasing time; scale bars: $10$ cm). \textbf{b}, Spectral decomposition of dynamics on the normal modes of the lattices, sorted by order of growing energies (grey: experiments; blue: numerics). The horizontal dashed lines indicate equipartition. The bottom color bars code for the symmetry class of the modes (Supplementary Information section 7). The gray area represents the $1$-$\sigma$ confidence interval on the experimental measurement. \textbf{c}, Sketch of the two most excited modes, which are not necessarily the lowest energy ones. \textbf{d}, Normal modes components of the active forces as a function of the normal modes components of the displacements (blue/red : projection on \textbf{c}, grey : other modes). The symbol size is set to represent to $1$-$\sigma$ confidence interval on the experimental measurement. \textbf{e}, Transition to the collective actuation regime: average oscillation frequency $\Omega$ as a function of $\pi$ (plain bullets: experiments; open circles: numerics). Background colors code for the dynamical regime (light blue: frozen and disordered; light green: heterogeneous (H); dark green: collective actuation). Triangular: $\pi_{FD} = 0.800$, $\pi_{CA} = 1.29$; kagome: $\pi_{FD} = 0.564$, $\pi_{CA} = 0.600$. The errorbars represent the $1$-$\sigma$ confidence intervals, inherited from the uncertainty on the microscopic parameters measurements. \textbf{f}, Individual oscillation frequencies $\omega_i$ illustrating the coexistence dynamics in experiments and numerical simulations; (triangular lattice: $\pi_{\text{exp}}/\pi_{\text{num}} = 1.22/1.09$; kagome lattice: $\pi_{\text{exp}}/\pi_{\text{num}} = 0.723/0.564$).}}
\label{fig:selection}
\vspace{-0.3cm}
\end{figure*}

If not hold, such an active solid adopts the translational and/or rotational rigid body motion dictated by the presence of zero modes (Movies 2,3), as reported in other theoretical models~\supercite{szabo2006phase, ferrante2013elasticity, menzel2013traveling}. Here we are interested in stable elastic solids, with no zero-modes. We therefore explore the emergence of collective dynamics in elastic lattices pinned at their edges. For both the triangular (Fig.\ref{fig:selection}-top) and the kagome (Fig.\ref{fig:selection}-bottom) lattice, we observe a regime where all the lattice nodes spontaneously break chiral symmetry and rotate around their equilibrium position in a collective steady state (Fig.\ref{fig:selection}-a and Movies 5,8).  This dynamical and chiral phase, which is reminiscent of oscillations in biological tissues~\supercite{vilfan2005oscillations, peyret2019sustained}, is clearly different from collective dynamics in active fluids~\supercite{Vicsek:2012ty, Bricard:2013jq} and rigid body motion in active solids~\supercite{menzel2013traveling,ferrante2013elasticity,briand2018spontaneously}.

The dynamics are best described when projected on the normal modes of the elastic structure sorted by order of growing energies. The dynamics condensate mostly on two modes (Fig.~\ref{fig:selection}-b), and describe a limit cycle driven by the misalignment of the polarity and the displacements (Fig.~\ref{fig:selection}-d). In the case of the triangular lattice, the selected modes are the two lowest energy ones. Interestingly, in the case of the kagome lattice, these are the fourth and fifth modes, not the lowest energy ones. 
For both lattices, the selected pair of degenerated modes are strongly polarized along one spatial direction; they are extended and the polarization of the modes in each pair are locally quasi-orthogonal (Fig.~\ref{fig:selection}-c). The numerical simulations confirm the experimental observations indicating that collective actuation is already present in the harmonic approximation and is not of inertial origin. It also allows for the observation of additional peaks in the spectrum, which belong to the same symmetry class as the two most actuated modes (Fig.~\ref{fig:selection}-b, Supplementary Information section 7). 
As we shall see below these properties are at the root of the selection principle of the actuated modes.

The transition to the collective actuation regime (Fig.~\ref{fig:selection}-e) is controlled by the elasto-active feedback. The larger it is, the more the particles reorient upon elastic deformations. Below a first threshold $\pi_{FD}$, the active solid freezes in a disordered state, with random polarizations and angular diffusion (Movies 4,7).  Beyond a second threshold $\pi_{CA}$, collective actuation sets in: synchronized oscillations take place and the noiseless dynamics follow a limit cycle, composed of several frequencies in rational ratio (Supplementary Information, section 10.2).
In between, the system is heterogeneous (Fig.~\ref{fig:selection}-f and Movie 6), with the oscillating dynamics being favored close to the center, while the frozen disordered regime invades the system layer by layer, from the edges towards the center, as $\pi$ decreases (Movie 11).

\begin{figure*}[t!]
\centering
\hspace*{-0.3cm}
\begin{tikzpicture}

\node[rotate=90] at (-3.8,0.8)
{\includegraphics[width=1.1\columnwidth]{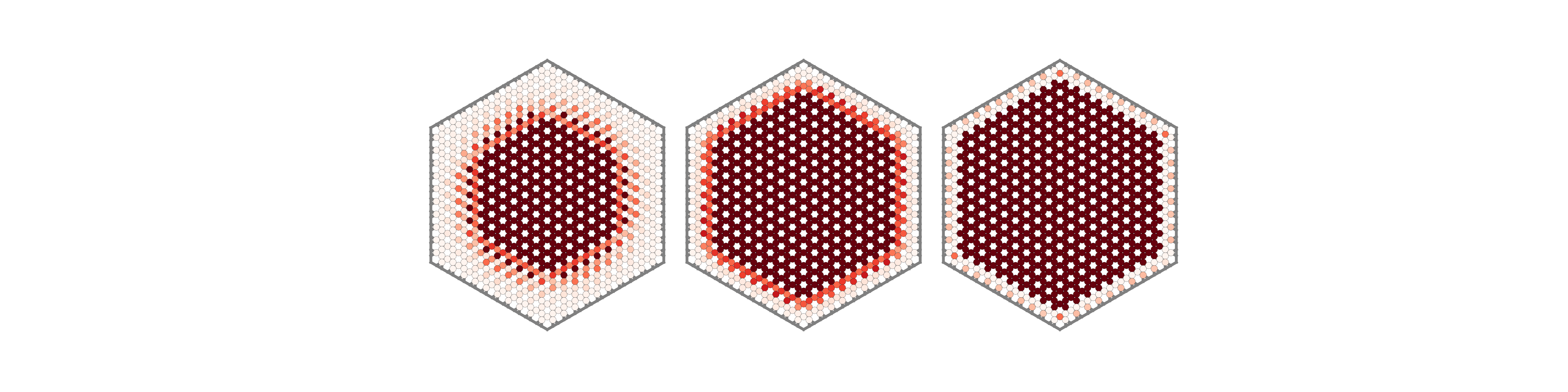}};

\node[rotate=90] at (-2.8,2.5) {\small $\pi = 30$};
\node[rotate=90] at (-2.8,1.0) {\small $\pi = 10$};
\node[rotate=90] at (-2.8,-0.5) {\small $\pi = 4.0$};

\node[rotate=90] at (-6.15,0.8)
{\includegraphics[width=1.1\columnwidth]{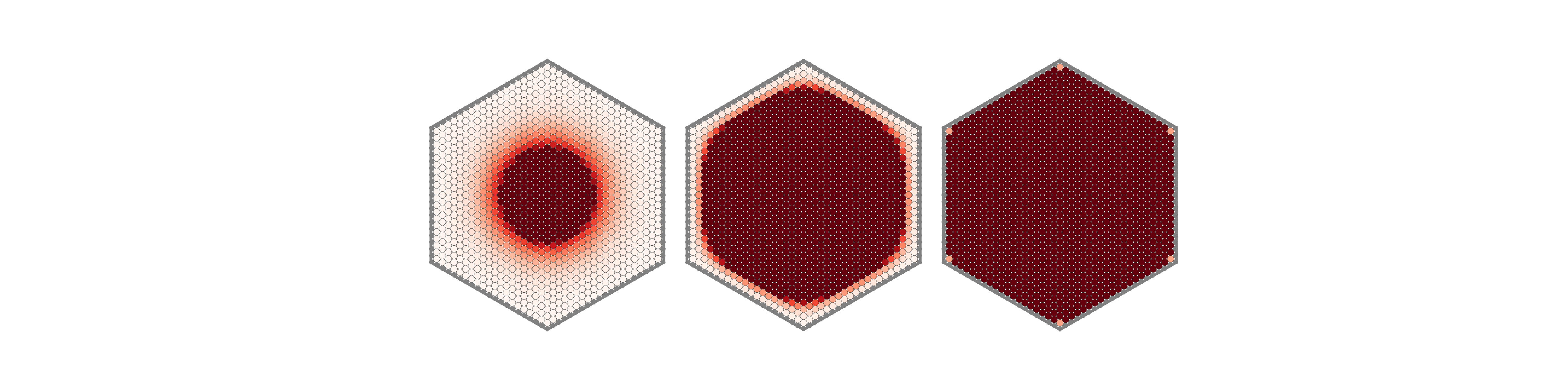}};

\node[rotate=90] at (-5.15,2.5) {\small $\pi = 25$};
\node[rotate=90] at (-5.15,1.0) {\small $\pi = 3.0$};
\node[rotate=90] at (-5.15,-0.5) {\small $\pi = 1.5$};

\fill[black] (-6.15,-0.61) circle (0.03);
\draw[->] (-6.15,-0.61) -- (-6.65,-1.47);
\draw[] (-6.64,-1.27) -- (-6.49,-1.36);
\node[] at (-6.75,-1.62) {\small $r$};
\node[] at (-6.75,-1.2) {\scriptsize $1$};

\node[rotate=-90] at (-4.9,-1.1) {\includegraphics[height=2.0cm]{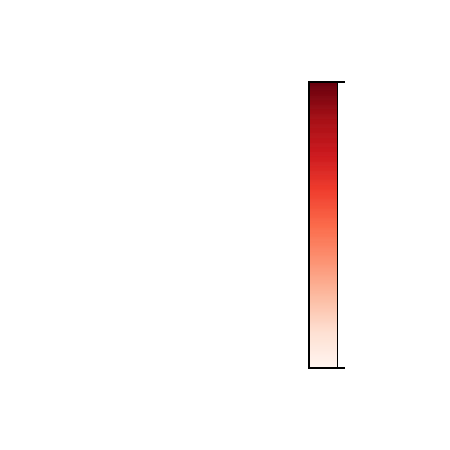}};
\node[] at (-5.52,-1.85) {\small $0$};
\node[] at (-4.05,-1.85) {\small $\omega_{\text{max}}$};


\node[] at (-3.5,5.0)
{\includegraphics[width=.25\columnwidth]{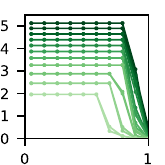}};

\node[] at (-6.1,5.0)
{\includegraphics[width=.25\columnwidth]{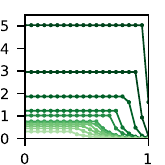}};

\node[rotate=90] at (-4.8,5.1) {\small $\omega_{i}$};
\node[] at (-3.4,3.75) {\small $r$};
\node[rotate=90] at (-7.4,5.1) {\small $\omega_{i}$};
\node[] at (-6.0,3.75) {\small $r$};

\draw[->] (-6.4,4.25) -- (-5.9,5.25);
\node[] at (-5.7,5.45) {\small $\pi\!\nearrow$};

\draw[->] (-3.4,4.55) -- (-2.75,5.7);
\node[] at (-3.85,4.5) {\small $\pi\!\nearrow$};

\node[rotate=90] at (3.75,3.97) {\small $\lambda_{1/2}$};
\node[] at (4.7,3.07) {\small $\pi$};

\node[] at (0.0,4.3) {\includegraphics[width=.47\columnwidth]{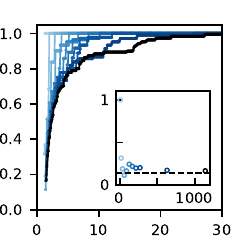}};
\node[] at (4.0,4.3) {\includegraphics[width=.47\columnwidth]{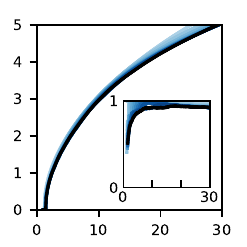}};
\node[] at (8.15,4.4) {\includegraphics[width=.465\columnwidth]{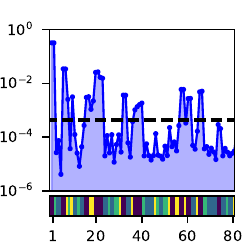}};

\node[rotate=90] at (3.84,-0.42) {\small $\lambda_{1/2}$};
\node[] at (4.77,-1.26) {\small $\pi$};

\node[] at (0.0,0.0) {\includegraphics[width=.47\columnwidth]{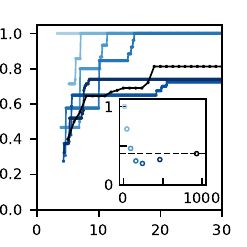}};
\node[] at (4.0,0.0) {\includegraphics[width=.47\columnwidth]{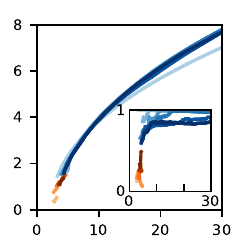}};
\node[] at (8.15,0.1) {\includegraphics[width=.465\columnwidth]{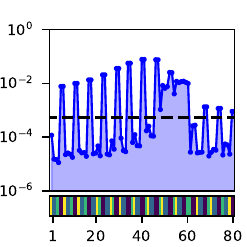}};


\draw[->] (-1.3,5.5) -- (-0.7,5.0);
\node[] at (-0.7,4.75) {\small $N\!\!\!\nearrow$};

\node[rotate=90] at (-2.3,0.2) {\small $f_{CA}$};
\node[] at (0.1,-2.1) {\small $\pi$};
\node[rotate=90] at (-0.35,-0.3) {\small $f_{CA}^{*}$};
\node[] at (0.6,-1.22) {\small $N$};
\node[rotate=90] at (-2.3,4.5) {\small $f_{CA}$};
\node[] at (0.1,2.2) {\small $\pi$};
\node[rotate=90] at (-0.45,4.0) {\small $f_{CA}^{*}$};
\node[] at (0.55,3.08) {\small $N$};

\node[rotate=90] at (1.9,0.2) {\small $\Omega_{CA}$};
\node[] at (4.1,-2.1) {\small $\pi$};
\node[rotate=90] at (1.9,4.5) {\small $\Omega_{CA}$};
\node[] at (4.1,2.2) {\small $\pi$};


\node[rotate=90] at (5.9,0.2) {\small $\langle \langle \boldsymbol{\varphi}_k | \boldsymbol{\hat{n}} \rangle^{2} \rangle_t$};
\node[] at (8.4,-2.1) {\small $k$ index};
\node[rotate=90] at (5.9,4.5) {\small $\langle \langle \boldsymbol{\varphi}_k | \boldsymbol{\hat{n}} \rangle^{2} \rangle_t$};
\node[] at (8.4,2.2) {\small $k$ index};


\node[] at (-6.75,6.2) {\small \textbf{a}};
\node[] at (-7.1,3.2) {\small \textbf{b}};
\node[] at (-1.3,6.2) {\small \textbf{c}};
\node[] at (2.7,6.2) {\small \textbf{d}};
\node[] at (7.0,6.2) {\small \textbf{e}};

 \end{tikzpicture}
 \vspace*{-2.35cm}
\caption{\small{\textbf{Large $N$ lattices.} \textbf{a} Radial distribution of the individual oscillation frequencies $\omega_i$ (left: triangular lattice, $N=1141$; right kagome lattice, $N=930$). Plots are color coded from light to dark green as $\pi$ increases (triangular lattice: for $\pi \in [1.5, 1.6, 1.7, 1.8, 1.9, 2.0, 2.5, 3.0, 5.0, 10, 30]$; kagome lattice: for $\pi \in [4.0, 5.0, 6.0, 7.0, 8.0, 9.0, 10, 11, 12, 13, 14]$). \textbf{b} Spatial distribution of $|\omega_i|$  (left: triangular lattice, $N=1141$; right kagome lattice, $N=930$), $\pi$ values as indicated.  \textbf{c}, Collective actuation fraction $f_{CA}$ as a function of $\pi$ for increasing $N$, color coded from light to dark blue (Inset: $f_{CA}$ at onset of collective actuation saturates to a finite value at large $N$). \textbf{d}, Collective oscillation frequency $\Omega_{CA}$ as a function of $\pi$ for increasing $N$, same color code (triangular lattices, $N = 7, 19, 37, 61, 91, 127, 169, 217, 271, 631, 1141$; kagome lattice, $N = 12, 42, 90, 156, 240, 462, 930$),  (Inset: condensation fraction on the selected symmetry class, $\lambda_{1/2}$ as a function of $\pi$ for increasing $N$). \textbf{e}, Spectral decomposition of the dynamics on the normal modes of the lattices, sorted by order of growing energies (only the first 80 modes are shown; both for $\pi=10$). For panels \textbf{c,d,e} : top row triangular lattice, bottom row kagome lattice.}}
\label{fig:Nsize}
\end{figure*}

Simulations with increasing values of $N$, while keeping the physical size $L$ constant (Methods), indicate that collective actuation subsists for large $N$ (Fig.~\ref{fig:Nsize}). The successive de-actuation steps converge toward a regular variation of the fraction of nodes activated in the center of the system, $f_{CA}$ (Fig.~\ref{fig:Nsize}-c-d and Movies 12,13). At the transition to the frozen disordered state, when $\pi=\pi_{FD}$, the fraction of actuated nodes drops discontinuously to zero, from a finite value $f_{CA}^*$, which decreases with $N$, but saturates at large $N$ (Fig.~\ref{fig:Nsize}-d).
In the case of the triangular lattices, the collective oscillation frequency, $\Omega$, measured in the region of collective actuation, decreases continuously to zero (Fig.~\ref{fig:Nsize}-d-top). This is however non generic : in the case of the kagome lattices, very close to the transition, the dynamics condensate on a different set of modes, pointing at the possible multiplicity of periodic solutions. The transition is essentially discontinuous. Most importantly the spectrum demonstrates that, inside the collective actuation regime, the symmetry class of modes that are selected is independent of the system size (Fig.~\ref{fig:Nsize}-e). The selection of the most actuated modes is again dictated by the geometry of the modes, and not only by their energies. In all cases the condensation level remains large, with a large condensation fraction $\lambda_{1/2}$ (see Methods) for a wide range of values of $\pi$ (Fig.~\ref{fig:Nsize}-d-inset).

Altogether our experimental and numerical findings demonstrate the existence of a selective and collective actuation in active solids. This new kind of collective behaviour specifically takes place because of the elasto-active feedback, that is the reorientation of the active units by the displacement field. The salient features of collective actuation are three-fold: (i) the transition from the disordered phase leads to a chiral phase with spontaneously broken symmetry; (ii) the actuated dynamics are not of inertial origin, take place on a few modes, not always the lowest energy ones, and therefore obey non-trivial selection rules; (iii) the transition follows a coexistence scenario, where the fraction of actuated nodes discontinuously falls to zero. In the remainder of the paper, we unveil the physical origins of these three attributes.

\begin{figure*}[t!]
\centering
\begin{tikzpicture}


\node[] at (-9.4,0.0)
{\includegraphics[width=4.26cm]{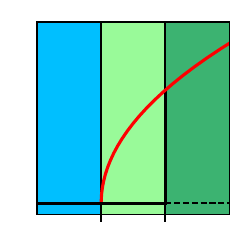}};

\node[] at (-9.7,-1.9) {\small $\omega_{0}^{2}$};
\node[] at (-8.5,-1.9) {\small $2\omega_{\text{min}}^{2}$};

\node[] at (-9.2,-2.3) {\small $\pi$};
\node[rotate=90] at (-11.3,0.2) {\small $\Omega$};

\node[rotate=90] at (-8.1,-0.65) {\scriptsize Collective};
\node[rotate=90] at (-7.8,-0.65) {\scriptsize Actuation};
\node[rotate=90] at (-9.6,0.88) {\scriptsize Coexistence};
\node[rotate=90] at (-10.7,0.95) {\scriptsize Disordered};

\draw[->, thick] (-8.68,-0.76) -- (-8.68,-0.3);
\node[] at (-10.35,-1.17) {\footnotesize $\boldsymbol{m} = 0$};
\node[rotate=35] at (-8.15,1.2) {\footnotesize $\color{red} \boldsymbol{m} = 1$};


\node[rotate=90] at (2.4,0.0) {\includegraphics[width=.37\columnwidth]{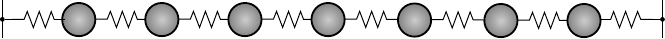}};
\node[rotate=-90] at (3.9,0.0) {\includegraphics[width=.45\columnwidth]{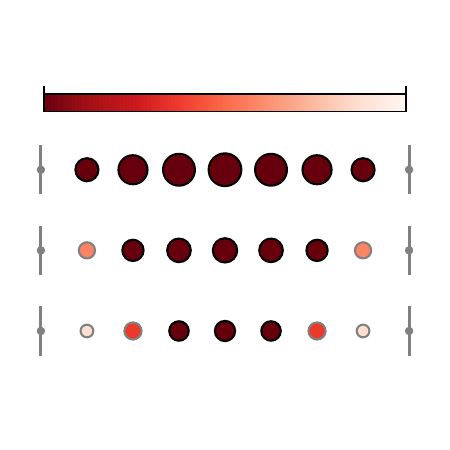}};

\node[] at (4.95,-1.8) {\small 0};
\node[] at (5.2,1.75) {\small $\omega_{\text{max}}$};

\draw[->] (2.85,-1.85) -- (4.55,-1.85);
\node[] at (3.8,-2.2) {\small $\pi\nearrow$};

\node[] at (0.0,0.0)
{\includegraphics[width=.5\columnwidth]{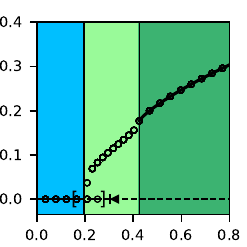}};

\node[] at (0.2,-2.3) {\small $\pi$};
\node[rotate=90] at (-2.4,0.2) {\small $\Omega$};

\node[rotate=90] at (1.3,-0.55) {\scriptsize Collective};
\node[rotate=90] at (1.6,-0.55) {\scriptsize Actuation};
\node[rotate=90] at (-0.48,0.7) {\scriptsize Heterogeneous};
\node[rotate=90] at (-1.3,0.5) {\scriptsize Frozen-Disordered};


\node[rotate=90] at (-0.9,-0.75) {\scriptsize $\pi^{\text{min}}_{c}$};
\node[rotate=45] at (-0.2,-0.85) {\scriptsize $\pi^{\text{max}}_{c}$};
\node[rotate=45] at (0.1,-0.85) {\scriptsize $\pi^{\text{upp}}$};


\node[anchor=south west,inner sep=0] at (-6.6,-2.2)
{\includegraphics[height=1.8cm]{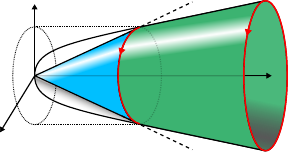}};

\node[] at (-5.55,-0.55) {\scriptsize $R = 1$};
\node[] at (-6.17,-0.25) {\scriptsize $u_x$};
\node[] at (-6.7,-2.07) {\scriptsize $u_y$};
\node[] at (-3.4,-1.1) {\small $\pi$};

\node[] at (-4.95,-2.1) {\scriptsize $\omega_{0}^{2}$};


\node[anchor=south west,inner sep=0] at (-7.2,0.25) {\includegraphics[height=1.53cm]{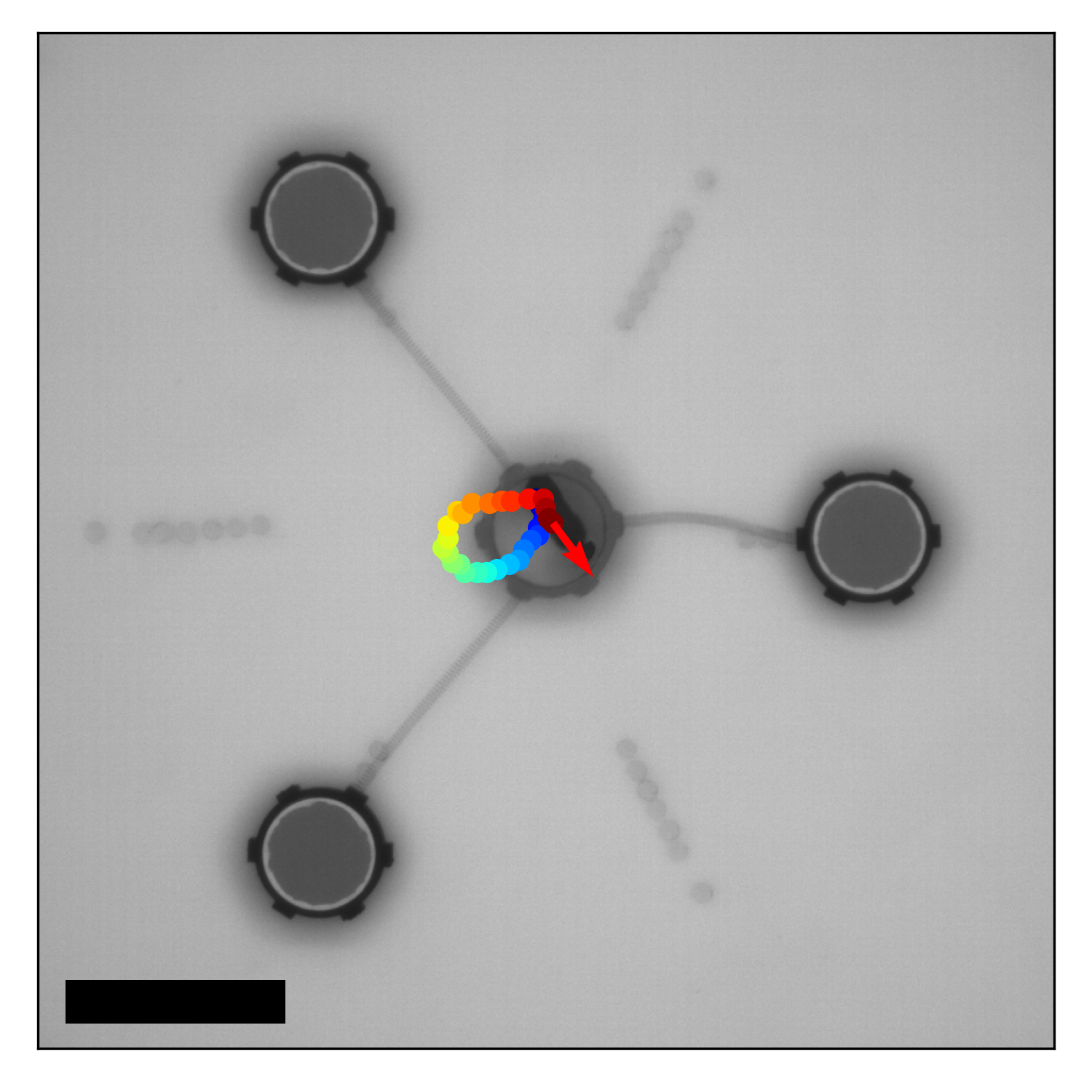}};
\node[anchor=south east,inner sep=0] at (-5.71,0.32) {\includegraphics[height=0.59cm]{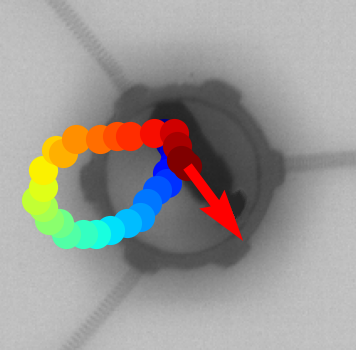}};

\draw[] (-5.72,0.33) -- (-5.72,0.91);
\draw[] (-5.72,0.91) -- (-6.32,0.91);
\draw[] (-6.32,0.91) -- (-6.32,0.33);
\draw[] (-6.32,0.33) -- (-5.72,0.33);




\node[anchor=south west,inner sep=0] at (-5.4,-0.08) {\includegraphics[height=1.94cm]{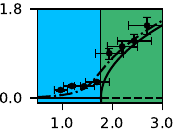}};

\node[] at (-3.85,-0.25) {\small $\pi$};
\node[rotate=90] at (-5.3,1.0) {\small $\Omega$};

\node[anchor=south west,inner sep=0] at (-10.95,2.1) {\small \textbf{a}};
\node[anchor=south west,inner sep=0] at (-7.2,2.1) {\small \textbf{b}};
\node[anchor=south west,inner sep=0] at (-4.9,2.1) {\small \textbf{c}};
\node[anchor=south west,inner sep=0] at (-6.8,-0.3) {\small \textbf{d}};
\node[anchor=south west,inner sep=0] at (-1.6,2.1) {\small \textbf{e}};
\node[anchor=south west,inner sep=0] at (2.35,2.1) {\small \textbf{f}};

 \end{tikzpicture}
 \vspace*{-0.7cm}
\caption{\small{\textbf{Mean field, single particle and 1$d$ lattices phase diagrams.} \textbf{a}, At the mean field level, the disordered, $\boldsymbol{m} = 0$, phase (black line) coexists with the fully polarized, $| \boldsymbol{m} | = 1$, chiral, $\Omega > 0$, phase (red line) for $\pi \in \left[ \omega_{0}^{2}, 2\omega_{\text{min}}^{2} \right]$. \textbf{b-c-d}, A single active unit connected to the three static vertices of a regular triangle (red arrow: polarity $\boldsymbol{\hat{n}}$; trajectories color coded from blue to red by increasing time; scale bar: $10$ cm;
Inset: zoom on the active unit) oscillates with  an average rotation frequency $\Omega$ increasing with $\pi$ (\textbf{c}); (($\bullet$): experimental data, (continuous line): analytical expression (Eq.~\ref{SP_th}), (dot-dashed line): numerical values in the presence of a bias; $\pi_c = 1.77$). The errorbars represent the $1$-$\sigma$ confidence intervals, inherited from the uncertainty on the microscopic parameters measurements. \textbf{d}, Phase space structure of the displacements: for $\pi<\omega_0^2$, an infinite set of marginal fixed points forms a circle of radius $R=\pi/\omega_0^2$; for $\pi>\omega_0^2$, all such fixed points are unstable and a limit cycle of radius $R=\left(\pi/\omega_0^2\right)^{1/2}$ branches off continuously. \textbf{e-f}, The collective actuation in a zero rest length chain of $N = 7$ nodes, pinned at both ends: \textbf{e}, average oscillation frequency $\Omega$ as a function of $\pi$ (continuous line: limit cycle found analytically; horizontal lines ($\Omega=0$): range of existence of only stable (continuous), only unstable (dashed) and coexisting stable and unstable (dot-dashed) fixed points; ($\circ$): numerical data; same background color as for Fig.~\ref{fig:selection}-\textbf{e}). $\pi_{c}^{\text{min}} = 0.152$, $\pi_{c}^{\text{max}} = 0.280$, $\pi_{c}^{\text{upp}} = 0.304$, $\pi_{FD} = 0.195$, $\pi_{CA} = 0.426$. \textbf{f}, Individual oscillation frequencies $\omega_i$ for increasing values of $\pi = [0.20, 0.33, 1.0]$, in the $N=7$ chain. Radii of the colored circles code for the average trajectory radius. Black, respectively gray, contours indicate $R_i \geq 1$ and $R_i \leq 1$.}}
\label{fig:theory}
\end{figure*}

At large scales, the dynamics of the displacement and polarization fields, ${\bf U}({\bf r},t)$ and ${\bf m}({\bf r},t)$, the local averages of, respectively, the microscopic displacements ${\bf u}_i$ and the polarizations ${\bf \hat{n}}_i$, are obtained from a coarse-graining procedure (see Supplementary Information, section 6) and read:
\begin{subequations} 
\vspace{-0.0cm}
\label{eq:cg}
\begin{align}
\partial_t{\boldsymbol{U}} &= \pi \boldsymbol{m} + \boldsymbol{F}_{e} \label{eq1:cg} \\
\partial_t{\boldsymbol{m}} &= (\boldsymbol{m}\!\times \!\partial_t{\boldsymbol{U}} )\!\times \!\boldsymbol{m}\!+\! \frac{1\!-\!\boldsymbol{m}^2}{2}\partial_t{\boldsymbol{U}} \!-\!D_r \boldsymbol{m},
 \label{eq2:cg} 
\end{align}
\end{subequations}
where the elastic force ${\bf F}_e\left[{\bf U}\right]$ is given by the choice of a constitutive relation and the relaxation term $-D_r {\bf m}$ results from the noise. Assuming linear elasticity, the frozen phase, in which the local random polarities and displacements average to ${\bf U}={\bf 0}$ and ${\bf m}={\bf 0}$, is stable for small elasto-active feedback. It becomes linearly unstable for $\pi>\pi^{cg}_c = 2\, \omega_{\rm min}^2$, where $\omega^2_{\rm min}$ is the smallest eigenvalue of the linear elastic operator (Supplementary Information, section 6.5). 
We then look for homogeneous solutions, assuming a condensation on two degenerated and spatially homogeneous modes, such that ${\bf F}_e = - \omega^2_0 {\bf U}$, with  $\omega^2_0$ the eigenfrequency of such modes. For $\pi>\omega_0^2$, we find a polarized chiral phase oscillating at a frequency
\begin{equation} \label{SP_th}
\Omega = \omega_{0} \sqrt{\pi - \omega_0^2},
\end{equation}
In the limiting case $D_r=0$, $|m|=1$ (Supplementary Information, section 6.6).
The resulting mean field phase diagram (Fig.~\ref{fig:theory}-a) thus captures the existence of the frozen and chiral phases and their phase space coexistence for a finite range of the elasto-active feedback. However, the disordered ${\bf m}={\bf 0}$, and the polarized chiral oscillating solutions being disconnected, the nature of the transition is controlled by inhomogeneous solutions, which cannot be investigated within perturbative approaches.

Alternatively, we turn ourselves to simpler geometries in which exact results can be obtained. A first important hint at the nature of the transition towards the chiral phase concerns the structure of the phase space, and is best understood from considering the dynamics of a single particle (Fig.~\ref{fig:theory}-b and Movies 9,10 and Supplementary Information section 5). Below $\pi_c=\omega_0^2$, the phase space for the displacements contains an infinite set of marginal fixed points, organized along a circle of radius $R=\pi/\omega_0^2$. At $\pi_c$, the escape rate of the polarity, away from its frozen orientation, becomes faster than the restoring dynamics of the displacement. As a result, the later permanently chase the polarity, and the stable rotation sets in. All fixed points become unstable at once; and a limit cycle of radius $R=\left(\pi/\omega_0^2\right)^{1/2}$ and oscillation frequency, $\Omega = \omega_{0} \sqrt{\pi - \omega_0^2}$, identical to the one obtained from the mean field approach, branches off continuously (Fig.~\ref{fig:theory}-c-d). Note that the oscillating dynamics does not arise from a Hopf bifurcation, but from the global bifurcation of a continuous set of fixed points into a limit circle.

Understanding how the nonlinear coupling of $N$ such elementary units leads to the selection mechanism of the actuated modes requires a more involved analysis.
One sees from Eqs.~(\ref{eq_dimensionless_noiseless_braket}) that any configuration  $\left(  \left\{ \boldsymbol{\hat{n}}_{i} \right\}, \left\{ \boldsymbol{u}_{i}  = \pi  \mathbb{M}_{ij}^{-1} \boldsymbol{\hat{n}}_{j} \right\} \right)$ is a fixed point of the dynamics.
In contrast with the one particle case, the linear destabilization threshold $\pi_c( \left\{ \boldsymbol{\hat{n}}_{i} \right\})$ depends on the fixed point configuration (Supplementary Information section 4.4). These thresholds are bounded $\pi_c^{\text{min}}\le \pi_c( \left\{ \boldsymbol{\hat{n}}_{i} \right\})   \le \pi_c^{\text{max}}$. A first fixed point becomes unstable for $\pi = \pi_c^{\text{min}} =\omega_\textrm{min}^2$, where  $\omega_\textrm{min}^2$ is the smallest eigenvalue of the dynamical matrix $\mathbb{M}$ (Supplementary Information section 4.5) and an upper bound for $\pi_c^{\text{max}}$ (Supplementary Information section 4.6) reads :
\begin{equation}
\mathcal \pi^{\text{upp}} =\min_{\{i,j\}} \left(\frac{\omega^2_i +\omega^2_j}{c(| \boldsymbol{\varphi}_{i} \rangle,| \boldsymbol{\varphi}_{j} \rangle)}\right),
\label{eq:upperbound}
\end{equation}
The function $c(\bullet,\bullet)$ only depends on the eigenvectors of $\mathbb{M}$, 
$\{ |\boldsymbol{\varphi}_{i} \rangle\}$. It is bounded between $0$ and $1$ and is maximal when the modes $|\boldsymbol{\varphi}_{i} \rangle$ and $|\boldsymbol{\varphi}_{j} \rangle$ are extended and locally orthogonal. More specifically, the pair of modes which dominates the dynamics, $\{ |\boldsymbol{\varphi}_{1} \rangle\, , \,|\boldsymbol{\varphi}_{2} \rangle\}$ for the triangular and $\{ |\boldsymbol{\varphi}_{4} \rangle\, , \,|\boldsymbol{\varphi}_{5} \rangle\}$ for the kagome lattice, are precisely the ones that optimize the bound. The construction of this bound is very general. It demonstrates that for any stable elastic structure, there is a strength of the elasto-active feedback above which the frozen dynamics is unstable and a dynamical regime must set in. It also captures the mode selection in the strongly condensed regime. Our findings about the linear stability of the fixed points for the triangular and kagome lattices are summarized in Extended Data Fig.1.

That some fixed points loose stability does not imply that collective actuation sets in: from these fixed points, the system can either slide to a neighboring stable fixed point or condensate on some dynamical attractor.
An exact theory to describe this condensation process is still missing in the general case, but can be formulated in the simpler, yet rich enough, case of a linear chain of $N$ active particles, fixed at both ends.
In the zero rest length limit of the springs, the rotational invariance of the dynamical equations ensures that the eigenvalues and eigenvectors of the dynamical matrix are degenerated by pairs of locally orthogonal modes. In such a situation, the limit cycle solution, corresponding to the collective actuation regime, is found analytically (Supplementary Information, section 9.2), leading to a precise transition diagram, illustrated here for $N = 7$ (Fig.\ref{fig:theory}-e).
When $\pi$ exceeds the threshold value $\pi_{CA}$, the limit cycle is stable. We have checked that it is the only stable periodic solution, up to $N = 20$.
If $\pi_{CA} \leq \pi \leq \pi_{c}^{\text{max}}$, it coexists with an infinite number of stable fixed points. The evolution of their respective basins of attraction can be largely understood by studying the $N=2$ case (Supplementary Information section 9.3.1, Fig. S9).
For $\pi<\pi_{CA}$, the dynamics leave the limit cycle and become heterogeneous. 

The physical origin of the spatial coexistence lies in the normalization constraint of the polarity field, $\lVert\boldsymbol{\hat{n}_i}\rVert=~1$, which translates into a strong constraint over the radii of rotation, namely $R_i\geq 1$ (Supplementary Information, section 9.2.3 and 9.2.4). Whenever $R_i$ becomes unity the polarity and displacement vectors become parallel, freezing the dynamics. 
The spatial distribution of the $R_i$ is set by that of the modes selected by the collective actuation, with particles closer to the boundaries having typically a smaller radius of rotation than the ones at the center. The threshold value $\pi_{CA}$, below which the dynamics leave the limit cycle, is precisely met when the particles at the boundary reach a radius of rotation $R=1$. For $\pi<\pi_{CA}$,
the competition between outer particles, which want to freeze, and the central particles, which want to cycle, leads to the sequential layer by layer de-actuation, illustrated in Fig.~\ref{fig:theory}-f for a linear chain with  $N=7$ and observed experimentally and numerically. The threshold value $\pi_{FD}$ is reached when, eventually, the remaining particles at the center freeze and the system discontinuously falls into the frozen disordered state.

Altogether we have shown that (i) the chiral phase takes its origin in the one-particle dynamics; (ii) the selection of modes results from the nonlinear elasto-active feedback, which connects the linear destabilization of the fixed points to the spatial extension and local orthogonality of pairs of modes; (iii) the spatial coexistence emerges from the normalization constraint of the polarity fields.

The role of noise, which was not considered in the numerical and theoretical analysis, is another matter of interest. In the frozen disordered regime, the noise is responsible for the angular diffusion of the polarities amongst the fixed points. In the collective actuation regime the noise level present in the experiment does not alter significantly the dynamics. Numerical simulations confirm that there is a sharp transition at a finite noise amplitude $D_c$, below which collective actuation is sustained (Extended Data Fig.2-a). For noise amplitude much lower than $D_c$, the noise merely reduces the mean angular frequency $\Omega$ (Extended Data Fig.2-b) . Closer to the transition, the noise allows for stochastic inversions of the direction of rotation, restoring the chiral symmetry. (Extended Data Fig.2-c).

Finally, it has been shown very recently, that non symmetrical interactions, together with non conservative dynamics, generically lead to chiral phases~\supercite{fruchart2021non}. 
Here, the polarity and displacement vectors of a single particle do experience non symmetrical interactions, the phase of the displacement chasing that of the polarity. Mapping the coarse grained equations to the most general equations one can write for rotationally symmetric vectorial order parameters~\supercite{fruchart2021non}, we find that the macroscopic displacement and polarity fields also couple non-symmetrically (Methods). This  suggests a possible description of the transition to collective actuation in terms of non reciprocal phase transitions.  If this were to be confirmed by a more involved analysis of the large scale dynamics, it would motivate the study of the disordered to chiral phase transition in active solids, which has not been addressed theoretically yet. In the same vein, one may ask whether the coarse-grained system shall obey standard or odd elasticity~\supercite{scheibner2020odd}. 

More generally, the recent miniaturization of autonomous active units~\supercite{miskin2020electronically} opens the path towards the extension of our design principle to the scale of material science. In this context, extending the relation between the structural design of active materials -- including the geometry and topology of the lattice, the presence of disorder, the inclusion of doping agents -- and their spontaneous actuation offers a wide range of perspectives.

\small
\printbibliography

\textbf{Supplementary Information} is available in the online version of the paper.

\textbf{Acknowledgements} P.B. was supported by a PhD grant from ED564 “Physique en Ile de France”. D.S. was supported by a Chateaubriand fellowship. G.D. acknowledges support from Fondecyt Grant No. 1210656 and ANID - Millenium Science Initiative Program - Code NCN17$\_$092. We are grateful to Michel Fruchart and Vincenzo Vitelli for fruitful discussions regarding non-reciprocity in elastic materials.

\textbf{Author Contributions} O.D., C.C. and G.D. conceived the project. P.B. and D.S. performed the experiments. P.B., D.S., and O.D. analyzed the experimental results.  P.B., V.D., O.D., C.H. and G.D. worked out the theory. All contributed to the writing of the manuscript. 

\textbf{Author Institutions} $^{1}$\textit{Gulliver UMR CNRS 7083, ESPCI Paris, Université PSL, 75005 Paris, France.} $^{2}$\textit{School of Physics and Astronomy, Tel-Aviv University, Tel Aviv 69978, Israel.} $^{3}$\textit{Instituto de F\'isica, Pontificia Universidad Cat\'olica de Chile, Casilla 306, Santiago, Chile.} $^{4}$\textit{ANID - Millenium Nucleus of Soft Smart Mechanical Metamaterials, Santiago, Chile.} $^{5}$\textit{Van der Waals-Zeeman Institute, Institute of Physics, Universiteit van Amsterdam, Science Park 904, 1098 XH Amsterdam, the Netherlands.} $^{6}$\textit{Univ Lyon, ENSL, CNRS, Laboratoire de Physique, F-69342 Lyon, France.}

\newpage
\onecolumn
\begin{center}
\textbf{\large Methods $\&$ Extended data: Selective and Collective Actuation in Active Solids} \\
\vspace{0.4cm}
P. Baconnier$^{1}$ , D. Shohat$^{1,2}$, C. Hernández López$^{3,4}$, C. Coulais$^{5}$, V. Démery$^{1,6}$, G. Düring$^{3,4}$, O. Dauchot$^{1}$ \\
\vspace{0.2cm}
\textit{$^{1}$Gulliver UMR CNRS 7083, ESPCI Paris, Université PSL, 75005 Paris, France.} \\
\textit{$^{2}$School of Physics and Astronomy, Tel-Aviv University, Tel Aviv 69978, Israel.} \\
\textit{$^{3}$Instituto de Física, Pontificia Universidad Católica de Chile, Casilla 306, Santiago, Chile.} \\
\textit{$^{4}$ANID - Millenium Nucleus of Soft Smart Mechanical Metamaterials, Santiago, Chile.} \\
\textit{$^{5}$Van der Waals-Zeeman Institute, Institute of Physics, Universiteit van Amsterdam, Science Park 904, 1098 XH
Amsterdam, the Netherlands.} \\
\textit{$^{6}$Univ Lyon, ENSL, CNRS, Laboratoire de Physique, F-69342 Lyon, France.}
\end{center}
\vspace{0.5cm}
\setcounter{equation}{0}
\setcounter{figure}{0}
\setcounter{table}{0}
\setcounter{page}{1}
\makeatletter
\renewcommand{\theequation}{M\arabic{equation}}
\renewcommand{\thefigure}{M\arabic{figure}}
\renewcommand{\thetable}{M\Roman{table}}
\renewcommand{\theHtable}{M\Roman{table}}   

\setcounter{section}{0} 
\renewcommand\thesection{\arabic{section}} 
\renewcommand{\theHsection}{\arabic{section}} 
\makeatletter
\renewcommand{\@seccntformat}[1]{\csname the#1\endcsname.\space}
\renewcommand{\appendixname}{} 
\def\p@section{} 
\def\p@subsection{} 
\def\thesection{\arabic{section}}
\renewcommand\thesubsection{\thesection.\arabic{subsection}} 
\renewcommand{\@seccntformat}[1]{\csname the#1\endcsname.\space} 
\makeatother



\paragraph{Data availability.} The experimental data generated in this study have been deposited in the Zenodo database under accession code \url{https://doi.org/10.5281/zenodo.6653906}. The numerical data that support the findings are available from the corresponding authors upon reasonable request.

\paragraph{Code availability.} All the codes supporting this study have been deposited in the Zenodo database under accession code \url{https://doi.org/10.5281/zenodo.6653906}.

\section{Methods} \label{section:methods}

\paragraph{Experiments.}
We use commercial HEXBUG nano$\circledR$ Nitro, highlighted in Fig. 1 of the main text. Their body length is 4 cm. The energy supply comes from a 1.5 V AG13/LR44 battery inside the robot that makes it able to vibrate and, as a consequence, to move thanks to its flexible curved legs. We embed these bugs in 3d printed cylindrical structures of 5 cm internal diameter, 3 mm thick, and 14 mm height (as the hexbugs themselves). These 3d printed annulus have 6 overhangs, that we pierced with a milling machine in order to hold the edges of the springs. Moreover, we set a thin PP plastic film on the top of the annulus to restrict the vertical motion of the hexbugs body, that we fix using commercial glue and a 3d printed 1 mm thick ring. The obtained elementary active elastic unit is shown in Fig. 1-b of the main text. These elementary components are connected by coil springs. We use two kind of springs: rigid springs RSC13 ($k \approx 120$ N/m, $l_0 \approx 3$ cm, external diameter $5$ mm) manufactured by Ets. Jean CHAPUIS; only for the rigid body motion experiments shown in Movies 2 and 3; and soft springs ($k \simeq 1$ N/m, $l_{0} \simeq 8$ cm, external diameter 5 mm) manufactured by Schweizer Federntechnik; for all the experiments discussed in the main text. We tune the springs stiffnesses by varying their length, the stiffness $k$ of a coil spring being inversely proportionnal to $l_{0}$, all other parameters held constants. For the triangular lattice, kagome lattice and single particle systems, the springs lengths are respectively $\{ 7.4, 6.6, 5.8, 5.0, 4.4, 3.6, 2.8 \}$ cm, $\{ 8.3, 7.3, 6.3, 5.5, 4.5, 3.5, 2.9, 2.2, 1.5 \}$ cm and $\{ 7.4, 6.7, 6.0, 5.3, 4.6, 3.9, 3.3, 2.6 \}$ cm. The transitions measured in Figs. 3 and 4 of the main text were performed with constant tension of the spring's lattices. We define the tension $\alpha$ as the ratio $l_{\text{s}}/l_{0}$, where $l_{0}$ is the unstressed length of the springs, and $l_{\text{s}}$ is the length of the springs in the stressed reference configuration. In the triangular lattice (resp. kagome lattice; resp. single particle lattice), springs are elongated by a factor $\alpha = 1.27$ (resp. $\alpha = 1.02$; resp. $\alpha = 1.29$). The XY table experiments of the Supplementary Information section 2.2.2 were done with a translating stage ModuFlat P30.

\paragraph{Data analysis.}
The dynamics of the active elastic structures are captured with a PixeLink PL-D734MU camera, at 40 frames per second, unless otherwise stated. The obtained movies are processed with Python. The annulus barycenters defines the active units' positions $\boldsymbol{r}_{i}(t)$. They are detected using a Hough circle transform restricted to the annulus radius as obtained from the movies. The hexbugs body directions are obtained by computing the first moments of the image restricted to each annulus, thresholded such that only the hexbugs bodies are detected. The proper orientations are then determined by integrating $\delta \boldsymbol{r}_{i}(t) . \boldsymbol{\hat{n}}_{i}(t)$ along the dynamics, where $\delta \boldsymbol{r}_{i}(t) = \boldsymbol{r}_{i}(t+\delta t) - \boldsymbol{r}_{i}(t)$, and determining its sign for each hexbug. The time average oscillation frequency $\omega_{i}$ of an active unit is measured by fitting the long-time behavior of $\langle \theta_{i}(t+\tau) - \theta_{i}(t) \rangle_{t}(\tau)$ with linear power law, where $\theta_{i}$ refers to the orientation of particle $i$. The collective oscillation frequency $\Omega = \frac{1}{N}\sum_i \langle\omega_i\rangle_t$.

\paragraph{Error estimates.}
Denoting the error on the detected positions $\Delta u$, and the annulus internal diameter $d$, we have $\Delta u/d \simeq 1/20$ (because the Hough circle transform does not find perfectly the barycenters). The angles are detected modulo a typical error of $\Delta \theta = 3^{\circ}$. Using the error estimates on the detected positions an orientations, we can also have the error estimates on the displacement/polarity field projections on the normal modes. They right $\Delta \left[ \langle \boldsymbol{\varphi}_k | \boldsymbol{\hat{n}} \rangle / \sqrt{N} \right] \simeq \Delta \theta \sqrt{Q_k}$ and $\Delta \left[ \langle \boldsymbol{\varphi}_k | \boldsymbol{u} \rangle / \sqrt{N} \right] \simeq \Delta u \sqrt{Q_k}$ where $Q_k = \left( \sum_i | \boldsymbol{\varphi}_k^i | \right)^2 / N$ is mode $k$'s participation ratio, bounded between $0$ and $1$. As an example, for a plane-wave mode $Q_k \simeq 0.67$, and we find $\Delta \left[ \langle \boldsymbol{\varphi}_k | \boldsymbol{\hat{n}} \rangle^2 / N \right] = 2 \frac{\langle \boldsymbol{\varphi}_k | \boldsymbol{\hat{n}} \rangle}{\sqrt{N}} \Delta \left[ \frac{\langle \boldsymbol{\varphi}_k | \boldsymbol{\hat{n}} \rangle}{\sqrt{N}} \right] \leq 2 \Delta \theta \sqrt{Q_k} \simeq 0.05$: the fraction of active force injected in a given mode is thus given modulo a typical $5\%$ error due to tracking inaccuracies.

\paragraph{Numerical simulations.}
We simulate the noiseless Eqs.~(1) from the main text with a Runge-Kutta method. The numerical curves shown in Figs. 2-a and 4-a were obtained from typically 20 random initial conditions for the polarities, and initial positions at their reference configuration, for each value of $\pi$. Temporal averaging takes place once a steady state is reached. The annealing simulations of Fig. 3 are initiated in the collective actuation regime. The value of $\pi$ is varied discontinuously, by steps of $0.1$, and data are time averaged once a new steady state is reached.  The values of $\pi_{c}^{\text{max}}$ were found through a numerical optimization process of $\pi_{c}( |\boldsymbol{\hat{n}} \rangle )$ (Supplementary Information section 4).

\section{Extended Data} \label{section:ext_data}

\paragraph{Fixed points stability analysis for the experimental structures} \label{section:fp_stab}
A given fixed point, i.e. any configuration of the polarity field, becomes unstable for $\pi > \pi_{c}$, where $\pi_{c}$ is given by Eq. (S36).
This stability threshold can be evaluated numerically for both experimental structures. We do so for one million configurations of the polarity field, sampled by drawing randomly and independently the orientations of each active units according to a uniform distribution in $ [0, 2\pi[$.
The results are shown in (Extended Data Figs. 1-b and d), and highlight the fraction of configurations that remain stable for a given value of $\pi$.

As expected we find no configurations destabilizing for $\pi < \pi_c^{\text{min}} = \omega_{\text{min}}^{2}$, where the first configurations, locally orthogonal to the lowest energy mode, become unstable. The upper bound obtained by evaluating Eq.~(S45) for all pairs of mode is not very sharp. However, we find that the pair of mode achieving the smallest bound coincides with the pair of mode on which the condensation takes place (Extended Data Figs. 1-a and c).

\paragraph{Large $N$ limit} \label{section:large_N}
We perform numerical simulations of  Eqs.~(1) of the main text for triangular, respectively kagome, lattices, increasing the number of active units, up to $N = 1141$, respectively $N=930$, while keeping $L$ constant, with a size $L$, such that the lowest energy modes have unity squared eigenfrequencies. The fraction of nodes activated in the center of the system, $f_{CA}$, is defined as the ratio of the number of active units rotating at least 90$\%$ of the maximum rotation frequency over the total number of active units. The collective actuation oscillation frequency $\Omega_{CA}$ is defined as the average of the individual rotation rates inside the collectively actuated region.
We quantify the condensation level by computing the averaged condensation fraction in the symmetry class $(1/2)$; $\lambda_{1/2}= \frac{1}{N_{1/2}} \sum_{i \in \left[1/2 \right]} \langle \langle \boldsymbol{\varphi}_{i} | \boldsymbol{\hat{n}} \rangle^2 \rangle_{t}$, where $\left[ 1/2 \right]$ refers to the modes belonging to the $(1/2)$ symmetry class, and $N_{1/2}$ is the number of modes involved. This quantity, bounded between $0$ and $1$, is the fraction of the dynamics condensed on the selected actuated modes (Fig. 3-d of the main text).

\paragraph{Role of the noise} \label{section:noise}
In the noiseless and overdamped framework, the collective actuation has been identified as an ordered dynamical regime, corresponding to a limit cycle. In the presence of noise, Eqs. (1) of the main text turn into coupled non-linear SDEs, that we simulate using an Euler method with fixed time step $\delta t=10^{-3}$, by adding a noisy angular contribution to each polarity, drawn independently from a gaussian distribution of zero mean and variance $2D\delta t$. We focus on the case of the triangular lattice.

We compute the collective absolute oscillation frequency, $|\Omega| = \frac{1}{N}\sum_i \langle|\omega_i|\rangle_t$, where $\langle|\omega_i|\rangle_t$ is measured by fitting the long-time behavior of $\langle | \theta_{i}(t+\tau) - \theta_{i}(t) | \rangle_{t}(\tau)$ with linear power law.
As seen from Extended Data Fig. 2-a there is a sharp transition at finite noise amplitude $D_c(\pi)$, below which collective actuation subsists.  For noise level much lower than $D_c(\pi)$, the presence of noise only reduces the  averaged angular frequency, (Extended Data Fig. 2-b).  Closer to the transition, the noise also allows for stochastic inversions of the direction of rotation, restoring the chiral symmetry.
(Extended Data Fig. 2-c)

\paragraph{Relation to non-reciprocal matter} \label{section:non_reciprocity}
It was recently shown that systems composed of microscopic degrees of freedom experiencing non symmetrical interactions are prone to develop chiral phases via a specific kind of transitions, which the authors called non-reciprocal~\cite{fruchart2021non}.

Our model system, composed of active units connected by elastic springs, is a priori a good candidate for the study of this physics. As a matter of fact its dynamics results from the coupling of $N$ polarity vectors $\boldsymbol{\hat{n}}_i$ and $N$ displacement vectors $\boldsymbol{u}_i$, where one can recognize two abstract species $A$ and $B$.
We first recall Eqs.~(S49), which govern the dynamics of a single particle in an harmonic potential:
\begin{subequations} \label{eq:chiral_parabola}
\renewcommand{\theequation}{\theparentequation.\arabic{equation}}
\begin{align}
 \dot{R}  &= \pi \cos(\theta - \varphi) - \omega_{0}^{2} R \label{eq1:chiral_parabola} \\
\dot{\varphi}  &= \frac{\pi}{R} \sin(\theta - \varphi) \label{eq2:chiral_parabola} \\
\dot{\theta} &= -\omega_{0}^{2}R \sin(\varphi - \theta) \label{eq3:chiral_parabola}
\end{align}
\end{subequations}
where $\varphi$ (resp. $\theta$) represents the angle of the displacement (resp. polarity) vector with respect to the $x$-axis, and where $R$ is the norm of the displacement vector. One sees that the phases $\varphi$ and $\theta$ are coupled non-symetrically, as $J_{\boldsymbol{\hat{n}} \rightarrow \boldsymbol{u}} = \pi/R \neq J_{\boldsymbol{u} \rightarrow \boldsymbol{\hat{n}}} = -\omega_{0}^{2}R$ in such a way that for $\pi>\pi_c$, the phase of the displacement vector chases that of the polarity.


It is therefore likely that the macroscopic dynamics for the order parameters associated to the mean polarization and mean displacement experience the kind of transition towards a chiral phase described in~\cite{fruchart2021non}. More precisely the coarse grained dynamics map on the general equations describing the dynamics of two vector order parameters ${\bf v}_a(t, x)$, which serve as the starting point in~\cite{fruchart2021non} :
\begin{equation} \label{eq:nonreciprocfields}
\partial_t {\bf v}_a = {\mathbb A}_{ab} {\bf v}_b + {\mathbb B}_{abcd} \left({\bf v}_b \cdot {\bf v}_c\right) {\bf v}_d+ \mathcal{O}(\nabla).
\end{equation}
They read
\begin{subequations}
\renewcommand{\theequation}{\theparentequation.\arabic{equation}}
\begin{align}
 \partial_t \boldsymbol{U} &= \pi \boldsymbol{m} + \mathcal{O}(\nabla),\\
 \partial_t \boldsymbol{m} &= \pi \ndfrac{1-\boldsymbol{m}^2}{2}\boldsymbol{m} + \mathcal{O}(\nabla).
\end{align}
\end{subequations}
where the non-zero coefficients, ${\mathbb A}_{um}=\pi$, ${\mathbb A}_{mm}=\pi/2$ and ${\mathbb B}_{mmmm}=-\pi/2$ are clearly non symmetric, opening the path to a so-called non-reciprocal phase transition.

Note, however, that the transition that we observe and study is not that from an ordered (anti-)aligned phase to the chiral one. It is that from the disordered phase to the chiral one. To our knowledge this type of transition has not been, yet, investigated at the theoretical level.

\begin{figure}[h!]
\captionsetup{format=plain}
\centering
\begin{tikzpicture}

\node[] at (0.0,0.0) {\includegraphics[width=\textwidth]{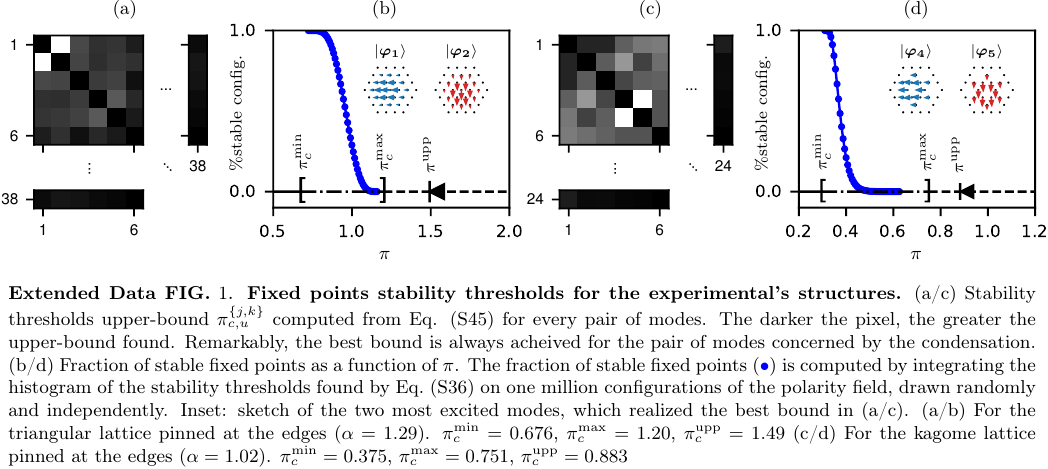}};

\end{tikzpicture}
\label{ED1}
\end{figure}

\begin{figure}[h!]
\captionsetup{format=plain}
\centering
\begin{tikzpicture}

\node[] at (0.0,0.0) {\includegraphics[width=\textwidth]{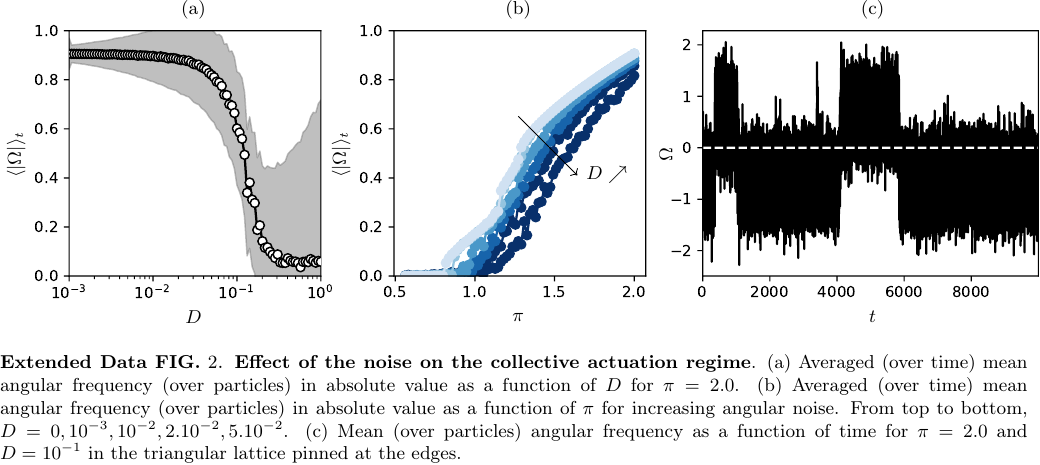}};

\end{tikzpicture}
\label{ED2}
\end{figure}

\newpage
\onecolumn
\begin{center}
\textbf{\large Supplementary Information: Selective and Collective Actuation in Active Solids} \\
\vspace{0.4cm}
P. Baconnier$^{1}$ , D. Shohat$^{1,2}$, C. Hernández López$^{3,4}$, C. Coulais$^{5}$, V. Démery$^{1,6}$, G. Düring$^{3,4}$, O. Dauchot$^{1}$ \\
\vspace{0.2cm}
\textit{$^{1}$Gulliver UMR CNRS 7083, ESPCI Paris, Université PSL, 75005 Paris, France.} \\
\textit{$^{2}$School of Physics and Astronomy, Tel-Aviv University, Tel Aviv 69978, Israel.} \\
\textit{$^{3}$Instituto de Física, Pontificia Universidad Católica de Chile, Casilla 306, Santiago, Chile.} \\
\textit{$^{4}$ANID - Millenium Nucleus of Soft Smart Mechanical Metamaterials, Santiago, Chile.} \\
\textit{$^{5}$Van der Waals-Zeeman Institute, Institute of Physics, Universiteit van Amsterdam, Science Park 904, 1098 XH
Amsterdam, the Netherlands.} \\
\textit{$^{6}$Univ Lyon, ENSL, CNRS, Laboratoire de Physique, F-69342 Lyon, France.}
\end{center}
\vspace{0.5cm}
\setcounter{equation}{0}
\setcounter{figure}{0}
\setcounter{table}{0}
\setcounter{page}{1}
\makeatletter
\renewcommand{\theequation}{S\arabic{equation}}
\renewcommand{\thefigure}{S\arabic{figure}}
\renewcommand{\thetable}{S\Roman{table}}
\renewcommand{\theHtable}{S\Roman{table}}   

\setcounter{section}{0} 
\renewcommand\thesection{\arabic{section}} 
\renewcommand{\theHsection}{\arabic{section}} 
\makeatletter
\renewcommand{\@seccntformat}[1]{\csname the#1\endcsname.\space}
\renewcommand{\appendixname}{} 
\def\p@section{} 
\def\p@subsection{} 
\def\thesection{\arabic{section}}
\renewcommand\thesubsection{\thesection.\arabic{subsection}} 
\renewcommand{\@seccntformat}[1]{\csname the#1\endcsname.\space} 
\makeatother

\normalsize
Here, we provide supplementary informations including movies that illustrate our main findings, details about experimental methods and data analysis and a comprehensive description of the theoretical model outlined in the main text, which accounts for the noiseless and over-damped dynamics of self-propelled  polar particles embedded in an elastic network. For the sake of clarity, this document is written in a self-consistent fashion, all the notations and definitions of the main text are explicitly re-defined.

\section{List of supplementary movies} \label{section:supp_videos}
The reader will find here a number of movies illustrating the frozen and collective actuation dynamics experimentally observed in the case of the single particle, the triangular lattice and the kagome lattice. We have also included one movie obtained from numerical simulations of the annealing dynamics from large values of $\pi$ to low values of $\pi$ in the case of the triangular lattice ($N=19$), and movie from numerical simulation of large triangular and kagome lattices which demonstrate that our observations are not limited to small systems. The first video illustrates the self-alignment of an active unit toward its direction of motion.

\begin{itemize}
 \item SI Movie 1: Alignment experiment. We impose a square motion to an active building block, and look at the response of the polarity. At each side of the square, the polarity aligns toward the new velocity vector. Acquired at 40 fps, displayed in real time.
 \item SI Movie 2: Experiment: translation in a triangular lattice ($N = 37$) with free boundary condition. Acquired at 30 fps, displayed in real time.
 \item SI Movie 3: Experiment: rotation dynamics in a triangular lattice ($N = 37$) with free boundary condition. Acquired at 30 fps, displayed in real time.
 \item SI Movie 4: Experiment: Frozen-Disordered dynamics in the triangular lattice ($N = 19$ active units). Acquired at 40 fps, displayed in real time.
 \item SI Movie 5: Experiment: Collective Actuation regime in the triangular lattice ($N = 19$ active units). Acquired at 40 fps, displayed in real time.
 \item SI Movie 6: Experiment: Heterogeneous regime in the triangular lattice ($N = 19$ active units). Acquired at 40 fps, displayed in real time.
 \item SI Movie 7: Experiment: Frozen-Disordered dynamics in the kagome lattice ($N = 12$ active units). Acquired at 40 fps, displayed in real time.
 \item SI Movie 8: Experiment: Collective Actuation dynamics in the kagome lattice ($N = 12$ active units). Acquired at 40 fps, displayed in real time.
 \item SI Movie 9: Experiment: Frozen dynamics in the single particle system. Acquired at 40 fps, displayed in real time.
 \item SI Movie 10: Experiment: Spontaneous oscillations in the single particle system. Acquired at 40 fps, displayed in real time.
 \item SI Movie 11: Numerical simulation : Annealing in $\pi$ in the triangular lattice ($N = 19$ active units), with the same tension as in the experiment of SI Video 6. The elasto-active coupling $\pi$ is decreased from the CA regime until the system finds a fixed point. Individual polarities are represented by a black arrow, springs are color-coded by stress state; an elongated spring turns red, while a compressed one turns blue.
 \item SI Movie 12: Numerical simulation of a large triangular lattice ($N = 1141$ active units), sized such that its lower energy modes have squared eigenfrequencies equal to unity. The system is initialized with zero displacement in every nodes and random initial condition for the polarities orientations, and $\pi = 2.0$. Polarities are shown as arrows colored by their orientations. Springs are represented in gray. The system quickly finds a collective actuation regime.
 \item SI Movie 13: Numerical simulation of a large kagome lattice ($N = 930$ active units), sized such that its lower energy modes have squared eigenfrequencies equal to unity. The system is initialized with zero displacement in every nodes and random initial condition for the polarities orientations, and $\pi = 10.0$. Polarities are shown as arrows colored by their orientations. Springs are represented in gray. The system quickly finds a collective actuation regime.
\end{itemize}

\section{Model} \label{section:model}
\subsection{General equations}

We consider $N$ active \emph{self-aligning} polar particles such as those described in \cite{weber2013long, lam2015self, dauchot2019dynamics}, connected by linear springs of stiffness $k$ and unstressed length $l_{0}$. For each active component, the activity takes the form of a force $\boldsymbol{F}\!_{a}= F_{0}\boldsymbol{\hat{n}}_{i}$ along the polarity $\boldsymbol{\hat{n}}_{i}$ of the particle. The equations describing the dynamics of such a system are
\begin{subequations} \label{eq_dimension}
\renewcommand{\theequation}{\theparentequation.\arabic{equation}}
\begin{align}
 m \frac{d\boldsymbol{v}_{i}}{dt} &= F_{0} \boldsymbol{\hat{n}}_{i} - \gamma \boldsymbol{v}_{i} + \sum_{j \in \partial i} k \left( |\boldsymbol{r}_{i} - \boldsymbol{r}_{j}| - l_{0} \right)\hat{\boldsymbol{e}}_{ij} \label{eq1:dimension} \\
 \tau \frac{d\boldsymbol{\hat{n}}_{i}}{dt} &= \zeta (\boldsymbol{\hat{n}}_{i} \times \boldsymbol{v}_{i}) \times \boldsymbol{\hat{n}}_{i} + \sqrt{2\alpha} \xi \boldsymbol{\hat{n}}_{i}^{\perp} \label{eq2:dimension}
\end{align}
\end{subequations}
where $m$ is the mass of the active particles, $\gamma$ the friction coefficient, $k$ the stiffness of the spring, and $\hat{\boldsymbol{e}}_{ij}$ the unit vector from $i$ to $j$. In the absence of confinement, the particles thus move with a cruise velocity $v_{0} = F_{0}/ \gamma$. The orientation dynamics Eq. (\ref{eq2:dimension}) contains the key ingredient, specific to the model, namely the presence of a self-aligning torque of the orientation $\boldsymbol{\hat{n}}_{i}$ towards the velocity $\boldsymbol{v}_{i}$. This torque originates from the fact that the dissipative force is not symmetric with respect to the propulsion direction $\boldsymbol{\hat{n}}_{i}$ when $\boldsymbol{v}_{i}$ is not aligned with $\boldsymbol{\hat{n}}_{i}$. This ingredient was shown to be at the root of frozen to orbiting dynamics for a single hexbug in a harmonic trap~\cite{dauchot2019dynamics}, as well as of the emergence of collective motion in a system of vibrated polar discs~\cite{weber2013long, lam2015self}. Finally, the orientation dynamics contains a delta-correlated gaussian noise $\xi(t)$ with zero mean and variance $\langle \xi (t) \xi (t') \rangle = \delta(t - t') ; \ \alpha / \tau^{2}$ is the rotational diffusion coefficient.
\newcommand{\A}{(0,0) circle (0.6)}
\newcommand{\B}{(2.0,1.6) circle (0.6)}
\newcommand{\C}{(-1.8,1.8) circle (0.6)}
\newcommand{\D}{(-0.6,-2.2) circle (0.6)}
\newcommand{\Ab}{(0,0) circle (0.03)}
\newcommand{\Bb}{(2.0,1.6) circle (0.03)}
\newcommand{\Cb}{(-1.8,1.8) circle (0.03)}
\newcommand{\Db}{(-0.6,-2.2) circle (0.03)}
\begin{figure}[h]
\centering
\begin{tikzpicture}
 \fill[cyan] \A;
 \fill[cyan] \B;
 \fill[cyan] \C;
 \fill[cyan] \D;
 \fill[black] \Ab;
 \fill[black] \Bb;
 \fill[black] \Cb;
 \fill[black] \Db;
 \draw (0,0) -- (2.0,1.6);
 \draw (0,0) -- (-1.8,1.8);
 \draw (0,0) -- (-0.6,-2.2);
 \draw[-{Latex[red]}, color=red, thick] (0,0) -- (-0.25,2.0);
 \draw[-{Latex[black]}] (0,0) -- (-0.134,1.06);
 \draw[-{Latex[black]}] (0,0) -- (0.86,0.7);
 \node [anchor=west] at (0.0,-0.25) {\small $\boldsymbol{r}_{i} = (x_{i},y_{i})$};
 \node [anchor=west] at (1.9,1.8) {\small $\boldsymbol{r}_{j} = (x_{j},y_{j})$};
 \node [anchor=west] at (0.8,0.4) {\small $\hat{\boldsymbol{e}}_{ij}$};
 \node [anchor=west] at (-0.060,1.0) {\small $\boldsymbol{\hat{n}}_{i}$};
 \node [anchor=west] at (-0.10,2.05) {\color{red} \normalsize $\boldsymbol{F}\!_{a}$};
\end{tikzpicture}
\caption{Notations}
\label{Notations_model}
\end{figure}

Rescaling length by $r_{0} = l_{0}$ and time by $t_{0} = \gamma /k$, the characteristic time of a damped spring, the dimensionless equations of motion read
\begin{subequations} \label{eq_dimensionless}
\renewcommand{\theequation}{\theparentequation.\arabic{equation}}
\begin{align}
 \tau_{v} \frac{d\boldsymbol{v}_{i}}{dt} &= \tilde{F}_{0}\boldsymbol{\hat{n}}_{i} -  \boldsymbol{v}_{i} + \sum_{j \in \partial i} \left( |\boldsymbol{r}_{i} - \boldsymbol{r}_{j}| - 1 \right)\hat{\boldsymbol{e}}_{ij} \label{eq1:dimensionless} \\
 \tau_{n} \frac{d\boldsymbol{\hat{n}}_{i}}{dt} &= (\boldsymbol{\hat{n}}_{i} \times \boldsymbol{v}_{i}) \times \boldsymbol{\hat{n}}_{i} + \sqrt{2D} \xi \boldsymbol{\hat{n}}_{i}^{\perp} \label{eq2:dimensionless}
\end{align}
\end{subequations}

with four parameters, $\tau_{v} = mk / \gamma ^{2}$, $\tau_{n} = \tau/(\zeta l_{0})$, $\tilde{F}_{0} = F_{0}/kl_{0}$ and $D = \alpha \gamma / k(\zeta l_{0})^{2}$. Note that $\tau_n = l_a/l_0$ and $\tilde F_0 = l_e/l_0$, where $l_a = \tau / \zeta$ is the alignment length, that is the length over which the particle must move to align its orientation onto its displacement, and $l_e=F_0/k$ is the elasto-active length, which is the distance that the active force can drive away the particle from its equilibrium position, given the elastic restoring force.

In the following, as well as in the main text, we consider the set of Eqs. (\ref{eq1:dimensionless}) and (\ref{eq2:dimensionless}) while taking the over-damped limit ($\tau_{v} \rightarrow 0$) in the position dynamics Eq. (\ref{eq1:dimensionless}) and the noiseless limit of the polarity dynamics Eq. (\ref{eq2:dimensionless}). This leads us to the equations:
\begin{subequations} \label{eq_dimensionless_overdamped}
\renewcommand{\theequation}{\theparentequation.\arabic{equation}}
\begin{align}
 \frac{d\boldsymbol{r}_{i}}{dt} &= \tilde{F}_{0}\boldsymbol{\hat{n}}_{i} + \boldsymbol{F}_{i}^{\textrm{el}} \label{eq1:dimensionless_overdamped} \\
 \tau_{n} \frac{d\boldsymbol{\hat{n}}_{i}}{dt} &= (\boldsymbol{\hat{n}}_{i} \times \boldsymbol{F}_{i}^{\textrm{el}}) \times \boldsymbol{\hat{n}}_{i} \label{eq2:dimensionless_overdamped}
\end{align}
\end{subequations}
with $\boldsymbol{F}_{i}^{\textrm{el}} = \sum_{j \in \partial i} \delta l_{ij} \hat{\boldsymbol{e}}_{ij}$ the dimensionless elastic force acting on particle $i$, and $\delta l_{ij} = |\boldsymbol{r}_{i} - \boldsymbol{r}_{j}| - 1$, $\tau_{n}$, $D$ and $\tilde{F}_{0}$ being the same quantities as before. In this limit, the torque in the equation for the orientation dynamics is aligning the orientation towards the elastic force acting on the active particle.

\subsection{Experimental measure of the microscopic parameters} \label{micro_measurements}

In this section, we describe three experiments, which we conducted to measure the parameters of the hexbugs dynamics. First, we evaluate the influence of inertia and the relevance of the overdamped limit. Then we measure the alignment length $l_{a}$ of one bug, our model's key parameter, as well as the angular noise $D_{\theta}$. Finally we measure the active force $F_{0}$ the hexbugs are able to exert.

\subsubsection{Inertia} \label{subsubsection:inertia}
\begin{figure}[h!]
\captionsetup{format=plain}
\centering
\begin{tikzpicture}

\node[anchor=south west,inner sep=0] at (8.0,0.0)
{\includegraphics[height=5.5cm]{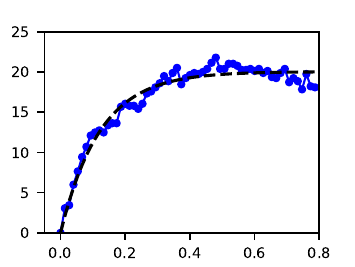}};

\node[anchor=south west,inner sep=0] at (4.0,0.2)
{\includegraphics[scale = 0.18, angle = -90]{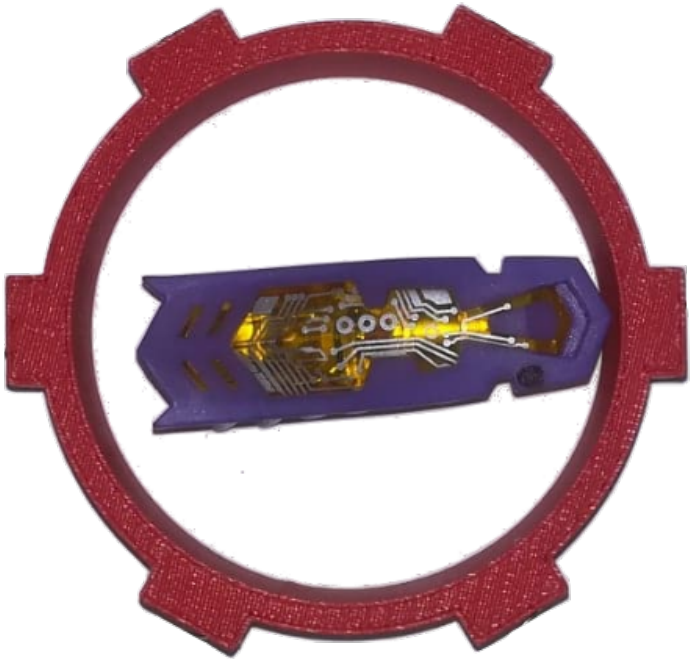}};

\node [] at (5.0,4.0) {\textbf{DCG}};
\draw[thick] (4.2,4.5) -- (5.8,4.5);
\draw[thick] (4.2,3.5) -- (5.8,3.5);
\draw[thick] (4.2,4.5) -- (4.2,3.5);
\draw[thick] (5.8,4.5) -- (5.8,3.5);

\draw[thick, dashed] (5.0,1.25) circle (0.3);

\draw[thick] (4.95,3.5) -- (4.95,1.25);
\draw[thick] (5.05,3.5) -- (5.05,1.25);

\node [] at (5.0,5.2) {\small (a)};

\node[] at (11.8,-0.2) {\small $t$ (s)};
\node[rotate=90] at (7.9,2.6) {\small $\langle V \rangle$ (cm/s)};
\node[anchor=west] at (10.2,1.2) {\small $v_{0} = 20.1 \pm 0.2$ cm/s};
\node[anchor=west] at (10.2,1.7) {\small $\tau_{d} = 0.12 \pm 0.01$ s};
\node [] at (11.8,5.2) {\small (b)};

\draw (4.95,2.05) -- (4.95,1.9);
\draw (5.05,2.05) -- (5.05,1.85);

\end{tikzpicture}
\caption{\small \textbf{Inertia measurement.} (a) Experimental active elastic unit powered by a DC generator that delivers a Heaviside signal of amplitude $1.5$ V at $t = 0$. The hexbug's polarity is fixed during the whole experiment. (b) Average speed over 20 realizations (\color{blue}\textbullet\color{black}), fitted by Eq. (\ref{dimension_acceleration}) (black dashed line) with a least square method.}
\label{inertia}
\end{figure}

We consider a single active particle, initially at rest, whose self-propulsion is switched on at $t = 0$ and whose orientation is fixed.
This process is realized by connecting the hexbugs power supply to a DC generator with thin iron wires through a hole pierced in the middle the the PP plastic film (Fig.~\ref{inertia}-a). The DC generator delivers a Heaviside signal of amplitude $1.5$ V at $t = 0$. We use two cardboard blocker on the sides of the hexbug to prevent its polarity to rotate with respect to the annulus orientation, setting the orientation $\boldsymbol{\hat{n}}$ essentially constant. The experiments were runt 20 times, acquired at 75 fps frame rate.
When the DC generator is switched on, the active unit accelerates, with its speed being given by
\begin{equation}
V(t) = v_{0}(1-e^{-t/\tau_{d}}),
\label{dimension_acceleration}
\end{equation}
where $v_{0} = F_{0}/\gamma$ is the cruise velocity, and $\tau_{d} = m/\gamma = \tau_{v}t_{0}$ is the acceleration time. The values $v_{0}=20.1\pm 0.2$ cm/s and $\tau_{d}=0.12 \pm 0.01$ s are obtained from a fit of the velocity averaged over the $20$ realizations. For timescales larger than $\tau_{d}$, such as those considered at the level of the collective dynamics, inertia can be safely neglected.

\subsubsection{Self-alignment} \label{subsubsection:self_alignment}

In order to quantify the self-alignment strength, we analyze the response of the polarity $\boldsymbol{\hat{n}}_{i}$ when imposing a square motion to the active unit using a XY-table. At each corner of the square, the orientation of the velocity $\boldsymbol{v}_{i}$ changes abruptly and the polarity aligns with the new imposed velocity. Snapshots of this process are shown in Fig.~1 of the main text for one corner of the square (Movie 1).
Denoting $\phi(t)$ the angle between the polarity and the velocity vector at any time,  Eq. (\ref{eq2:dimension}) reads:
\begin{equation} \label{alignement_properties:langevin}
 \tau \frac{d \phi}{dt} = - V \zeta \sin(\phi) + \sqrt{2 \alpha} \xi
\end{equation}
Where $V$ is the imposed speed, and where we consider a fixed velocity vector. In the absence of noise, the solution to the initial condition problem with $\phi(t=0) = \phi_{0} = 90^{\circ}$,  is
\begin{equation} \label{alignment}
\tan\left(\frac{\phi(t)}{2}\right)= \tan\left(\frac{\phi_{0}}{2}\right)\exp \left(-\frac{t}{\tau_{a}}\right),
\end{equation}
where the alignement time is $\tau_{a} = l_{a}/V$.
In the presence of noise, the Fokker-Planck equation associated to Eq.~(\ref{alignement_properties:langevin}) is

\begin{equation} \label{alignement_properties:fokker_planck}
 \frac{\partial P}{\partial t}(\phi, t) = \frac{\partial}{\partial \phi} \left( \frac{V\zeta}{\tau} \sin(\phi) P(\phi, t) \right) + \frac{\partial^{2}}{\partial \phi^{2}}\left(\frac{\alpha}{\tau^{2}} P(\phi, t)\right),
\end{equation}

where $P(\phi, t)$ is the probability distribution of the angle $\phi$ at time $t$.
The stationary probability density $P_{ss}(\phi)$ satisfies:

\begin{equation}
 \frac{d^{2}P_{ss}}{d \phi^{2}}(\phi) + \sin(\phi) \frac{V\zeta\tau}{\alpha} \frac{d P_{ss}}{d \phi}(\phi) + \cos(\phi) \frac{V\zeta\tau}{\alpha} P_{ss}(\phi) = 0,
\end{equation}

which has the following solution:

\begin{equation}
 P_{ss}(\phi) = \mathcal{N} \exp \left(\frac{V\zeta\tau}{\alpha} \cos (\phi) \right).
\end{equation}

where $\mathcal{N}$ is a normalization factor.
In the vicinity of $\phi = 0$, this distribution is a Gaussian with a standard deviation

\begin{equation} \label{sigma_value}
\sigma = \sqrt{\frac{\alpha}{\tau \zeta V}} = \sqrt{\frac{D_{\theta}l_{a}}{V}},
\end{equation}

where $D_{\theta} = \alpha/ \tau^{2}$ is the angular diffusion coefficient.

\begin{figure}
\captionsetup{format=plain}
\centering
\begin{tikzpicture}
\node[anchor=south west,inner sep=0] at (0.0,-0.12)
{\includegraphics[height=5.7cm]{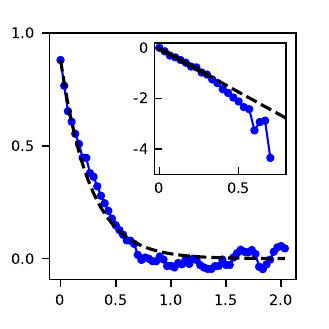}};
\node[anchor=south west,inner sep=0] at (5.8,0)
{\includegraphics[height=5.7cm]{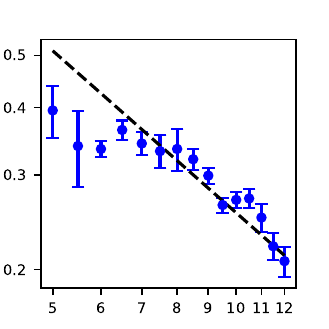}};
\node[anchor=south west,inner sep=0] at (11.5,0)
{\includegraphics[height=5.7cm]{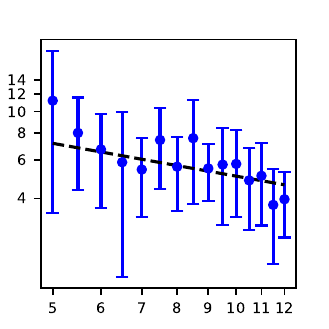}};

\node[] at (3.9,2.0) {\small $t$ (s)};
\node[rotate=90] at (1.9,3.9) {\small $\log\Bigg[ \overline{ \frac{ \tan(\phi/2) }{ \tan(\phi_{0}/2) } } \Bigg]$};

\node[rotate=-25] at (4.3,4.2) {\small $-1/\tau_{a}$};

\node[] at (3.2,0.0) {\small $t$ (s)};
\node[rotate=90] at (-0.1,3.0) {\small $\overline{\tan(\phi/2)}$};
\node[] at (3.0,5.5) {\small (a)};

\node[] at (8.8,0.0) {\small $V$ (cm/s)};
\node[anchor=south west] at (8.2,4.5) {\small $l_{a} = 2.5\pm 0.3$ cm};
\node[rotate=90] at (5.6,3.0) {\small $\tau_{a}$ (s)};
\node[] at (8.7,5.5) {\small (b)};

\node[] at (14.5,0.0) {\small $V$ (cm/s)};
\node[rotate=90] at (11.4,3.0) {\small $\sigma$ ($^{\circ}$)};
\node[anchor=south west] at (12.8,4.5) {\small $D_{\theta} = 1.75\pm 0.15$ $\text{rad}^{2}$s$^{-1}$};
\node[] at (14.5,5.5) {\small (c)};

\end{tikzpicture}
\caption{\small \textbf{Self-alignment experiments.} (a) Average misalignment $\overline{\tan(\phi/2)}(t) = \sum_{i} \tan(\phi_{i}/2)(t) / N$ of 10 independent realizations with $V = 10$ cm/s (\color{blue}\textbullet\color{black}), superposed with Eq. (\ref{alignment}) (black dashed line). Inset: $\tau_{a}$ is measured by fitting the short times of $\overline{\tan\left({\phi}(t)/2\right)/\tan\left({\phi}_{0}/2\right)}$ with an exponential decay. (b) Alignment time $\tau_{a}$ as a function of the imposed speed $V$ (\color{blue}\textbullet\color{black}). Vertical errorbars are given by the 1-$\sigma$ confidence intervals. We show an inverse function (black dotted line) which prefactor gives an estimate of the alignment length $l_{a}$ (c) Averaged standard deviation $\sigma$ of missalignment as a function of the imposed speed $V$ (\color{blue}\textbullet\color{black}). It is measured by averaging the standard deviations of missalignments for 10 realizations at a given speed. Data are analysed after two associated $\tau_{a}$ to considered only the stationnary distributions. Vertical errorbars are given by the standard deviation of the standard deviations for each speed. We fit the data with Eq. (\ref{sigma_value}) (black dashed line).}
\label{alignment_properties:data}
\end{figure}

Experimentally, we set the motion speed from 5 cm/s to 12 cm/s (upper limit of the XY table), by steps of 0.5 cm/s and perform 10 independent realizations for each speed value, that we acquire at 40 fps. The average response for a speed $V=10$ cm/s is shown in Fig. \ref{alignment_properties:data}(a), and illustrate the transitory regime.

The alignment time, $\tau_{a}$, is obtained by fitting the initial decay of $\tan\left(\frac{\phi(t)}{2}\right)/\tan\left(\frac{\phi_0}{2}\right)$, averaged over the realizations, and plotted as a function of the imposed speed $V$ on Fig. \ref{alignment_properties:data}(b). For large enough speed, where the relative importance of the noise is weaker, $\tau_{a}$ decreases as $1/V$, in agreement with Eq. (\ref{alignment}), and we extract an alignment length $l_{a} = 2.5 \pm 0.3$ cm.

As seen from Fig. \ref{alignment_properties:data}(c), the standard deviation of the misalignment fluctuations, measured in the steady regime, is observed to decay is a way that is consistent with the prediction of Eq. (\ref{sigma_value}). This allows us to extract an angular diffusion constant $D_{\theta} =1.75 \pm 0.15$ $\text{rad}^{2}s^{-1}$.

\subsubsection{Active force}
To evaluate the active force we restrict the motion of an elementary active elastic unit by putting it in a sufficiently thin rectangular channel, and we fix the hexbug polarity so that it always points in the long direction of the arena, as shown in Fig. 1 of the main text.
The active unit is attached to one end of the channel by a spring. As activity is switched on, the hexbug moves in the forward direction up to the point where the elastic force balances the active one. As we know the spring's stiffness, the extension of the spring in the steady state gives a measure of the active force $F_0=43 \pm 3$ mN. The uncertainty is given by the standard deviation of the measure on 5 different hexbugs. Having extracted $F_0$, we can obtain the elastic length $l_e=F_0/k$ for the springs of different stiffness that we use. Together with $l_a$, we are therefore in position to have the experimental value for $\pi=l_e/l_a$, the central control parameter of the experiment.

\section{The harmonic approximation} \label{section:harmonic approximation}

\subsection{Formulation of the harmonic approximation using Bra-ket notations}
As we shall see below, it will be useful to recast Eqs. (\ref{eq_dimensionless_overdamped}) using bra-ket notations:
\begin{subequations} \label{eq:dimensionless_braket}
\renewcommand{\theequation}{\theparentequation.\arabic{equation}}
\begin{align}
 |\dot{\boldsymbol{r}}\rangle &= \tilde{F}_{0}|\boldsymbol{\hat{n}}\rangle + |\boldsymbol{F}^\textrm{el}\rangle \label{eq1:dimensionless_braket} \\
 \tau_{n} |\dot{\theta}\rangle &= \mathbb{K}|\boldsymbol{F}^\textrm{el}\rangle \label{eq2:dimensionless_braket}
\end{align}
\end{subequations}
where $\langle i | \boldsymbol{a} \rangle = \boldsymbol{a}_i$, and where the matrix $\mathbb{K}$ has dimension $N\times 2N$, with elements $\langle i | \mathbb{K} | j \rangle = \boldsymbol{\hat{n}}_i^{\perp} \delta_{ij}$, which explicitly depends on the polarity field configuration $| \boldsymbol{\hat{n}} \rangle$.
We also introduce the displacement field $| \boldsymbol{u} \rangle$, defined as the displacement with respect to the reference configuration $| \boldsymbol{R} \rangle$, $| \boldsymbol{r} \rangle = | \boldsymbol{R} \rangle + | \boldsymbol{u} \rangle$, as done in the Cauchy-Born theory of elastic solids \cite{alexander1998amorphous}.

The harmonic approximation consists in linearizing the elastic force for small gradients of displacement \cite{alexander1998amorphous}:
\begin{equation} \label{harmonic_approx}
| \boldsymbol{F}^\textrm{el} \rangle = - \mathbb{M} | \boldsymbol{u} \rangle
\end{equation}
where $\mathbb{M}$ is called the dynamical matrix, which is square, real and symmetric.
Thus one can find a complete orthonormal basis in which it is diagonal, namely the normal modes $| \boldsymbol{\varphi}_k \rangle$, with corresponding eigenvalues $\omega_{k}^{2}$, the squared frequencies of vibrations, also called the modes' energies. By convention, we sort the eigenvalues from the smallest to the largest $\omega_{1}^{2} \leq ... \leq \omega_{i}^{2} \leq \dots \leq \omega_{dN}^{2}$, with $N$ the number of particles, and $d=2$ the space dimensin.
Note that in an overdamped elastic model, the denomination \textit{squared eigenfrequencies} is misleading as the projection of the displacement field on the mode $| \boldsymbol{\varphi}_k \rangle$ does not oscillate at frequency $\omega_{k}$, but relaxes to the reference configuration on a typical time $1/\omega_{k}^{2}$.
The structures considered in this work are mechanically stable \cite{lubensky2015phonons}: $\mathbb{M}$ is positive definite, and all normal modes have a finite energy $\omega_{k}^{2} > 0$.

Combining Eqs. (\ref{eq:dimensionless_braket}) and (\ref{harmonic_approx}) and rescaling  $\boldsymbol{u} \rightarrow \tau_{n}\boldsymbol{u}$, we end up with a system of equations :
\begin{subequations} \label{eq:dimensionless_noiseless_braket_theta}
\renewcommand{\theequation}{\theparentequation.\arabic{equation}}
\begin{align}
 |\dot{\boldsymbol{u}}\rangle &= \pi |\boldsymbol{\hat{n}}\rangle - \mathbb{M} |\boldsymbol{u}\rangle, \label{eq1:dimensionless_noiseless_braket_theta} \\
 |\dot{\theta}\rangle &= - \mathbb{K} \mathbb{M} |\boldsymbol{u}\rangle, \label{eq2:dimensionless_noiseless_braket_theta}
\end{align}
\end{subequations}
or equivalently
\begin{subequations} \label{eq:dimensionless_noiseless_braket_n}
\renewcommand{\theequation}{\theparentequation.\arabic{equation}}
\begin{align}
 |\dot{\boldsymbol{u}}\rangle &= \pi |\boldsymbol{\hat{n}}\rangle - \mathbb{M} |\boldsymbol{u}\rangle \label{eq1:dimensionless_noiseless_braket_u} \\
 |\dot{\boldsymbol{n}}\rangle &= - \mathbb{K}^{T} \mathbb{K} \mathbb{M} |\boldsymbol{u}\rangle \label{eq2:dimensionless_noiseless_braket_n}
\end{align}
\end{subequations}
where $\pi = \tilde{F}_{0}/\tau_{n} = F_{0}/kl_{a} = l_{e}/l_{a}$ is the only dimensionless parameter of the problem.

\subsection{Projection on the normal modes}
Decomposing the displacement and polarity fields:
\begin{subequations}
\renewcommand{\theequation}{\theparentequation.\arabic{equation}}
\begin{align}
  | \boldsymbol{u} \rangle &= \sum_{k} a_{k}^{u} | \boldsymbol{\varphi}_k \rangle \\
  | \boldsymbol{\hat{n}} \rangle &= \sum_{k} a_{k}^{n} | \boldsymbol{\varphi}_k \rangle,
\end{align}
\end{subequations}
the equations of motion (\ref{eq:dimensionless_noiseless_braket_n}) translate into
\begin{subequations}\label{eq:proj}
\renewcommand{\theequation}{\theparentequation.\arabic{equation}}
\begin{align}
\frac{d a_{k}^{u}}{dt} &= \pi a_{k}^{n} - \omega_{k}^{2} a_{k}^{u},\label{eq:proj_u}\\
\frac{d a_{k}^{n}}{dt} &= - \sum_{lpq} \omega_{q}^{2} \Gamma_{pqlk} a_{q}^{u} a_{l}^{n} a_{p}^{n},\label{eq:proj_n}
\end{align}
\end{subequations}
where we have introduced the coupling coefficients
\begin{equation}
 \Gamma_{pqlk}
 = \sum_{i} \boldsymbol{\varphi}_k^i\cdot \left[ ( \boldsymbol{\varphi}_p^i \times \boldsymbol{\varphi}_q^i) \times \boldsymbol{\varphi}_l^i\right] = \sum_{i} \left[ \boldsymbol{\varphi}_p^i\times \boldsymbol{\varphi}_q^i \right]  \cdot \left[ \boldsymbol{\varphi}_l^i \times \boldsymbol{\varphi}_k^i \right],
\end{equation}
with $\boldsymbol{\varphi}_k^i=\langle i|\boldsymbol{\varphi}_k\rangle$, the component of the mode $| \boldsymbol{\varphi}_k \rangle$ on the node $i$.
One notices the strong nonlinearity of the second equation, inherited from the self-alignment dynamics of the polarity. The coupling coefficients $\Gamma_{pqlk}$ are antisymmetric under the exchanges $p \leftrightarrow q$ and $l\leftrightarrow k$, and symmetric under the exchange $(p, q) \leftrightarrow (l, k)$.
This implies for instance that $\sum_{pl} \Gamma_{pqlk} a_{l}^{n} a_{p}^{n}$ is symmetric under the exchange $k \leftrightarrow q$.

In addition to the dynamical equation, the normalization condition $|\boldsymbol{\hat{n}}_i|=1$ for all $i$ implies that the $2N$ polarity coefficients $a_k^n$ belong to a $N$-dimensional manifold isomorphic to the $N$-torus.
Since the normalization condition implies that $\sum_i \boldsymbol{\hat{n}}_i^2= \sum_{k} {a_{k}^{n}}^{2} = N$, this manifold is included in the $(2N-1)$-sphere of radius $\sqrt{N}$.

\section{Fixed points stability analysis} \label{section:fixed_point_stability_th}

\subsection{An infinite set of fixed points}\label{}

Equilibrium configurations of Eqs.~(\ref{eq:dimensionless_noiseless_braket_theta}) are given by:
\begin{subequations}
 \begin{align}
 \pi | \boldsymbol{\hat{n}} \rangle - \mathbb{M} | \boldsymbol{u} \rangle &= 0 \label{eq:fixed_point_dis}\\
  \mathbb{K} \mathbb{M} | \boldsymbol{u} \rangle &= 0\label{eq:fixed_point_pol}
 \end{align}
\end{subequations}
Eq.~(\ref{eq:fixed_point_dis}) imposes $|\boldsymbol{u}\rangle = \pi \mathbb{M}^{-1} | \boldsymbol{\hat{n}} \rangle$; then Eq.~(\ref{eq:fixed_point_pol}) is always satisfied since $\mathbb{K} |\boldsymbol{\hat{n}} \rangle = 0$ by construction: to any configuration of the polarity field $| \boldsymbol{\hat{n}} \rangle$ corresponds the fixed point $\left\{ |\bsu\rangle = \pi \mathbb{M}^{-1} | \boldsymbol{\hat{n}} \rangle, | \boldsymbol{\hat{n}} \rangle\right\}$.
The set of fixed points is thus isomorphic to the $N$-torus.

\subsection{Dynamics linearized around a given fixed point}\label{}

To study the stability of a given fixed point $\left\{ |\boldsymbol{u}^{0}\rangle, | \boldsymbol{\hat{n}}^{0}\rangle \right\}$ we consider small perturbations $| \boldsymbol{\hat{n}} \rangle = | \boldsymbol{\hat{n}}^{0} \rangle + |\boldsymbol{\delta \hat{n}} \rangle$ and $| \boldsymbol{u} \rangle = | \boldsymbol{u}^{0} \rangle + | \boldsymbol{\delta u} \rangle$, where $\boldsymbol{\hat{n}}_{i}^{0} = (\cos{\theta_{i}^{0}}, \sin{\theta_{i}^{0}})$ and $\boldsymbol{\delta \hat{n}}_{i} = (-\sin{\theta_{i}^{0}}, \cos{\theta_{i}^{0}})\delta \theta_{i} = \boldsymbol{\hat{n}}_{i}^{0\perp}\delta \theta_{i} = \langle i | \mathbb{K}_{0}^{T} | \delta \theta \rangle$. Linearizing Eqs. (\ref{eq:dimensionless_noiseless_braket_theta}) one gets:
\begin{subequations} \label{eq:dimensionless_stability1}
\renewcommand{\theequation}{\theparentequation.\arabic{equation}}
\begin{align}
 |\dot{\boldsymbol{\delta u}}\rangle &= - \mathbb{M} | \boldsymbol{\delta u}\rangle + \pi \mathbb{K}_{0}^{T} |\delta \theta \rangle \\
 |\dot{\delta \theta} \rangle &= - \mathbb{K}_{0} \mathbb{M} | \boldsymbol{\delta u} \rangle - \pi \delta \mathbb{K} | \boldsymbol{\hat{n}}^{0} \rangle
\end{align}
\end{subequations}
Since $\boldsymbol{\delta \hat{n}}_{i}^{\perp} = (- \cos{\theta_{i}^{0}}, -\sin{\theta_{i}^{0}})\delta \theta_{i} = -\boldsymbol{\hat{n}}_{i}^{0} \delta \theta_{i}$, we use the contraction $\delta \mathbb{K} | \boldsymbol{\hat{n}}^{0} \rangle = - | \delta \theta \rangle$. Finally, rescaling $t \rightarrow \pi^{-1} t$ leads to the following system:
\begin{subequations} \label{eq:dimensionless_stability1}
\renewcommand{\theequation}{\theparentequation.\arabic{equation}}
\begin{align}
 |\dot{\boldsymbol{\delta u}}\rangle &= - \pi^{-1} \mathbb{M} | \boldsymbol{\delta u}\rangle + \mathbb{K}_{0}^{T} |\delta \theta \rangle \\
 |\dot{\delta \theta} \rangle &= - \pi^{-1} \mathbb{K}_{0} \mathbb{M} | \boldsymbol{\delta u} \rangle + | \delta \theta \rangle
 \end{align}
\end{subequations}
Therefore the stability of the configuration $| \boldsymbol{\hat{n}}^{0} \rangle$ is encoded in the $3N$ eigenvalues of the matrix
\begin{equation} \label{D_matrix}
 \mathbb{D} = \begin{pmatrix}
-\pi^{-1} \mathbb{M} & \mathbb{K}_{0}^{T} \\
- \pi^{-1} \mathbb{K}_{0} \mathbb{M} & \mathbb{I}
\end{pmatrix}
\end{equation}
The matrix $\mathbb{D}$ depends on the parameter $\pi$, the network geometry, and the equilibrium configuration of the polarities encoded in the matrix $\mathbb{K}_{0}$.
In the following we drop the subscript $0$, but one should remember that $\mathbb{K}$ depends on the configuration of the polarities.

\subsection{Properties of the spectrum valid for all fixed points}
Consider the eigenvector $|\Psi\rangle=\left(|\boldsymbol{b} \rangle, | \boldsymbol{c} \rangle\right)$
of the matrix $\mathbb{D}$  with eigenvalue $\lambda$, then:
\begin{subequations}
\begin{align}
  - \pi^{-1} \mathbb{M} | \boldsymbol{b} \rangle + \mathbb{K}^{T} | \boldsymbol{c} \rangle &= \lambda | \boldsymbol{b} \rangle, \label{eq:D_eigenval_eq1}\\
  -\pi^{-1} \mathbb{K} \mathbb{M} | \boldsymbol{b} \rangle + | \boldsymbol{c} \rangle &= \lambda | \boldsymbol{c} \rangle.\label{eq:D_eigenval_eq2}
\end{align}
\end{subequations}
Multiplying Eq.~(\ref{eq:D_eigenval_eq1}) by $\mathbb{K}$ and noting that $\mathbb{K}\mathbb{K}^T=\mathbb{I}$ leads to:
\begin{equation}\label{eq:D_eigenval_eq3}
- \pi^{-1} \mathbb{K} \mathbb{M} | \boldsymbol{b} \rangle + | \boldsymbol{c} \rangle = \lambda \mathbb{K} | \boldsymbol{b} \rangle.
\end{equation}
Comparing with Eq.~(\ref{eq:D_eigenval_eq2}), we obtain that either $\mathbb{K} | \boldsymbol{b} \rangle = | \boldsymbol{c} \rangle$ or $\lambda = 0$.
\begin{itemize}
\item
First, we consider the case $\lambda = 0$.
From Eq.~(\ref{eq:D_eigenval_eq3}), $|\boldsymbol{c} \rangle = \pi^{-1} \mathbb{K} \mathbb{M} |\boldsymbol{b} \rangle$; using this relation in Eq.~(\ref{eq:D_eigenval_eq1}) leads to
\begin{equation}
\left( \mathbb{I} - \mathbb{K}^{T}\mathbb{K} \right) \mathbb{M} | \boldsymbol{b} \rangle = 0.
\end{equation}
This means that $\mathbb{M} | \boldsymbol{b} \rangle$ must be an eigenvector of $\mathbb{K}^{T}\mathbb{K}$ with eigenvalue 1.
The operator $\mathbb{K}^{T}\mathbb{K}$ is the projector on the space spanned by $(|\boldsymbol{\hat{n}}_{i}^{\perp} \rangle)_i$: it has $N$ eigenvectors $ | \boldsymbol{\kappa}_{i} \rangle = | \boldsymbol{\hat{n}}_{i} \rangle = \boldsymbol{\hat{n}}_{i} | i \rangle$ with eigenvalue 0 and
$N$ eigenvectors $ | \boldsymbol{\kappa}_{i} \rangle = | \boldsymbol{\hat{n}}_{i}^{\perp} \rangle = \boldsymbol{\hat{n}}_{i}^{\perp} | i \rangle$ with eigenvalue 1.

Hence, for any equilibrium configuration, there are $N$ eigenvectors with eigenvalue 0, given by
\begin{subequations}
\begin{align}
|\boldsymbol{b} \rangle &= \mathbb{M}^{-1} \boldsymbol{\hat{n}}_{i}^{\perp} | i \rangle \\
|\boldsymbol{c}\rangle &=  \pi^{-1} |i\rangle.
\end{align}
\end{subequations}
These eigenvectors span the tangent space of the $N$ dimensional fixed points manifold.
We also note that as a consequence the  equilibrium configurations are all marginally stable.

\item Second, we consider the case $\mathbb{K} | \boldsymbol{b} \rangle = | \boldsymbol{c} \rangle$.
Inserting this relation in Eq.~(\ref{eq:D_eigenval_eq1}), we obtain
\begin{equation}
\left(-\pi^{-1}\mathbb{M}+\mathbb{K}^T\mathbb{K} \right)|\boldsymbol{b}\rangle = \lambda|\boldsymbol{b}\rangle.
\end{equation}
$\lambda$ should thus be an eigenvalue of the symmetric matrix
\begin{equation}
\tilde \bbD = -\pi^{-1}\bbM+\bbK^T\bbK.
\end{equation}
\end{itemize}
Since $\tilde{\bbD}$ is symmetric, $\lambda$ is real; hence, the spectrum of $\bbD$, $\Spec(\bbD)$, is real and is given by
\begin{equation}
\Spec(\bbD)=\{0\}\cup\Spec \left(\tilde\bbD \right).
\end{equation}
Since the eigenvalues of $\bbM$ are bounded between $\omega_\text{min}^2$ and $\omega_\text{max}^2$, and the eigenvalues of $\bbK^T\bbK$ are 0 and 1, the eigenvalues of $\tilde\bbD$ are bounded by
\begin{equation}\label{eq:spectrum_bounds}
-\frac{\omega_\text{max}^2}{\pi} \leq\Spec\left(\tilde\bbD \right)\leq 1-\frac{\omega_\text{min}^2}{\pi}.
\end{equation}
When $\pi\to 0$, we see from Eq.~(\ref{eq:spectrum_bounds}) that $\Spec(\tilde \bbD)\to-\infty$.
When $\pi\to\infty$, $\tilde \bbD\to\bbK^T\bbK$, which has eigenvalues $0$ and $1$ with $N$ associated eigenvectors each.

\subsection{Stability threshold of a given fixed point}
\label{sub:stability_given_fixed_point}

A given fixed point is stable if $\Spec(\tilde\bbD)\leq 0$, which is equivalent to the fact that for any vector $|\bsb\rangle$,
\begin{equation}
\langle\bsb|\tilde\bbD|\bsb\rangle\leq 0.
\end{equation}
With the explicit expression of $\tilde\bbD$, this reads
\begin{equation}
\langle\bsb|-\pi^{-1}\bbM+\bbK^T\bbK|\bsb\rangle\leq 0.
\end{equation}
We now project $|\bsb\rangle$ on the eigenvectors of $\bbM$; denoting $b_k=\langle \boldsymbol{\varphi}_k|\bsb\rangle$, this reads
\begin{equation}
\sum_{jk} b_j b_k\left(-\pi^{-1}\omega_j\omega_k+\langle \boldsymbol{\varphi}_j|\bbK^T\bbK| \boldsymbol{\varphi}_k\rangle\right)\leq 0.
\end{equation}
Now defining $\tilde b_k=\omega_kb_k$, this becomes
\begin{equation}
\sum_{jk} \tilde b_j \tilde b_k\left(-\pi^{-1}+\frac{\langle \boldsymbol{\varphi}_j|\bbK^T\bbK| \boldsymbol{\varphi}_k\rangle}{\omega_j\omega_k}\right)\leq 0.
\end{equation}
Introducing the matrix
\begin{equation}\label{eq:def_mat_L}
\mathbb{L}_{jk}=\frac{\langle \boldsymbol{\varphi}_j|\bbK^T\bbK|\boldsymbol{\varphi}_k\rangle}{\omega_j\omega_k},
\end{equation}
the stability condition reads
\begin{equation}
\Spec \left(-\pi^{-1}\bbI+\bbL \right)\leq 0.
\end{equation}
But $\Spec \left(-\pi^{-1}\bbI+\bbL \right)=-\pi^{-1}+\Spec(\bbL)$.
Finally, the fixed point $| \boldsymbol{\hat{n}} \rangle$ is stable if
\begin{equation}\label{eq:stability_threshold_n}
\pi\leq\pi_c \left(| \boldsymbol{\hat{n}} \rangle \right)=\frac{1}{\max\Spec\left(\bbL\left(| \boldsymbol{\hat{n}} \rangle\right)\right)}.
\end{equation}

\begin{figure}[h!]
\centering
\captionsetup{format=plain}
 \begin{tikzpicture}

 \draw[dashed] (0.2*\sideFPSCartoon, 4*\stepFPSCartoon+0.6) -- (0.2*\sideFPSCartoon, 0*\stepFPSCartoon-0.6);

 \draw[->, thick] (-1*\sideFPSCartoon, 0*\stepFPSCartoon) -- (1*\sideFPSCartoon, 0*\stepFPSCartoon);

 \draw[-, thick] (0.2*\sideFPSCartoon, 0*\stepFPSCartoon-0.1) -- (0.2*\sideFPSCartoon, 0*\stepFPSCartoon+0.1);
 \node[] at (0.2*\sideFPSCartoon-0.2, 0*\stepFPSCartoon+0.3) {\small $0$};
  \draw[-] (0.2*\sideFPSCartoon+0.1, 0*\stepFPSCartoon+0.1) -- (0.2*\sideFPSCartoon-0.1, 0*\stepFPSCartoon-0.1);
 \draw[-] (0.2*\sideFPSCartoon+0.1, 0*\stepFPSCartoon-0.1) -- (0.2*\sideFPSCartoon-0.1, 0*\stepFPSCartoon+0.1);

 \draw[-, thick] (0.8*\sideFPSCartoon, 0*\stepFPSCartoon-0.1) -- (0.8*\sideFPSCartoon, 0*\stepFPSCartoon+0.1);
 \node[] at (0.8*\sideFPSCartoon, 0*\stepFPSCartoon+0.3) {\small $1$};
 \draw[-] (0.8*\sideFPSCartoon+0.1, 0*\stepFPSCartoon+0.1) -- (0.8*\sideFPSCartoon-0.1, 0*\stepFPSCartoon-0.1);
 \draw[-] (0.8*\sideFPSCartoon+0.1, 0*\stepFPSCartoon-0.1) -- (0.8*\sideFPSCartoon-0.1, 0*\stepFPSCartoon+0.1);

 \node[] at (\sideFPSCartoon+0.3,0*\stepFPSCartoon) {\small $\lambda$};

 \node[] at (-1.3*\sideFPSCartoon,0*\stepFPSCartoon) {\small $\pi \rightarrow +\infty$};

 \draw[->, red] (0.2*\sideFPSCartoon+0.4, 0*\stepFPSCartoon+0.4) arc (130:140:2.0);
 \node[anchor=west] at (0.2*\sideFPSCartoon+0.35, 0*\stepFPSCartoon+0.5) {\small $\color{red}{2N}$};

 \draw[->, red] (0.8*\sideFPSCartoon+0.4, 0*\stepFPSCartoon+0.4) arc (130:140:2.0);
 \node[anchor=west] at (0.8*\sideFPSCartoon+0.35, 0*\stepFPSCartoon+0.5) {\small $\color{red}{N}$};

 \node[] at (\sideFPSCartoon+0.9,0*\stepFPSCartoon) {\small (e)};

 \draw[->, thick] (-1*\sideFPSCartoon, 1*\stepFPSCartoon) -- (1*\sideFPSCartoon, 1*\stepFPSCartoon);

 \draw[-, thick] (0.2*\sideFPSCartoon, 1*\stepFPSCartoon-0.1) -- (0.2*\sideFPSCartoon, 1*\stepFPSCartoon+0.1);
 \node[] at (0.2*\sideFPSCartoon-0.2, 1*\stepFPSCartoon+0.3) {\small $0$};
  \draw[-] (0.2*\sideFPSCartoon+0.1, 1*\stepFPSCartoon+0.1) -- (0.2*\sideFPSCartoon-0.1, 1*\stepFPSCartoon-0.1);
 \draw[-] (0.2*\sideFPSCartoon+0.1, 1*\stepFPSCartoon-0.1) -- (0.2*\sideFPSCartoon-0.1, 1*\stepFPSCartoon+0.1);

 \draw[-, thick] (0.8*\sideFPSCartoon, 1*\stepFPSCartoon-0.1) -- (0.8*\sideFPSCartoon, 1*\stepFPSCartoon+0.1);
 \node[] at (0.8*\sideFPSCartoon, 1*\stepFPSCartoon+0.3) {\small $1$};

 \draw[-] (0.3*\sideFPSCartoon+0.1, 1*\stepFPSCartoon+0.1) -- (0.3*\sideFPSCartoon-0.1, 1*\stepFPSCartoon-0.1);
 \draw[-] (0.3*\sideFPSCartoon+0.1, 1*\stepFPSCartoon-0.1) -- (0.3*\sideFPSCartoon-0.1, 1*\stepFPSCartoon+0.1);

 \node[anchor=west] at (0.3*\sideFPSCartoon-0.2,1*\stepFPSCartoon-0.4) {\small $\color{red}{\lambda_{max} > 0}$};
 \node[] at (\sideFPSCartoon+0.3,1*\stepFPSCartoon) {\small $\lambda$};

 \node[] at (-1.3*\sideFPSCartoon,1*\stepFPSCartoon) {\small $\pi > \pi_{c}(| \boldsymbol{\hat{n}} \rangle)$};

 \draw[->, red] (0.2*\sideFPSCartoon+0.4, 1*\stepFPSCartoon+0.4) arc (130:140:2.0);
 \node[anchor=west] at (0.2*\sideFPSCartoon+0.35, 1*\stepFPSCartoon+0.5) {\small $\color{red}{N}$};

 \draw[-, thick] (-0.4*\sideFPSCartoon, 1*\stepFPSCartoon-0.1) -- (-0.4*\sideFPSCartoon, 1*\stepFPSCartoon+0.1);
 \node[] at (-0.4*\sideFPSCartoon-0.2, 1*\stepFPSCartoon+0.4) {\small $-\omega_{\text{max}}^{2}/\pi$};

 \draw[-] (-0.2*\sideFPSCartoon+0.1, 1*\stepFPSCartoon+0.1) -- (-0.2*\sideFPSCartoon-0.1, 1*\stepFPSCartoon-0.1);
 \draw[-] (-0.2*\sideFPSCartoon+0.1, 1*\stepFPSCartoon-0.1) -- (-0.2*\sideFPSCartoon-0.1, 1*\stepFPSCartoon+0.1);

 \draw[-] (-0.12*\sideFPSCartoon+0.1, 1*\stepFPSCartoon+0.1) -- (-0.12*\sideFPSCartoon-0.1, 1*\stepFPSCartoon-0.1);
 \draw[-] (-0.12*\sideFPSCartoon+0.1, 1*\stepFPSCartoon-0.1) -- (-0.12*\sideFPSCartoon-0.1, 1*\stepFPSCartoon+0.1);

 \draw[-] (-0.04*\sideFPSCartoon+0.1, 1*\stepFPSCartoon+0.1) -- (-0.04*\sideFPSCartoon-0.1, 1*\stepFPSCartoon-0.1);
 \draw[-] (-0.04*\sideFPSCartoon+0.1, 1*\stepFPSCartoon-0.1) -- (-0.04*\sideFPSCartoon-0.1, 1*\stepFPSCartoon+0.1);

 \draw[-] (0.04*\sideFPSCartoon+0.1, 1*\stepFPSCartoon+0.1) -- (0.04*\sideFPSCartoon-0.1, 1*\stepFPSCartoon-0.1);
 \draw[-] (0.04*\sideFPSCartoon+0.1, 1*\stepFPSCartoon-0.1) -- (0.04*\sideFPSCartoon-0.1, 1*\stepFPSCartoon+0.1);

 \draw[-] (0.14*\sideFPSCartoon+0.1, 1*\stepFPSCartoon+0.1) -- (0.14*\sideFPSCartoon-0.1, 1*\stepFPSCartoon-0.1);
 \draw[-] (0.14*\sideFPSCartoon+0.1, 1*\stepFPSCartoon-0.1) -- (0.14*\sideFPSCartoon-0.1, 1*\stepFPSCartoon+0.1);

 \node[] at (\sideFPSCartoon+0.9,1*\stepFPSCartoon) {\small (d)};

 \draw[->, thick] (-1*\sideFPSCartoon, 2*\stepFPSCartoon) -- (1*\sideFPSCartoon, 2*\stepFPSCartoon);

 \draw[-, thick] (0.2*\sideFPSCartoon, 2*\stepFPSCartoon-0.1) -- (0.2*\sideFPSCartoon, 2*\stepFPSCartoon+0.1);
 \node[] at (0.2*\sideFPSCartoon-0.2, 2*\stepFPSCartoon+0.3) {\small $0$};
  \draw[-] (0.2*\sideFPSCartoon+0.1, 2*\stepFPSCartoon+0.1) -- (0.2*\sideFPSCartoon-0.1, 2*\stepFPSCartoon-0.1);
 \draw[-] (0.2*\sideFPSCartoon+0.1, 2*\stepFPSCartoon-0.1) -- (0.2*\sideFPSCartoon-0.1, 2*\stepFPSCartoon+0.1);

 \draw[-, thick] (0.8*\sideFPSCartoon, 2*\stepFPSCartoon-0.1) -- (0.8*\sideFPSCartoon, 2*\stepFPSCartoon+0.1);
 \node[] at (0.8*\sideFPSCartoon, 2*\stepFPSCartoon+0.3) {\small $1$};

 \node[] at (\sideFPSCartoon+0.3,2*\stepFPSCartoon) {\small $\lambda$};

 \node[] at (-1.3*\sideFPSCartoon,2*\stepFPSCartoon) {\small $\pi = \pi_{c}(| \boldsymbol{\hat{n}} \rangle)$};

 \draw[->, red] (0.2*\sideFPSCartoon+0.4, 2*\stepFPSCartoon+0.4) arc (130:140:2.0);
 \node[anchor=west] at (0.2*\sideFPSCartoon+0.35, 2*\stepFPSCartoon+0.5) {\small $\color{red}{N+1}$};

 \draw[-, thick] (-0.6*\sideFPSCartoon, 2*\stepFPSCartoon-0.1) -- (-0.6*\sideFPSCartoon, 2*\stepFPSCartoon+0.1);
 \node[] at (-0.6*\sideFPSCartoon-0.2, 2*\stepFPSCartoon+0.4) {\small $-\omega_{\text{max}}^{2}/\pi$};

  \draw[-] (-0.4*\sideFPSCartoon+0.1, 2*\stepFPSCartoon+0.1) -- (-0.4*\sideFPSCartoon-0.1, 2*\stepFPSCartoon-0.1);
 \draw[-] (-0.4*\sideFPSCartoon+0.1, 2*\stepFPSCartoon-0.1) -- (-0.4*\sideFPSCartoon-0.1, 2*\stepFPSCartoon+0.1);

 \draw[-] (-0.3*\sideFPSCartoon+0.1, 2*\stepFPSCartoon+0.1) -- (-0.3*\sideFPSCartoon-0.1, 2*\stepFPSCartoon-0.1);
 \draw[-] (-0.3*\sideFPSCartoon+0.1, 2*\stepFPSCartoon-0.1) -- (-0.3*\sideFPSCartoon-0.1, 2*\stepFPSCartoon+0.1);

 \draw[-] (-0.2*\sideFPSCartoon+0.1, 2*\stepFPSCartoon+0.1) -- (-0.2*\sideFPSCartoon-0.1, 2*\stepFPSCartoon-0.1);
 \draw[-] (-0.2*\sideFPSCartoon+0.1, 2*\stepFPSCartoon-0.1) -- (-0.2*\sideFPSCartoon-0.1, 2*\stepFPSCartoon+0.1);

 \draw[-] (-0.1*\sideFPSCartoon+0.1, 2*\stepFPSCartoon+0.1) -- (-0.1*\sideFPSCartoon-0.1, 2*\stepFPSCartoon-0.1);
 \draw[-] (-0.1*\sideFPSCartoon+0.1, 2*\stepFPSCartoon-0.1) -- (-0.1*\sideFPSCartoon-0.1, 2*\stepFPSCartoon+0.1);

 \draw[-] (0.1*\sideFPSCartoon+0.1, 2*\stepFPSCartoon+0.1) -- (0.1*\sideFPSCartoon-0.1, 2*\stepFPSCartoon-0.1);
 \draw[-] (0.1*\sideFPSCartoon+0.1, 2*\stepFPSCartoon-0.1) -- (0.1*\sideFPSCartoon-0.1, 2*\stepFPSCartoon+0.1);

 \node[] at (\sideFPSCartoon+0.9,2*\stepFPSCartoon) {\small (c)};

 \draw[->, thick] (-1*\sideFPSCartoon, 3*\stepFPSCartoon) -- (1*\sideFPSCartoon, 3*\stepFPSCartoon);

 \draw[-, thick] (0.2*\sideFPSCartoon, 3*\stepFPSCartoon-0.1) -- (0.2*\sideFPSCartoon, 3*\stepFPSCartoon+0.1);
 \node[] at (0.2*\sideFPSCartoon-0.2, 3*\stepFPSCartoon+0.3) {\small $0$};
  \draw[-] (0.2*\sideFPSCartoon+0.1, 3*\stepFPSCartoon+0.1) -- (0.2*\sideFPSCartoon-0.1, 3*\stepFPSCartoon-0.1);
 \draw[-] (0.2*\sideFPSCartoon+0.1, 3*\stepFPSCartoon-0.1) -- (0.2*\sideFPSCartoon-0.1, 3*\stepFPSCartoon+0.1);

 \draw[-, thick] (0.8*\sideFPSCartoon, 3*\stepFPSCartoon-0.1) -- (0.8*\sideFPSCartoon, 3*\stepFPSCartoon+0.1);
 \node[] at (0.8*\sideFPSCartoon, 3*\stepFPSCartoon+0.3) {\small $1$};

 \node[] at (\sideFPSCartoon+0.3,3*\stepFPSCartoon) {\small $\lambda$};

 \node[] at (-1.3*\sideFPSCartoon,3*\stepFPSCartoon) {\small $0 < \pi < \pi_{c}(| \boldsymbol{\hat{n}} \rangle)$};

 \draw[->, red] (0.2*\sideFPSCartoon+0.4, 3*\stepFPSCartoon+0.4) arc (130:140:2.0);
 \node[anchor=west] at (0.2*\sideFPSCartoon+0.35, 3*\stepFPSCartoon+0.5) {\small $\color{red}{N}$};

 \draw[-, thick] (-0.8*\sideFPSCartoon, 3*\stepFPSCartoon-0.1) -- (-0.8*\sideFPSCartoon, 3*\stepFPSCartoon+0.1);
 \node[] at (-0.8*\sideFPSCartoon-0.2, 3*\stepFPSCartoon+0.4) {\small $-\omega_{\text{max}}^{2}/\pi$};

 \draw[-] (-0.6*\sideFPSCartoon+0.1, 3*\stepFPSCartoon+0.1) -- (-0.6*\sideFPSCartoon-0.1, 3*\stepFPSCartoon-0.1);
 \draw[-] (-0.6*\sideFPSCartoon+0.1, 3*\stepFPSCartoon-0.1) -- (-0.6*\sideFPSCartoon-0.1, 3*\stepFPSCartoon+0.1);

 \draw[-] (-0.5*\sideFPSCartoon+0.1, 3*\stepFPSCartoon+0.1) -- (-0.5*\sideFPSCartoon-0.1, 3*\stepFPSCartoon-0.1);
 \draw[-] (-0.5*\sideFPSCartoon+0.1, 3*\stepFPSCartoon-0.1) -- (-0.5*\sideFPSCartoon-0.1, 3*\stepFPSCartoon+0.1);

 \draw[-] (-0.4*\sideFPSCartoon+0.1, 3*\stepFPSCartoon+0.1) -- (-0.4*\sideFPSCartoon-0.1, 3*\stepFPSCartoon-0.1);
 \draw[-] (-0.4*\sideFPSCartoon+0.1, 3*\stepFPSCartoon-0.1) -- (-0.4*\sideFPSCartoon-0.1, 3*\stepFPSCartoon+0.1);

 \draw[-] (-0.3*\sideFPSCartoon+0.1, 3*\stepFPSCartoon+0.1) -- (-0.3*\sideFPSCartoon-0.1, 3*\stepFPSCartoon-0.1);
 \draw[-] (-0.3*\sideFPSCartoon+0.1, 3*\stepFPSCartoon-0.1) -- (-0.3*\sideFPSCartoon-0.1, 3*\stepFPSCartoon+0.1);

 \draw[-] (-0.1*\sideFPSCartoon+0.1, 3*\stepFPSCartoon+0.1) -- (-0.1*\sideFPSCartoon-0.1, 3*\stepFPSCartoon-0.1);
 \draw[-] (-0.1*\sideFPSCartoon+0.1, 3*\stepFPSCartoon-0.1) -- (-0.1*\sideFPSCartoon-0.1, 3*\stepFPSCartoon+0.1);

 \draw[-] (0.05*\sideFPSCartoon+0.1, 3*\stepFPSCartoon+0.1) -- (0.05*\sideFPSCartoon-0.1, 3*\stepFPSCartoon-0.1);
 \draw[-] (0.05*\sideFPSCartoon+0.1, 3*\stepFPSCartoon-0.1) -- (0.05*\sideFPSCartoon-0.1, 3*\stepFPSCartoon+0.1);

 \draw[color=red,decorate,decoration={brace,raise=0.1cm}](-0.6*\sideFPSCartoon-0.1, 3*\stepFPSCartoon+0.1) -- (0.05*\sideFPSCartoon+0.1, 3*\stepFPSCartoon+0.1) node[above=0.2cm,pos=0.5] {$\color{red}{2N}$};

 \node[] at (\sideFPSCartoon+0.9,3*\stepFPSCartoon) {\small (b)};

 \draw[->, thick] (-1*\sideFPSCartoon, 4*\stepFPSCartoon) -- (1*\sideFPSCartoon, 4*\stepFPSCartoon);

 \draw[-, thick] (0.2*\sideFPSCartoon, 4*\stepFPSCartoon-0.1) -- (0.2*\sideFPSCartoon, 4*\stepFPSCartoon+0.1);
 \node[] at (0.2*\sideFPSCartoon-0.2, 4*\stepFPSCartoon+0.3) {\small $0$};
 \draw[-] (0.2*\sideFPSCartoon+0.1, 4*\stepFPSCartoon+0.1) -- (0.2*\sideFPSCartoon-0.1, 4*\stepFPSCartoon-0.1);
 \draw[-] (0.2*\sideFPSCartoon+0.1, 4*\stepFPSCartoon-0.1) -- (0.2*\sideFPSCartoon-0.1, 4*\stepFPSCartoon+0.1);

 \draw[-, thick] (0.8*\sideFPSCartoon, 4*\stepFPSCartoon-0.1) -- (0.8*\sideFPSCartoon, 4*\stepFPSCartoon+0.1);
 \node[] at (0.8*\sideFPSCartoon, 4*\stepFPSCartoon+0.3) {\small $1$};

 \node[] at (\sideFPSCartoon+0.3,4*\stepFPSCartoon) {\small $\lambda$};

 \node[] at (-1.3*\sideFPSCartoon,4*\stepFPSCartoon) {\small $\pi \rightarrow 0$};

 \draw[->, red] (0.2*\sideFPSCartoon+0.4, 4*\stepFPSCartoon+0.4) arc (130:140:2.0);
 \node[anchor=west] at (0.2*\sideFPSCartoon+0.35, 4*\stepFPSCartoon+0.5) {\small $\color{red}{N}$};

 \node[] at (-1.0*\sideFPSCartoon+0.2,4*\stepFPSCartoon+0.2) {\normalsize $\color{red}{\Leftarrow}$};
 \node[] at (-1.0*\sideFPSCartoon+0.25,4*\stepFPSCartoon+0.5) {\small $\color{red}{2N}$};

 \node[] at (\sideFPSCartoon+0.9,4*\stepFPSCartoon) {\small (a)};

 \end{tikzpicture}
 \caption{\small \textbf{Eigenvalue spectrum for an arbitrary fixed point $\{ |\boldsymbol{u}\rangle = \pi \mathbb{M}^{-1} | \boldsymbol{\hat{n}} \rangle, | \boldsymbol{\hat{n}} \rangle \}$.} The fixed point is stable for $\pi < \pi_{c}(| \boldsymbol{\hat{n}} \rangle)$, where $\pi_{c}(| \boldsymbol{\hat{n}} \rangle)$ is given by Eqs. (\ref{eq:stability_threshold_n}).
 (a) $\pi \rightarrow 0$, $N$ zero eigenvalues and the 2$N$ which are left are strictly negative, given by $-\omega_{k}^{2}/\pi$.
 (b) $0 < \pi < \pi_{c}(| \boldsymbol{\hat{n}} \rangle)$, $N$ zero eigenvalues, and the 2$N$ which are left are strictly negative.
 (c) $\pi = \pi_{c}(| \boldsymbol{\hat{n}} \rangle)$, $N+1$ zero eigenvalues, and the 2$N-1$ which are left are strictly negative.
 (d) $\pi > \pi_{c}(| \boldsymbol{\hat{n}} \rangle)$, the greatest eigenvalue is strictly positive.
 (e) $\pi \rightarrow +\infty$, $2N$ zero eigenvalues, $N$ one eigenvalues.}
 \label{fp_stability}
\end{figure}

\subsection{First linear destabilization}\label{sec:lower_bound_stab_threshold}
We first determine a lower bound of the stability thresholds and then show that this bound is sharp.

Let $\pi_{c}^{\text{min}}$ be the smallest value of $\pi$ for which there exists an unstable configuration.
We thus have $\pi_c \left(| \boldsymbol{\hat{n}} \rangle \right)\geq \pi_{c}^{\text{min}}$ for all $| \boldsymbol{\hat{n}} \rangle$.
From Eq.~(\ref{eq:spectrum_bounds}) there can be a positive eigenvalue only if $\pi>\omega_\text{min}^2$, hence we have
\begin{equation}\label{eq:stability_threshold_lb}
\pi_c \left(| \boldsymbol{\hat{n}} \rangle \right)\geq \pi_{c}^{\text{min}} = \omega_\text{min}^2.
\end{equation}

We now exhibit a configuration $| \boldsymbol{\hat{n}}_\text{min}\rangle$ that does destabilize at $\omega_\text{min}^2$.
Consider the eigenmode of $\bbM$ associated to the eigenvalue $\omega_1^2=\omega_\text{min}^2$, $| \boldsymbol{\varphi}_1\rangle$;
and a configuration $| \boldsymbol{\hat{n}}_\text{min}\rangle$ where the orientation $\boldsymbol{\hat{n}}_i$ is orthogonal to $\boldsymbol{\varphi}_1^i$ for every particle $i$.
For this configuration, $\bbL_{11}=\omega_\text{min}^{-2}$, hence $\max\Spec(\bbL)\geq \omega_\text{min}^{-2}$ and $\pi_c(| \boldsymbol{\hat{n}} \rangle)\leq \omega_\text{min}^2$.
With the lower bound (\ref{eq:stability_threshold_lb}), we conclude that $\pi_c^{\text{min}} = \pi_c(|\boldsymbol{\hat{n}}_\text{min}\rangle)= \omega_\text{min}^2$: the lower bound (\ref{eq:stability_threshold_lb}) is sharp and the first configuration to destabilize is related to the lowest energy mode in a simple way.

\subsection{Upper bound of the stability thresholds}\label{}
We don't have an explicit analytical expression for the largest destabilization threshold, $\pi_c^{\text{max}}$,  but we can determine an upper bound $\pi_c^{\text{upp}}$ of it above which there exists no stable fixed point.

To do so, we look for a $| \boldsymbol{\hat{n}} \rangle$-independent lower bound of the maximal eigenvalue of $\tilde \bbD$.
We use  the restriction of the matrix $\tilde \bbD$ to the two modes $j$ and $k$, which is a $2\times 2$ matrix that we denote $\tilde \bbD_{\{j,k\}}$.
We have
\begin{equation}
\max\Spec\left(\tilde \bbD\left(| \boldsymbol{\hat{n}} \rangle\right)\right)
\geq \max\Spec\left(\tilde \bbD_{\{j,k\}}\left(|\boldsymbol{\hat n}\rangle\right)\right)
=\frac{\tilde \bbD_{jj}+\tilde \bbD_{kk}}{2}+\sqrt{\frac{(\tilde \bbD_{jj}-\tilde \bbD_{kk})^2}{4}+\tilde \bbD_{jk}^2}
\geq \frac{\tilde \bbD_{jj}+\tilde \bbD_{kk}}{2}.
\end{equation}

Explicitly,
\begin{align}
\frac{\tilde \bbD_{jj}+\tilde \bbD_{kk}}{2} &= -\frac{\omega_j^2+\omega_k^2}{2\pi}+\frac{1}{2}\left(\langle \boldsymbol{\varphi}_j|\bbK^T\bbK| \boldsymbol{\varphi}_j\rangle+\langle \boldsymbol{\varphi}_k|\bbK^T\bbK| \boldsymbol{\varphi}_k\rangle \right)\\
&=-\frac{\omega_j^2+\omega_k^2}{2\pi}+\frac{1}{2}\sum_i \left[( \boldsymbol{\varphi}_j^i\times\boldsymbol{\hat{n}}_i)^2+( \boldsymbol{\varphi}_k^i\times\boldsymbol{\hat{n}}_i)^2 \right]
\end{align}
Now we have to minimize the term in the sum over the orientations $\boldsymbol{\hat{n}}_i$.
This amounts to find the minimal eigenvalue of the matrix
\begin{equation}
\boldsymbol{\varphi}_j^i\big( \boldsymbol{\varphi}_j^i\big)^T + \boldsymbol{\varphi}_k^i \big(\boldsymbol{\varphi}_k^i\big)^T = \begin{pmatrix}
\big( \varphi_{j,x}^i \big)^2 + \big( \varphi_{k,x}^i \big)^2 & \big( \varphi_{j,x}^i \big) \big( \varphi_{j,y}^i \big) + \big( \varphi_{k,x}^i \big) \big( \varphi_{k,y}^i \big)\\
\big( \varphi_{j,x}^i \big) \big(\varphi_{j,y}^i \big) + \big( \varphi_{k,x}^i \big) \big( \varphi_{k,y}^i \big) & \big( \varphi_{j,y}^i \big)^2 + \big( \varphi_{k,y}^i \big)^2
\end{pmatrix}
= \begin{pmatrix}
c_{11} & c_{12} \\ c_{12} & c_{22}
\end{pmatrix},
\end{equation}
which is
\begin{align}
\lambda_\mathrm{min}&= \frac{1}{2} \left[c_{11}+c_{22} - \sqrt{\left(c_{11}-c_{22} \right)^2+4 c_{12}^2}\right]\\
&=\frac{1}{2}\left[\left(\boldsymbol{\varphi}_{j}^i \right)^2+\left(\boldsymbol{\varphi}_{k}^i \right)^2 -\left(\left[\left(\boldsymbol{\varphi}_{j}^i \right)^2+\left(\boldsymbol{\varphi}_{k}^i \right)^2 \right]^2-4\left[\boldsymbol{\varphi}_{j}^i\times \boldsymbol{\varphi}_{k}^i \right]^2 \right)^{1/2}\right].
\end{align}
Using the fact that the modes are normalized, we finally get the bound
\begin{equation}
\max\Spec\left(\tilde \bbD\left(|\boldsymbol{\hat{n}}\rangle\right)\right) \geq -\frac{\omega_j^2+\omega_k^2}{2\pi}+\frac{1}{2} \left[1- \frac{1}{2}\sum_i \left(\left[\left(\boldsymbol{\varphi}_{j}^i \right)^2+\left(\boldsymbol{\varphi}_{k}^i \right)^2 \right]^2-4\left[\boldsymbol{\varphi}_{j}^i\times \boldsymbol{\varphi}_{k}^i \right]^2\right)^{1/2}\right],
\end{equation}

All the fixed points are unstable when this bound is positive, which happens for
\begin{equation}\label{bound_nonorthogonal}
\pi\geq \pi_{c,u}^{\{j,k\}} = \frac{\omega^2_j +\omega^2_k}{c(| \boldsymbol{\varphi}_{j} \rangle,| \boldsymbol{\varphi}_{k} \rangle)},
\end{equation}
with
\begin{equation}
c(| \boldsymbol{\varphi}_{j} \rangle,| \boldsymbol{\varphi}_{k} \rangle) = 1- \frac{1}{2}\sum_i \left(\left[\left(\boldsymbol{\varphi}_{j}^i \right)^2+\left(\boldsymbol{\varphi}_{k}^i \right)^2 \right]^2-4\left[\boldsymbol{\varphi}_{j}^i\times \boldsymbol{\varphi}_{k}^i \right]^2 \right)^{1/2}
\end{equation}

Finally the bound for $\pi$, above which there exists no stable fixed point is
\begin{equation}
\mathcal \pi_c^{\text{upp}} =\min_{\{j,k\}} \left(\frac{\omega^2_j +\omega^2_k}{c(| \boldsymbol{\varphi}_{j} \rangle,| \boldsymbol{\varphi}_{k} \rangle)}\right),
\label{eq:upperbound}
\end{equation}
Note that the function $c(\bullet,\bullet)$ is bounded between $0$, when $j=k$ and $1$, when the pair of modes $(| \boldsymbol{\varphi}_{j} \rangle,| \boldsymbol{\varphi}_{k} \rangle)$ are locally orthogonal and of the same norm.

\section{One particle dynamics} \label{section:single_particle}

The aim of this section is to study the different dynamical regimes and fixed points of Eqs.~(\ref{eq:proj}) in the case of a system of one particle in dimension $d = 2$, which consequently has two eigenmodes.
We note these eigenmodes $| \boldsymbol{\varphi}_1 \rangle$ and $| \boldsymbol{\varphi}_2 \rangle$ (respectively along $\boldsymbol{\hat{x}}$ and $\boldsymbol{\hat{y}}$), with corresponding eigenvalues $\omega_{1}^{2}$ and $\omega_{2}^{2}$.
We decompose $| \bsu\rangle = a_{1}^{u}(t) | \boldsymbol{\varphi}_1 \rangle + a_2^u(t) | \boldsymbol{\varphi}_2 \rangle$ and $| \boldsymbol{\hat{n}} \rangle = a_{1}^{n}(t) | \boldsymbol{\varphi}_1 \rangle + a_2^n(t) | \boldsymbol{\varphi}_2 \rangle$.
The fact that there is only one particle simplifies the problem:
\begin{itemize}
\item There is only one normalization condition ${a_{1}^{n}}^{2} + {a_{2}^{n}}^{2} = 1$.
\item The only non-zero coupling coefficients are

\begin{equation}
 \nonumber
 \Gamma_{1212} = -\Gamma_{2112} = -\Gamma_{1221} = \Gamma_{2121} = \Gamma = \sum_{i} \left( \boldsymbol{\varphi}_1^i \times \boldsymbol{\varphi}_2^i \right)^{2} = 1.
\end{equation}

\end{itemize}
\subsection{Governing ODEs}
\subsubsection{General case}

Using the above simplification in Eqs.~(\ref{eq:proj}), we find the ODEs governing the amplitude of the displacement and polarities on each mode:
\begin{subequations} \label{general_dynamical_system_first}
\renewcommand{\theequation}{\theparentequation.\arabic{equation}}
\begin{align}
 \dot{a}_{1}^{u}  &= \pi a_{1}^{n}  - \omega_{1}^{2} a_{1}^{u}, \\
\dot{a}_{2}^{u}  &= \pi a_{2}^{n} - \omega_{2}^{2} a_{2}^{u}, \\
\dot{a}_{1}^{n} &= - \left(\omega_{1}^{2} a_{1}^{u} a_{2}^{n} - \omega_{2}^{2}a_2^u a_{1}^{n}\right) a_2^n, \\
\dot{a}_{2}^{n} &= \left(\omega_{1}^{2}a_1^u a_{2}^{n} - \omega_{2}^{2} a_{2}^{u} a_{1}^{n}\right)a_1^n.
\end{align}
\end{subequations}
%

%
\subsubsection{Degenerate case}
In the degenerate case, $\omega_{1}^{2} = \omega_{2}^{2} = \omega_{0}^{2}$, it is more convenient to use polar coordinates. We introduce $R$, $\varphi$ and $\theta$ such that $a_{1}^{u} = R\cos(\varphi)$, $a_{2}^{u} = R\sin(\varphi)$, $a_{1}^{n} = \cos(\theta)$ and $a_{2}^{n} = \sin(\theta)$.
Using $\gamma = \theta - \varphi$, Eqs. (\ref{general_dynamical_system_first}) become:

\begin{subequations} \label{general_dynamical_system_first_polar_gamma}
\renewcommand{\theequation}{\theparentequation.\arabic{equation}}
\begin{align}
 \dot{R}  &= \pi \cos(\gamma) - \omega_{0}^{2} R, \label{eq1:general_dynamical_system_first_polar_gamma}\\
\dot{\varphi}  &= \frac{\pi}{R} \sin(\gamma), \label{eq2:general_dynamical_system_first_polar_gamma}\\
\dot{\gamma} &= \left( \omega_{0}^{2}R - \frac{\pi}{R} \right) \sin(\gamma). \label{eq3:general_dynamical_system_first_polar_gamma}
\end{align}
\end{subequations}

\subsection{Fixed Points}

\subsubsection{General case}

We use the polar angle of the polarity $\theta$, such that $a_1^n=\cos(\theta)$ and $a_2^n=\sin(\theta)$.
The fixed points are given by
\begin{subequations}
\begin{align}
\omega_1^2 a_1^u & = \pi\cos(\theta_0),\\
\omega_2^2 a_2^u & = \pi\sin(\theta_0),
\end{align}
\end{subequations}
for any orientation $\theta_0$.
The stability of the fixed points can be determined with Eq.~(\ref{eq:stability_threshold_n}).
The matrix $\mathbb{L}$ (Eq.~(\ref{eq:def_mat_L})) reads
\begin{equation}
\mathbb{L} = \begin{pmatrix}
\frac{\sin(\theta_0)^2}{\omega_1^2} & -\frac{\cos(\theta_0)\sin(\theta_0)}{\omega_1\omega_2} \\
-\frac{\cos(\theta_0)\sin(\theta_0)}{\omega_1\omega_2} & \frac{\cos(\theta_0)^2}{\omega_2^2}
\end{pmatrix};
\end{equation}
its eigenvalues are 0 and $\frac{\sin(\theta_0)^2}{\omega_1^2}+\frac{\cos(\theta_0)^2}{\omega_2^2}$, so that this state is stable for
\begin{equation}
\pi\leq \pi_c(\theta_0)=\frac{\omega_1^2 \omega_2^2}{\omega_2^2 \sin(\theta_0)^2 + \omega_1^2 \cos(\theta_0)^2}.
\end{equation}

\subsubsection{Degenerate case}
In the degenerate case, the fixed points are given by $R=\pi/\omega_0^2$, $\phi=\theta=\theta_0$. The rotational symmetry ensures that they are all equivalent and stable for
\begin{equation}
\pi\leq \pi_c=\omega_0^2.
\end{equation}

\subsection{Orbiting solutions in the degenerate case}
Orbiting solutions are defined by $\dot R=0$, $\dot\gamma=0$ and $\dot\varphi=\Omega\neq 0$.
From Eqs. (\ref{general_dynamical_system_first_polar_gamma}), we obtain
\begin{subequations}\label{frequency_amplitude_general}
\begin{align}
R & = \frac{\sqrt{\pi}}{\omega_0},\\
\gamma & = \arccos \left(\frac{\omega_0}{\sqrt{\pi}} \right),\\
\Omega & = \omega_0\sqrt{\pi-\omega_0^2}.
\end{align}
\end{subequations}
This solution exists for $\pi>\omega_0^2$, i.e., when the fixed points are unstable.
This solution corresponds to the one found in Ref.~\cite{dauchot2019dynamics}, where $\omega_0=1$.

\subsection{Influence of the bias}

\vspace{-0.2cm}
\begin{figure}[t!]
\captionsetup{format=plain}
\centering
\begin{tikzpicture}

\node[anchor=south west,inner sep=0] at (3.0,-0.5)
{\includegraphics[height=5.5cm]{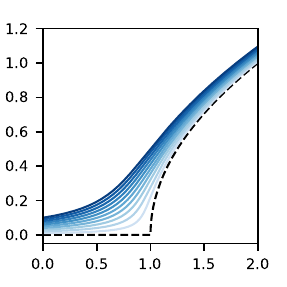}};
\node[anchor=south west,inner sep=0] at (9.0,-0.5)
{\includegraphics[height=5.5cm]{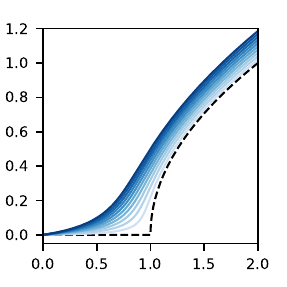}};


\node[rotate=90] at (2.8,2.3) {\small $\Omega$};
\node[rotate=90] at (8.8,2.3) {\small $\Omega$};

\node[] at (5.85,-0.45) {\small $\pi$};
\node[] at (11.85,-0.45) {\small $\pi$};

\node[] at (5.85,4.8) {\small (a)};
\node[] at (11.85,4.8) {\small (b)};

\draw[->] (6.7,1.7) -- (5.7,2.4);
\draw[->] (12.7,1.7) -- (11.7,2.4);

\node[anchor=south east] at (5.8,2.4) {\small $\omega_{B} \nearrow$};
\node[anchor=south east] at (11.8,2.4) {\small $\tilde{\omega}_{B} \nearrow$};

\end{tikzpicture}
\caption{
\small \textbf{Biased single parabola.}
(a) $\Omega$ as a function of $\pi$ for increasing dimensionless biases $\omega_{B} \in [0.01, 0.02, 0.03, 0.04, 0.05, 0.06, 0.07, 0.08, 0.09, 0.1]$ (blue solid lines) compared to the unbiased case of Eq. (\ref{frequency_amplitude_general}) (black dashed line).
(b) $\Omega$ as a function of $\pi$ for increasing dimensional bias $\tilde\omega_{B} \in [0.01, 0.02, 0.03, 0.04, 0.05, 0.06, 0.07, 0.08, 0.09, 0.1]$ (blue solid lines) compared to the unbiased case of Eq. (\ref{frequency_amplitude_general}) (black dashed line). Here we use $l_a/v_0 = 1$.
}
\label{imperfect_bifurcation}
\end{figure}

The hexbugs used in experiments can be biased, that is they may preferentially turn to the right or to the left. This is due to both fabrication imperfections and intrinsic asymmetry brought by the rotating motor.
This bias can be taken into account by adding a constant torque $\tilde\omega_B$ in the equation describing the dynamics of the particle polarity. Eq.~(\ref{eq3:general_dynamical_system_first_polar_gamma}) becomes
\begin{equation}
\label{eq:bias}
\dot{\gamma} = \left( \omega_{0}^{2}R - \frac{\pi}{R} \right) \sin(\gamma) + \omega_B,
\end{equation}
where $\omega_B=t_0 \tilde\omega_B$ is the dimensionless bias, with the characteristic time $t_0 = \gamma/k$.

Looking again for orbiting solution, Eq.~(\ref{eq1:general_dynamical_system_first_polar_gamma}) and Eq.~(\ref{eq2:general_dynamical_system_first_polar_gamma}) lead to the angular velocity
\begin{equation}
\Omega = \omega_0 \sqrt{\left(\frac{\pi}{R \omega_0}\right)^2-\omega_0^2}.
\end{equation}
and, after substitution, Eq.~(\ref{eq:bias}) reads
\begin{equation}
\left(\rho-1\right)^2 \left(\rho-\frac{\pi}{\omega_0^2} \right)+\frac{\omega_B^2}{\omega_0^4}\rho=0.
\end{equation}
where we have introduced the variable $\rho = R^2\omega_0^2/\pi$, which satisfies the following conditions: by definition $\rho\geq 0$; from the angular velocity $\rho\leq\pi/\omega_0^2$, and assuming that $\sin(\gamma)$ and $\omega_B$ are positive implies that $\rho\leq 1$. Only the smallest solution of the last equation satisfies these conditions; the corresponding angular velocity is shown on Fig.~\ref{imperfect_bifurcation}a.

In Fig 4-c of the main text, we plot the angular velocity as a function of $\pi$, and compare it to the result of the above calculation. To do so, one must consider that, experimentally,
$\pi = F_0/k l_a$ is varied by changing $k$ (see \hyperref[section:methods]{Methods}) and keeping all other experimental parameters constant. Doing so, the characteristic time $t_0 = \gamma/k$ also varies, so that the dimensionless bias $\omega_B$ is not constant. To obtain the curve corresponding to the experimental data, one must therefore compute the angular velocity at constant dimensional bias $\tilde\omega_B = \omega_B k/\gamma$, namely at constant $\omega_B v_0/ \pi l_a$, as illustrated on
Fig.~\ref{imperfect_bifurcation}b.

\section{Coarse-grained model} \label{section:coarse_grained}

\subsection{Continuous fields}

Let's remind the microscopic equations written in the main text, and equivalent to Eqs. (\ref{eq:dimensionless_noiseless_braket_n}):
\begin{subequations}
\label{eq:main}
\begin{align}
 \dot{\boldsymbol{u}}_{i} &= \pi \boldsymbol{\hat{n}}_{i} - \mathbb{M}_{ij} \boldsymbol{u}_{j} \label{eq1:main} \\
 \dot{\boldsymbol{n}}_{i} &= (\boldsymbol{\hat{n}}_{i} \times \dot{\boldsymbol{u}}_{i} ) \times \boldsymbol{\hat{n}}_{i}, \label{eq2:main}
\end{align}
\end{subequations}
Now, we want to coarse-grain these equations to be able to describe the continuous limit of active elastic materials. Instead of looking at a discrete elasticity problem, we consider a 2$d$ continuous elastic sheet, defined by the deformation field $\boldsymbol{U}(\boldsymbol{r},t)$, and with embedded activity. The orientation of the particles is described by a polarization field $\boldsymbol{m}(\boldsymbol{r},t)$, which can be understood as the mean over the mesoscopic scale of the polarity vectors. Thus we can get rid of the normalization condition the discrete formulation requires, and consider a polarization field with varying amplitude at any point of the sheet, and whose orientation and amplitude are governed by the elastic forces. First we define the average over the mesoscopic scale :
\begin{equation}
\nonumber
 \rho(\boldsymbol{r},t)\boldsymbol{m}(\boldsymbol{r},t) = \frac{1}{S}\int_{v(\boldsymbol{r})} \boldsymbol{\hat{n}}(\boldsymbol{r},t) d\boldsymbol{r} = \sum_{i \in v(\boldsymbol{r})} \boldsymbol{\hat{n}}_{i}(t) \delta(\boldsymbol{r}_{i} - \boldsymbol{r})
\end{equation}

\begin{equation}
\nonumber
 \boldsymbol{U}(\boldsymbol{r},t) = \frac{1}{S} \int_{v(\boldsymbol{r})} \boldsymbol{u}(\boldsymbol{r},t) d\boldsymbol{r} = \sum_{i \in v(\boldsymbol{r})} \boldsymbol{u}_{i}(t) \delta(\boldsymbol{r}_{i} - \boldsymbol{r})
\end{equation}

where $v(\boldsymbol{r})$ is a disk of small radius, centered at position $\boldsymbol{r}$ and of surface $S$; and where $\rho(\boldsymbol{r},t)$ is the surface density of active force. Note that for a particle $i$ inside $v(\boldsymbol{r})$, the local fields equal the average value plus the fluctuations, thus $\boldsymbol{\hat{n}}_{i}(t) = \boldsymbol{m}(\boldsymbol{r},t) + \delta \boldsymbol{m}_{i}(\boldsymbol{r},t)$ and $\boldsymbol{u}_{i}(t) = \boldsymbol{U}(\boldsymbol{r},t) + \delta \boldsymbol{U}_{i}(\boldsymbol{r},t)$, where $\langle \delta \boldsymbol{m}_{i}(\boldsymbol{r},t) \rangle_{v(\boldsymbol{r})} = \langle \delta \boldsymbol{U}_{i}(\boldsymbol{r},t) \rangle_{v(\boldsymbol{r})} = 0$. For the rest of this derivation, we consider the density of active force constant in time and space, equal at $\rho_{0}$. Moreover, we consider this average density equals to unity, as it simply rescales activity. Within such a framework, the normalization of the polarity vectors, $|\boldsymbol{\hat{n}}_i| = 1$, translates into the constraint $|\boldsymbol{m}(\boldsymbol{r},t)| \leq 1$ for the polarization.

\subsection{Strain dynamics}

The continuous limit of Eq. (\ref{eq1:main}) is simply obtained by averaging over $v(\boldsymbol{r})$, which is trivial as this equation is linear
\begin{eqnarray}
 \partial_{t} \boldsymbol{U} (\boldsymbol{r},t) &=& \pi \boldsymbol{m} + \frac{1}{S} \int_{v(\boldsymbol{r})} \boldsymbol{f_{el}}(\boldsymbol{r},t) d\boldsymbol{r} \\
 \partial_{t} \boldsymbol{U} (\boldsymbol{r},t) &=& \pi \boldsymbol{m} (\boldsymbol{r},t) + \boldsymbol{F_{el}}(\boldsymbol{r},t)
\end{eqnarray}
where, assuming the average over the local elastic forces leads to the Hooke's law for the continuum elastic force,  $\boldsymbol{F}_{el} = \text{div} \sigma = - \mathbb{L}\boldsymbol{U}(\boldsymbol{r},t)$, with :
\begin{eqnarray} \label{elastic_force_cont}
 \sigma &=& \frac{E}{1+\nu} \left( \varepsilon + \frac{\nu}{1 - 2\nu} \text{Tr}(\varepsilon)\mathbb{I} \right) \\
 \varepsilon &=& \frac{1}{2}\left( \nabla \boldsymbol{U} + \nabla \boldsymbol{U}^{t}\right)
\end{eqnarray}
where $E$ and $\nu$ are respectively the Young modulus and the Poisson ratio of the elastic material.

The displacements field dynamics is composed of a driving term along the polarization field direction, and a relaxation term, which is a second order derivative in space of the displacement field. The elastic term thus smoothes the displacement field on length scales smaller than $l^{\star}$, obtained from the balance between the two terms:
\begin{equation}
 l^{\star} \simeq 1/\sqrt{\pi}
\end{equation}

As a consequence, for a coarse-graining length smaller than $l^{\star}$, we can safely ignore the fluctuation of the displacement field. This considerably simplifies the coarse-graining of the polarity dynamics.

\subsection{Polarity dynamics}

Rewriting the dynamics for the polarity Eq. (\ref{eq2:main}), using the projector to the normal of $\boldsymbol{\hat{n}}$
\begin{equation} \label{polarity_proj}
 \dot{\boldsymbol{n}}_{i} = ( \mathbb{I} - \boldsymbol{\hat{n}}_{i} \otimes \boldsymbol{\hat{n}}_{i} ) \dot{\boldsymbol{u}}_{i}.
\end{equation}
Ignoring the fluctuations of the displacement field, we find:
\begin{equation}
 \partial_{t} \boldsymbol{m} = ( \mathbb{I} - \langle \boldsymbol{\hat{n}}_i \otimes \boldsymbol{\hat{n}}_i \rangle ) \partial_{t} \boldsymbol{U}.
\end{equation}
Now, we want to express the average $\langle \boldsymbol{\hat{n}}_i \otimes \boldsymbol{\hat{n}}_i \rangle$ as a function of the macroscopic field $\boldsymbol{m}$. By symmetry (in particular, from invariance by rotation), there are only two terms allowed:
\begin{equation} \label{closure}
 \langle \boldsymbol{\hat{n}}_i \otimes \boldsymbol{\hat{n}}_i \rangle = \phi(m) \mathbb{I} + \psi(m) \boldsymbol{m} \otimes \boldsymbol{m},
\end{equation}
where $\phi(m)$ and $\psi(m)$ are two functions of $m$, which must satisfy one additional constraint: since $\Tr(\boldsymbol{\hat{n}}_i \otimes \boldsymbol{\hat{n}}_i ) = 1$, one must have for any distribution of orientations:
\begin{equation} \label{trace_constraint}
 \Tr \langle \boldsymbol{\hat{n}}_i \otimes \boldsymbol{\hat{n}}_i \rangle = 1.
\end{equation}
Eventually, the functions $\phi(m)$ and $\psi(m)$ depend on the distribution of the orientations.
The limiting cases $m=0$ and $m=1$ follow from  Eqs~(\ref{closure})  and (\ref{trace_constraint}):
\begin{itemize}
 \item $m = 0 \Rightarrow \phi(0) = 1/2$ (from Eq. (\ref{trace_constraint})).
 \item $m = 1 \Rightarrow \psi(1) = 1$ and $\phi(1) = 0$ (from the equality of all polarity vectors).
\end{itemize}

As a simple ansatz, we write $\langle \boldsymbol{\hat{n}}_i \otimes \boldsymbol{\hat{n}}_i \rangle$ as the only second order polynomial in $m$ that is compatible with the constraints
above:
\begin{equation}
 \langle \boldsymbol{\hat{n}}_i \otimes \boldsymbol{\hat{n}}_i \rangle = \frac{1 - m^{2}}{2}\mathbb{I} + \boldsymbol{m} \otimes \boldsymbol{m},
\end{equation}
We finally obtain
\begin{equation}
 \mathbb{I} - \langle \boldsymbol{\hat{n}}_i \otimes \boldsymbol{\hat{n}}_i \rangle = \frac{1+m^{2}}{2} \mathbb{I} - \boldsymbol{m} \otimes \boldsymbol{m} = \frac{1-m^{2}}{2} \mathbb{I} + m^{2} (\mathbb{I} - \hat{\boldsymbol{m}} \otimes \hat{\boldsymbol{m}}).
\end{equation}
where
\begin{equation}
 m^{2} (\mathbb{I} - \hat{\boldsymbol{m}} \otimes \hat{\boldsymbol{m}}) \boldsymbol{A} = (\boldsymbol{m} \times \boldsymbol{A}) \times \boldsymbol{m},
\end{equation}
hence
\begin{equation}
 \partial_{t} \boldsymbol{m} = (\boldsymbol{m} \times \partial_{t} \boldsymbol{U}) \times \boldsymbol{m} + \frac{1 - m^{2}}{2} \partial_{t} \boldsymbol{U}
\end{equation}

Altogether, the coarse grained equations read:

\begin{subequations} \label{eq:coarse_grained}
\renewcommand{\theequation}{\theparentequation.\arabic{equation}}
\begin{align}
 \partial_{t} \boldsymbol{U} &= \pi \boldsymbol{m} + \boldsymbol{F_{el}} \label{eq1:coarse_grained} \\
 \partial_{t}\boldsymbol{m} &= ( \boldsymbol{m} \times \boldsymbol{F_{el}} ) \times \boldsymbol{m} + \frac{1 - \boldsymbol{m}^{2}}{2} \partial_{t} \boldsymbol{U} \label{eq2:coarse_grained}
\end{align}
\end{subequations}

\subsection{Final form with polarization relaxation}
Additionally considering angular noise in the microscopic polarity dynamics (Eq. \ref{eq2:dimensionless_noiseless_braket_n}), we find that the polarization dynamics (Eq. \ref{eq2:coarse_grained}) is modified. Let's consider the following microscopic polarity dynamics:
\begin{equation}
\dot{\boldsymbol{n}}_{i} = \sqrt{2D} \xi_i \boldsymbol{\hat{n}}_i^{\perp}
\end{equation}
where $\xi_i$ are i.i.d. gaussian random variables with zero mean and variance $\delta(t - t')$. It is possible to exactly coarse-grain this equation, following the approach of \cite{lam2015polar}. We find:

\begin{equation}
 \partial_t \boldsymbol{m}(\boldsymbol{r},t) = - D \boldsymbol{m}(\boldsymbol{r},t)
\end{equation}

Thus the final form for the coarse-grained equations, considering microscopic angular noise:

\begin{subequations} \label{eq:coarse_grained_relax}
\renewcommand{\theequation}{\theparentequation.\arabic{equation}}
\begin{align}
 \partial_{t} \boldsymbol{U} &= \pi \boldsymbol{m} + \boldsymbol{F_{el}} \label{eq1:coarse_grained_relax} \\
 \partial_{t}\boldsymbol{m} &= ( \boldsymbol{m} \times \boldsymbol{F_{el}} ) \times \boldsymbol{m} + \frac{1 - \boldsymbol{m}^{2}}{2} \left[ \pi \boldsymbol{m} + \boldsymbol{F_{el}} \right] - D_r \boldsymbol{m} \label{eq2:coarse_grained_relax}
\end{align}
\end{subequations}

where $D_r = D$ is inherited from the particles' angular diffusion coefficient, and contributes to the relaxation of the polarization toward zero.

\subsection{Disordered phase}

In the absence of the relaxation term $(D_r=0)$, any field $\boldsymbol{m}(\boldsymbol{r},t)$ such that the elastic forces locally balance the activity ($\boldsymbol{F}_{el} = - \pi \boldsymbol{m}$) is a fixed point.  However any small amount of noise, microscopic or effectively coming from the coarse-graining procedure, will induce a non zero relaxation term $(D_r>0)$. In that case, any stationary field with $m\neq 0$, relaxes to the only remaining fixed point $\boldsymbol{U} = \boldsymbol{m} = {\bf 0}$.

The linearized equations around this state read:
\begin{subequations} \label{eq:coarse_grained_relax_disordered_linear}
\renewcommand{\theequation}{\theparentequation.\arabic{equation}}
\begin{empheq}{align}
 \partial_{t} \delta \boldsymbol{U} &= \pi \delta\boldsymbol{m} + \boldsymbol{F_{el}}\left[ \delta \boldsymbol{U} \right] \label{eq1:coarse_grained_relax_disordered_linear} \\
 \partial_{t} \delta \boldsymbol{m} &=  \frac{1}{2} \left( \pi \delta\boldsymbol{m} + \boldsymbol{F_{el}}\left[ \delta \boldsymbol{U} \right] \right) - D_{r} \delta \boldsymbol{m} \label{eq2:coarse_grained_relax_disordered_linear}
\end{empheq}
\end{subequations}

If $\delta \boldsymbol{U} (\boldsymbol{r},t) = \delta a(t) \boldsymbol{\phi}(\boldsymbol{r})$ and $\delta \boldsymbol{m}(\boldsymbol{r},t) = \delta b(t) \boldsymbol{\phi}(\boldsymbol{r})$, where $\boldsymbol{\phi}$ is an eigenmode of $\boldsymbol{F_{el}}$ such that $\boldsymbol{F_{el}} \left[ \boldsymbol{\phi} \right] = - \omega_{k}^{2} \boldsymbol{\phi}$, and where $\delta a$ and $\delta b$ are small quantities, we get:

\begin{equation}
 \frac{d}{dt} \left[\begin{array}{@{}c@{}}
    \delta a(t) \\
    \delta b(t)
    \end{array} \right] = \begin{pmatrix}
    -\omega_{k}^{2} & \pi \\
    -\omega_{k}^{2}/2 - & \pi/2 - D_r
  \end{pmatrix} \cdot \left[\begin{array}{@{}c@{}}
    \delta a(t) \\
    \delta b(t)
    \end{array} \right]
\end{equation}

\begin{figure}[t!]
\hspace*{-0.5cm}
\centering
\begin{tikzpicture}

\node[] at (0.0,0.0) {\includegraphics[height=6.0cm]{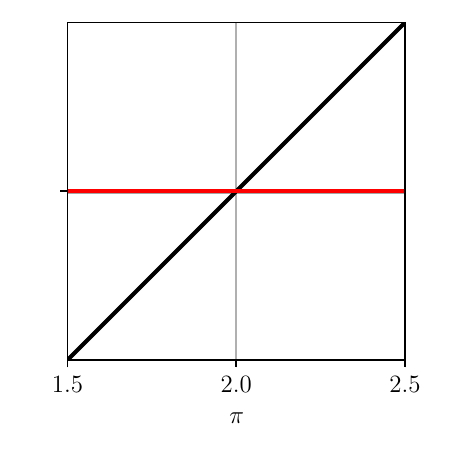}};
\node[] at (5.6,0.0) {\includegraphics[height=6.0cm]{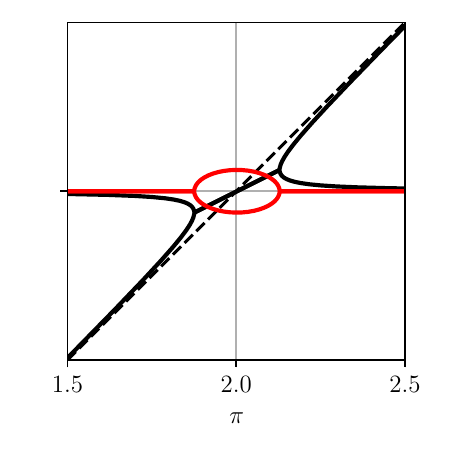}};
\node[] at (11.2,0.0) {\includegraphics[height=6.0cm]{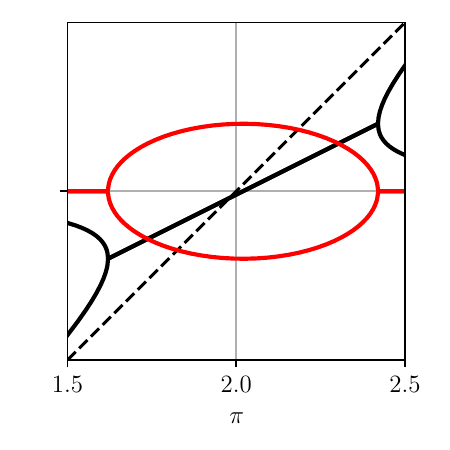}};

\node[] at (-1.7,2.3) {\small (a)};
\node[] at (3.9,2.3) {\small (b)};
\node[] at (9.5,2.3) {\small (c)};

\draw[] (11.45,0.45) -- (11.45,-1.8);


\node[] at (11.75,-1.55) {\small $\pi_c$};

\end{tikzpicture}
\vspace*{-0.6cm}
\caption{\small{\textbf{Disordered phase instability with long relaxation.} Solutions of Eq. (\ref{noisy_eigenproblem}) as a function of the elasto-active feedback $\pi$, for $\omega_{\text{min}}^{2} = 1$. Black (resp. red) curves represent the real (resp. imaginary) parts of the solutions. (a) $D_r = 0$, (b) $D_r = 10^{-3}$, (c) $D_r = 10^{-2}$.}}
\label{fig:eigenproblem_disordered}
\end{figure}

Note that expanding $ \boldsymbol{\phi}(\boldsymbol{r}) $ in the Fourier basis, the eigenfrequencies would be wavenumber-dependent because the elastic force only depends on the gradients of displacement. The solutions $\lambda$ of the eigenvalue problem satisfy:
\begin{equation} \label{noisy_eigenproblem}
 \lambda^{2} - \lambda(\pi/2 - \omega_{k}^{2} - D_r) + D_r\omega_{k}^{2} = 0
\end{equation}
and are represented on Fig.~\ref{fig:eigenproblem_disordered}.

For $D_r=0$, one finds two real eigenvalues $\lambda = 0$ and $\lambda = \pi/2 - \omega_k^2$ (Fig. \ref{fig:eigenproblem_disordered}a). For $\pi< \min_{k} (2 \omega_k^2) = 2 \omega_{\text{min}}^{2}$, the fixed point is marginally stable. For $\pi>2 \omega_{\text{min}}^2$, it is unstable and the dynamics grow along the lowest energy elastic mode. The coarse-grained description does not contain the non trivial selection observed in discrete systems.

For $D_r>0$, one finds that the nature of the bifurcation is modified (Fig. \ref{fig:eigenproblem_disordered}b). In the relevant small noise regime, the polarization relaxes much slower than the elastic modes $D_{r} \ll \omega_{k}^{2}$ and one finds that:
\begin{itemize}
 \item when $|\frac{\pi}{2} - \omega_{\text{min}}^{2}| > 2\sqrt{D_{r} \omega_{\text{min}}^{2}}$ (far enough from noiseless threshold), the two eigenvalues are real, with same sign:
 \begin{equation}
  \lambda = \frac{1}{2}\left( \frac{\pi}{2} - \omega_{\text{min}}^{2} \right) \pm \frac{1}{2}\sqrt{ \left( \frac{\pi}{2} - \omega_{\text{min}}^{2} \right)^{2} - 4D_{r}\omega_{\text{min}}^{2} }
 \end{equation}
 \item when $|\frac{\pi}{2} - \omega_{\text{min}}^{2}| < 2\sqrt{D_{r} \omega_{\text{min}}^{2}}$ (close enough from noiseless threshold), the two eigenvalues are complex conjugate, with same real part sign:
 \begin{equation}
  \lambda = \frac{1}{2}\left( \frac{\pi}{2} - \omega_{\text{min}}^{2} \right) \pm \frac{i}{2}\sqrt{ 4D_{r}\omega_{\text{min}}^{2} - \left( \frac{\pi}{2} - \omega_{\text{min}}^{2} \right)^{2} }
 \end{equation}
 Thus at threshold $\pi = 2\omega_{\text{min}}^{2}$, the imaginary part of the eigenvalues writes $\pm i \sqrt{D_{r}\omega_{\text{min}}^{2}}$. When $\pi=\pi_c$, the fixed point turns unstable via a Hopf bifurcation
\end{itemize}

Note that beyond first order in $D_r$, the instability threshold ($\pi_c=2 \omega_\text{min}^2$ in the noiseless case) is shifted to larger values of $\pi$ : large noise stabilizes the disordered fixed point (Fig. \ref{fig:eigenproblem_disordered}c). The oscillation frequency resulting from this Hopf bifurcation is finite at the bifurcation, with an amplitude proportional to $D_r^{1/2}$. It decreases when $\pi$ is increased further above the instability threshold. With no surprise the linear destabilization properties tell us very little about the disconnected non linear dynamics describing the collective actuation regime.

\subsection{Homogeneous phases}
We are here interested in describing the physics in the bulk of the material, far from the boundaries, where collective actuation concentrates. In line with the physics of the triangular lattice, let's assume the dynamics condensate on two degenerated modes, which far from the boundaries are homogeneous and akin to the two perpendicular translation modes. By convention, they write:  $\langle i | \boldsymbol{\varphi}_{1} \rangle = \boldsymbol{e}_{x}$ and $ \langle i | \boldsymbol{\varphi}_{2} \rangle = \boldsymbol{e}_{y}$.

In this context and neglecting the contributions from all the other modes, the elastic force are homogeneous $\boldsymbol{F}^{el}\left[ \boldsymbol{U} \right] = - \omega_{0}^{2} \boldsymbol{U}$, where $\omega_0$ is the eigenfrequency of the selected pair of modes. Note that for a fully connected graph of nodes, a mean field realization of the dynamics, the above description is exact, with $\omega_0=\omega_{\rm min}$. The coarse-grained equations then read:
\begin{subequations} \label{eq:coarse_grained_homoforce}
\renewcommand{\theequation}{\theparentequation.\arabic{equation}}
\begin{empheq}{align}
 \partial_{t} \boldsymbol{U} &= \pi \boldsymbol{m} - \omega_{0}^{2} \boldsymbol{U} \label{eq1:coarse_grained_homoforce} \\
\partial_{t}\boldsymbol{m} &= - \omega_{0}^{2} ( \boldsymbol{m} \times \boldsymbol{U} ) \times \boldsymbol{m} + \frac{1 - \boldsymbol{m}^{2}}{2} \partial_{t} \boldsymbol{U} - D_{r} \boldsymbol{m} \label{eq2:coarse_grained_homoforce}
\end{empheq}
\end{subequations}
We introduce the angles $\varphi$ and $\theta$, respectively the angle of the displacement $\boldsymbol{U}$ with respect to the $x$-axis, and the angle of the polarization $\boldsymbol{m}$ with respect to the $x$-axis; and the norms $R$ and $m$ of the vectors $\boldsymbol{U}$ and $\boldsymbol{m}$. Once again, we denote $\gamma = \theta - \varphi$. These variables obey the following dynamical equations:
\begin{subequations} \label{eq:coarse_grained_homoforce_polar}
\renewcommand{\theequation}{\theparentequation.\arabic{equation}}
\begin{empheq}{align}
 \partial_{t} R &= \pi m \cos\gamma - \omega_{0}^{2} R \label{eq1:coarse_grained_homoforce_polar} \\
R \partial_{t}\varphi &= \pi m \sin\gamma \label{eq2:coarse_grained_homoforce} \\
 m \partial_{t}\theta &= \frac{1 + m^{2}}{2} \omega_{0}^{2} R \sin\gamma \label{eq3:coarse_grained_homoforce} \\
m\partial_{t}m &= \frac{1 - m^{2}}{2} \left[ \pi m^{2} - \omega_{0}^{2} R m \cos\gamma \right] - D_{r} m^{2} \label{eq4:coarse_grained_homoforce_polar}
\end{empheq}
\end{subequations}
\subsubsection{Fixed point}
Due to the presence of the relaxation term in Eq. (\ref{eq4:coarse_grained_homoforce_polar}), the only fixed point of Eqs. (\ref{eq:coarse_grained_homoforce_polar}) expresses as ($R_{0} = 0$, $m_{0} = 0$).

The linearized equations around the fixed point are:
\begin{equation}
\begin{pmatrix}
 \dot{\delta R} \\
 \dot{\delta m}
\end{pmatrix}
=
 \begin{pmatrix}
  -\omega_{0}^{2} & \pi \\
  -\omega_{0}^{2}/2 & \pi/2 - D_r
 \end{pmatrix}
 \begin{pmatrix}
 \delta R \\
 \delta m
\end{pmatrix},
\end{equation}
and the eigenvalue problem reduces to solve the polynomial:
\begin{equation} \label{noisy_eigenproblem_homo}
 \lambda^{2} - \lambda(\pi/2 - \omega_{0}^{2} - D_r) + D_r\omega_{0}^{2} = 0
\end{equation}
Thus we recover Eq. (\ref{noisy_eigenproblem}), with $\omega_{\text{min}}^{2} = \omega_{0}^{2}$; and the disordered fixed point is stable for $\pi < 2\omega_{0}^{2}$, unstable otherwise.
\subsubsection{Oscillating solution}
Now we look for oscillating solutions of Eqs. (\ref{eq:coarse_grained_homoforce_polar}) at frequency $\Omega > 0$. It boils down to solve the equation for the amplitude of the polarization:

\begin{equation} \label{m_amplitude}
 D_r m = \frac{1 - m^{2}}{2} m^{2} (\pi - \frac{2\omega_{0}^{2}}{1 + m^{2}})
\end{equation}
\paragraph{Without relaxation.} If $D_r = 0$, the only solution of Eq. (\ref{m_amplitude}) giving $\Omega > 0$ is $m = 1$. The non-linear saturation vanishes and one recovers the equations for the single particle system, which predict a polarized solution with oscillation at frequency $\Omega$, amplitude $R_{0}$ in displacements, and phase shift $\gamma_{0}$ between polarity and velocity vectors, such that:

 \begin{subequations} \label{eq:homogeneous_chiral}
\renewcommand{\theequation}{\theparentequation.\arabic{equation}}
\begin{empheq}{align}
 m &= 1, \label{eq1:homogeneous_chiral} \\
 R_{0} &= \sqrt{\pi}/\omega_{0}, \label{eq2:homogeneous_chiral} \\
 \cos\gamma_{0} &= 1/R_{0}, \label{eq3:homogeneous_chiral} \\
 \Omega &= \omega_{0}\sqrt{\pi - \omega_{0}^{2}}, \label{eq4:homogeneous_chiral}
\end{empheq}
\end{subequations}

 when $\pi>\omega_0^2$.

\paragraph{With relaxation.} The presence of a small relaxation rate of the polarization amplitude ($D_r = \varepsilon \ll 1$) modifies the picture. Assuming $m = 1 - \delta m$, we find that, for $\pi>\omega_0^2$, $\delta m = \frac{\varepsilon}{\pi - \omega_{0}^{2}}$ and
\begin{subequations} \label{eq:homogeneous_chiral_noise}
\renewcommand{\theequation}{\theparentequation.\arabic{equation}}
\begin{empheq}{align}
 m &= 1 - \delta m, \label{eq0:homogeneous_chiral_noise} \\
 R_{0} &= \frac{\sqrt{\pi}}{\omega_{0}}, \label{eq2:homogeneous_chiral_noise} \\
 \cos\gamma_{0} &= \frac{\omega_{0}}{\sqrt{\pi}}(1 + \frac{1}{2}\delta m), \label{eq3:homogeneous_chiral_noise} \\
 \Omega &= \omega_{0}\sqrt{\pi - \omega_{0}^{2}}\left[ 1- \delta m \left( 1 + \frac{1}{2} \frac{\omega_{0}^{2}}{\pi - \omega_{0}^{2}} \right) \right], \label{eq4:homogeneous_chiral_noise}
\end{empheq}
\end{subequations}
As expected, a noisy microscopic dynamics decreases both the polarization and the phase shift $\gamma_{0}$. These two effect balance, resulting in an unmodified value for $R_0$. The oscillation frequency $\Omega$ also decreases with $D_r$.

\section{N particles dynamics : symmetry considerations} \label{subsubsection:symmetry}
Symmetry considerations contribute significantly to the mode selection in two ways. First, together with the normalization condition of the polarization on each mode of the elastic structure, it imposes the selection of some specific modes. Second, it selects the modes which can be nonlinearly actuated, starting from a given pair of modes.

The first step is to sort the modes according to the class of symmetry of the elastic structure of interest, here the triangular and kagome lattices and the linear chain.

\subsection{Normal modes sorted by class of symmetry in $D_{6}$ geometry}
The symmetry group of the triangular and kagome lattices with hexagonal boundaries is the dihedral group $D_{6}$.
It is generated by the rotation $\tau$ of angle $\pi/3$ and a reflection $\sigma$ (say, of axis $y=0$), which satisfy $\tau^{6} = 1$ and $\sigma^{2} = 1$.
The eigenvalues of $\sigma$ are $\pm 1$.
The eigenvalues of $\tau$ are $\exp(ik\pi/3)$ for $k \in \left\{ -2,\dots 3 \right\}$:
\begin{equation}
 \nonumber
 \text{Spec}_{D_{6}}(\tau)
 = \left( 1, e^{i\pi/3}, e^{-i\pi/3}, e^{2i\pi/3}, e^{-2i\pi/3}, -1 \right)
\end{equation}
The eigenmodes associated to the complex eigenvalues are complex and come in pairs: to a mode $| \boldsymbol{\varphi}_{+} \rangle$ with eigenvalue $e^{i n\pi/3}$ is associated a mode $| \boldsymbol{\varphi}_{-} \rangle$ with eigenvalue $e^{-i n\pi/3}$ and with the same energy.
These two modes can be combined into two real modes $| \boldsymbol{\varphi}_{1} \rangle$ and $| \boldsymbol{\varphi}_{2} \rangle$ with the same energy as $| \boldsymbol{\varphi}_{\pm} \rangle$.
$| \boldsymbol{\varphi}_{1} \rangle$ and $| \boldsymbol{\varphi}_{2} \rangle$ are not eigenvectors of $\tau$, but the 2-dimensional space that they span is stable under the action of $\tau$.
The action of $\tau$ on these modes is characterized by $\langle \boldsymbol{\varphi}_{1}|\tau| \boldsymbol{\varphi}_{1}\rangle=\langle \boldsymbol{\varphi}_{2}|\tau| \boldsymbol{\varphi}_{2}\rangle$, which is the real part of the eigenvalue of $| \boldsymbol{\varphi}_{\pm} \rangle$.
Hence, the symmetry of a normal mode $| \boldsymbol{\varphi}_{k} \rangle$ is characterized by two real numbers,
\begin{subequations}
\renewcommand{\theequation}{\theparentequation.\arabic{equation}}
\begin{align}
\langle \boldsymbol{\varphi}_{k}|\tau| \boldsymbol{\varphi}_{k}\rangle & \in \{1, 1/2, -1/2, -1\},\\
\langle \boldsymbol{\varphi}_{k}|\sigma| \boldsymbol{\varphi}_{k}\rangle & \in \{1, -1\}.
\end{align}
\end{subequations}

\subsection{Normal modes sorted by class of symmetry in $D_{2}$ geometry}
The symmetry group of the line is the dihedral group $D_{2}$. It is generated by the rotation $\tau$ of angle $\pi$ and a reflection $\sigma$ (say, of axis $y=0$). They satisfy $\tau^{2} = 1$, $\sigma^{2} = 1$.
Here, the symmetry of a normal mode $| \boldsymbol{\varphi}_{k} \rangle$ is thus characterized by two real numbers,
\begin{subequations}
\renewcommand{\theequation}{\theparentequation.\arabic{equation}}
\begin{align}
\langle \boldsymbol{\varphi}_{k}|\tau| \boldsymbol{\varphi}_{k}\rangle & \in \{1, -1\},\\
\langle \boldsymbol{\varphi}_{k}|\sigma| \boldsymbol{\varphi}_{k}\rangle & \in \{1, -1\}.
\end{align}
\end{subequations}

\subsection{Symmetry constraint on the mode selection.}
The normalization of the polarization, or in other words the fact that the active forces are of constant modulus at every node imposes that the set of activated nodes, as a whole, must contain non zero amplitude displacements on every single node. For instance, in the case of the triangular lattice, the only modes for which the displacement is non zero on the central node are those which are eigenvectors of $\tau$, the rotation in the dihedral group $D_{6}$, with eigenvalues $e^{\pm i\pi/3}$. The dynamics must therefore have a finite projection on these modes.
This demonstrate that the collective actuation regimes encountered in the triangular lattice pinned at the edges will necessarily include actuations of the $(1/2, \pm 1)$ modes, as indeed observed. Similarly the collective actuation of the line pinned at the edges with an odd number of free nodes includes actuations of the $(-1, \pm 1)$ modes.

The nonlinear couplings then control the transfer of energy between the different symmetry classes. The nonlinearities that are central to the present work, come from the elasto-active feedback. Also they are the only one present in the numerical simulations of the harmonic dynamics. These nonlinearities are best expressed in Eq.~(\ref{eq:proj_n}), which describes the polarity dynamics projected on the normal modes:
\begin{equation}
\nonumber
 \frac{d a_{k}^{n}}{dt} = - \sum_{lpq} \omega_{q}^{2} \Gamma_{pqlk} a_{q}^{u} a_{l}^{n} a_{p}^{n}
\end{equation}
We see that a mode with eigenvalue $\lambda$ with respect to the rotation operator can receive energy from modes with eigenvalues $\lambda'$ only if they satisfy the following relationship $\lambda = \lambda'^{3}$ (blue squares in Fig. \ref{couplings_table}).
Both in the triangular and kagome lattice cases, the condensed modes, which belong to the symmetry class $(1/2, \pm 1)$ only couple to themselves and to the modes belonging to the symmetry class $(-1, \pm 1)$. This observation explains the precise selection of the secondary peaks in the residual pattern of actuation in Figs.~2-b and 3-e of the main text.

In the experimental system, there are large deformations of the springs for which the harmonic approximation is not valid and nonlinear elastic coupling between the modes also  arises. Here also the symmetries restrict the possible couplings : the energy cannot flow from a mode $|\boldsymbol{\varphi}\rangle$ which is symmetric with respect to a symmetry $g$ ($\Gamma_{g}|\boldsymbol{\varphi}\rangle = |\boldsymbol{\varphi}\rangle$) to a mode which is antisymmetric ($\Gamma_{g}|\boldsymbol{\varphi}\rangle = -|\boldsymbol{\varphi}\rangle$). On the contrary, the energy can flow from an antisymmetric mode to a symmetric one (black squares in Fig. \ref{couplings_table}).

%
%
%
\begin{figure}[h!]
 \captionsetup{format=plain}
 \centering
 \begin{tikzpicture}
\fill[black] (7.0*\step,7.0*\step) -- (8.0*\step,7.0*\step) -- (8.0*\step,8.0*\step) -- (7.0*\step,7.0*\step) -- cycle;
\fill[black] (6.0*\step,7.0*\step) -- (7.0*\step,7.0*\step) -- (7.0*\step,8.0*\step) -- (6.0*\step,7.0*\step) -- cycle;
\fill[black] (5.0*\step,7.0*\step) -- (6.0*\step,7.0*\step) -- (6.0*\step,8.0*\step) -- (5.0*\step,7.0*\step) -- cycle;
\fill[black] (3.0*\step,7.0*\step) -- (4.0*\step,7.0*\step) -- (4.0*\step,8.0*\step) -- (3.0*\step,7.0*\step) -- cycle;
\fill[black] (2.0*\step,7.0*\step) -- (3.0*\step,7.0*\step) -- (3.0*\step,8.0*\step) -- (2.0*\step,7.0*\step) -- cycle;
\fill[black] (1.0*\step,7.0*\step) -- (2.0*\step,7.0*\step) -- (2.0*\step,8.0*\step) -- (1.0*\step,7.0*\step) -- cycle;
\fill[cyan] (1.0*\step,7.0*\step) -- (1.0*\step,8.0*\step) -- (2.0*\step,8.0*\step) -- (1.0*\step,7.0*\step) -- cycle;
\fill[black] (2.0*\step,6.0*\step) -- (3.0*\step,6.0*\step) -- (3.0*\step,7.0*\step) -- (2.0*\step,6.0*\step) -- cycle;
\fill[cyan] (2.0*\step,6.0*\step) -- (2.0*\step,7.0*\step) -- (3.0*\step,7.0*\step) -- (2.0*\step,6.0*\step) -- cycle;
\fill[black] (3.0*\step,5.0*\step) -- (4.0*\step,5.0*\step) -- (4.0*\step,6.0*\step) -- (3.0*\step,5.0*\step) -- cycle;
\fill[cyan] (3.0*\step,5.0*\step) -- (3.0*\step,6.0*\step) -- (4.0*\step,6.0*\step) -- (3.0*\step,5.0*\step) -- cycle;
\fill[black] (4.0*\step,4.0*\step) -- (5.0*\step,4.0*\step) -- (5.0*\step,5.0*\step) -- (4.0*\step,4.0*\step) -- cycle;
\fill[cyan] (4.0*\step,4.0*\step) -- (4.0*\step,5.0*\step) -- (5.0*\step,5.0*\step) -- (4.0*\step,4.0*\step) -- cycle;
\fill[black] (5.0*\step,3.0*\step) -- (6.0*\step,3.0*\step) -- (6.0*\step,4.0*\step) -- (5.0*\step,3.0*\step) -- cycle;
\fill[cyan] (5.0*\step,3.0*\step) -- (5.0*\step,4.0*\step) -- (6.0*\step,4.0*\step) -- (5.0*\step,3.0*\step) -- cycle;
\fill[black] (6.0*\step,2.0*\step) -- (7.0*\step,2.0*\step) -- (7.0*\step,3.0*\step) -- (6.0*\step,2.0*\step) -- cycle;
\fill[cyan] (6.0*\step,2.0*\step) -- (6.0*\step,3.0*\step) -- (7.0*\step,3.0*\step) -- (6.0*\step,2.0*\step) -- cycle;
\fill[black] (7.0*\step,1.0*\step) -- (8.0*\step,1.0*\step) -- (8.0*\step,2.0*\step) -- (7.0*\step,1.0*\step) -- cycle;
\fill[cyan] (7.0*\step,1.0*\step) -- (7.0*\step,2.0*\step) -- (8.0*\step,2.0*\step) -- (7.0*\step,1.0*\step) -- cycle;
\fill[black] (8.0*\step,0.0*\step) -- (9.0*\step,0.0*\step) -- (9.0*\step,1.0*\step) -- (8.0*\step,0.0*\step) -- cycle;
\fill[cyan] (8.0*\step,0.0*\step) -- (8.0*\step,1.0*\step) -- (9.0*\step,1.0*\step) -- (8.0*\step,0.0*\step) -- cycle;
\fill[black] (5.0*\step,7.0*\step) -- (6.0*\step,7.0*\step) -- (6.0*\step,8.0*\step) -- (5.0*\step,7.0*\step) -- cycle;
\fill[cyan] (5.0*\step,7.0*\step) -- (5.0*\step,8.0*\step) -- (6.0*\step,8.0*\step) -- (5.0*\step,7.0*\step) -- cycle;
\fill[black] (6.0*\step,3.0*\step) -- (7.0*\step,3.0*\step) -- (7.0*\step,4.0*\step) -- (6.0*\step,3.0*\step) -- cycle;
\fill[black] (7.0*\step,3.0*\step) -- (8.0*\step,3.0*\step) -- (8.0*\step,4.0*\step) -- (7.0*\step,3.0*\step) -- cycle;
\fill[black] (6.0*\step,6.0*\step) -- (7.0*\step,6.0*\step) -- (7.0*\step,7.0*\step) -- (6.0*\step,6.0*\step) -- cycle;
\fill[cyan] (6.0*\step,6.0*\step) -- (6.0*\step,7.0*\step) -- (7.0*\step,7.0*\step) -- (6.0*\step,6.0*\step) -- cycle;
\fill[black] (7.0*\step,5.0*\step) -- (8.0*\step,5.0*\step) -- (8.0*\step,6.0*\step) -- (7.0*\step,5.0*\step) -- cycle;
\fill[cyan] (7.0*\step,5.0*\step) -- (7.0*\step,6.0*\step) -- (8.0*\step,6.0*\step) -- (7.0*\step,5.0*\step) -- cycle;
\fill[black] (8.0*\step,4.0*\step) -- (9.0*\step,4.0*\step) -- (9.0*\step,5.0*\step) -- (8.0*\step,4.0*\step) -- cycle;
\fill[cyan] (8.0*\step,4.0*\step) -- (8.0*\step,5.0*\step) -- (9.0*\step,5.0*\step) -- (8.0*\step,4.0*\step) -- cycle;
\fill[black] (4.0*\step,7.0*\step) -- (5.0*\step,7.0*\step) -- (5.0*\step,8.0*\step) -- (4.0*\step,7.0*\step) -- cycle;
\fill[black] (4.0*\step,6.0*\step) -- (5.0*\step,6.0*\step) -- (5.0*\step,7.0*\step) -- (4.0*\step,6.0*\step) -- cycle;
\fill[black] (4.0*\step,5.0*\step) -- (5.0*\step,5.0*\step) -- (5.0*\step,6.0*\step) -- (4.0*\step,5.0*\step) -- cycle;
\fill[black] (8.0*\step,7.0*\step) -- (9.0*\step,7.0*\step) -- (9.0*\step,8.0*\step) -- (8.0*\step,7.0*\step) -- cycle;
\fill[black] (8.0*\step,6.0*\step) -- (9.0*\step,6.0*\step) -- (9.0*\step,7.0*\step) -- (8.0*\step,6.0*\step) -- cycle;
\fill[black] (8.0*\step,5.0*\step) -- (9.0*\step,5.0*\step) -- (9.0*\step,6.0*\step) -- (8.0*\step,5.0*\step) -- cycle;
\fill[black] (8.0*\step,3.0*\step) -- (9.0*\step,3.0*\step) -- (9.0*\step,4.0*\step) -- (8.0*\step,3.0*\step) -- cycle;
\fill[black] (8.0*\step,2.0*\step) -- (9.0*\step,2.0*\step) -- (9.0*\step,3.0*\step) -- (8.0*\step,2.0*\step) -- cycle;
\fill[black] (8.0*\step,1.0*\step) -- (9.0*\step,1.0*\step) -- (9.0*\step,2.0*\step) -- (8.0*\step,1.0*\step) -- cycle;
\draw[help lines, step=\step, thick, color=black] (0,0) grid (9*\step,9*\step);
\draw[color=black, thick] (1*\step,7*\step) -- (2*\step,8*\step);
\draw[color=black, thick] (1*\step,6*\step) -- (3*\step,8*\step);
\draw[color=black, thick] (1*\step,5*\step) -- (4*\step,8*\step);
\draw[color=black, thick] (1*\step,4*\step) -- (5*\step,8*\step);
\draw[color=black, thick] (1*\step,3*\step) -- (6*\step,8*\step);
\draw[color=black, thick] (1*\step,2*\step) -- (7*\step,8*\step);
\draw[color=black, thick] (1*\step,1*\step) -- (8*\step,8*\step);
\draw[color=black, thick] (1*\step,0*\step) -- (9*\step,8*\step);
\draw[color=black, thick] (2*\step,0*\step) -- (9*\step,7*\step);
\draw[color=black, thick] (3*\step,0*\step) -- (9*\step,6*\step);
\draw[color=black, thick] (4*\step,0*\step) -- (9*\step,5*\step);
\draw[color=black, thick] (5*\step,0*\step) -- (9*\step,4*\step);
\draw[color=black, thick] (6*\step,0*\step) -- (9*\step,3*\step);
\draw[color=black, thick] (7*\step,0*\step) -- (9*\step,2*\step);
\draw[color=black, thick] (8*\step,0*\step) -- (9*\step,1*\step);

\node[] at (0.5*\step, 8.5*\step)
{\scriptsize $\begin{pmatrix}
C_{6} \\
\sigma
\end{pmatrix}$};
\node[] at (0.5*\step, 0.5*\step) {\tiny $\begin{pmatrix}
1 \\
1
\end{pmatrix}$};

\node[] at (0.5*\step, 1.5*\step) {\tiny $\begin{pmatrix}
1 \\
-1
\end{pmatrix}$};

\node[] at (0.5*\step, 2.5*\step) {\tiny $\begin{pmatrix}
-1 \\
1
\end{pmatrix}$};

\node[] at (0.5*\step, 3.5*\step) {\tiny $\begin{pmatrix}
-1 \\
-1
\end{pmatrix}$};

\node[] at (0.5*\step, 4.5*\step) {\tiny $\begin{pmatrix}
-0.5 \\
1
\end{pmatrix}$};

\node[] at (0.5*\step, 5.5*\step) {\tiny $\begin{pmatrix}
-0.5 \\
-1
\end{pmatrix}$};

\node[] at (0.5*\step, 6.5*\step) {\tiny $\begin{pmatrix}
0.5 \\
1
\end{pmatrix}$};

\node[] at (0.5*\step, 7.5*\step) {\tiny $\begin{pmatrix}
0.5 \\
-1
\end{pmatrix}$};

\node[] at (8.5*\step, 8.5*\step) {\tiny $\begin{pmatrix}
1 \\
1
\end{pmatrix}$};

\node[] at (7.5*\step, 8.5*\step) {\tiny $\begin{pmatrix}
1 \\
-1
\end{pmatrix}$};

\node[] at (6.5*\step, 8.5*\step) {\tiny $\begin{pmatrix}
-1 \\
1
\end{pmatrix}$};

\node[] at (5.5*\step, 8.5*\step) {\tiny $\begin{pmatrix}
-1 \\
-1
\end{pmatrix}$};

\node[] at (4.5*\step, 8.5*\step) {\tiny $\begin{pmatrix}
-0.5 \\
1
\end{pmatrix}$};

\node[] at (3.5*\step, 8.5*\step) {\tiny $\begin{pmatrix}
-0.5 \\
-1
\end{pmatrix}$};

\node[] at (2.5*\step, 8.5*\step) {\tiny $\begin{pmatrix}
0.5 \\
1
\end{pmatrix}$};

\node[] at (1.5*\step, 8.5*\step) {\tiny $\begin{pmatrix}
0.5 \\
-1
\end{pmatrix}$};

\fill[black] (-4.25*\step,1.0*\step) -- (-2.75*\step,1.0*\step) -- (-2.75*\step,2.5*\step) -- (-4.25*\step,1.0*\step) -- cycle;

\fill[cyan] (-4.25*\step,1.0*\step) -- (-4.25*\step,2.5*\step) -- (-2.75*\step,2.5*\step) -- (-4.25*\step,1.0*\step) -- cycle;
1
\draw[color=black, thick] (-4.25*\step,2.5*\step) -- (-2.75*\step,2.5*\step);
\draw[color=black, thick] (-2.75*\step,2.5*\step) -- (-2.75*\step,1.0*\step);
\draw[color=black, thick] (-4.25*\step,1.0*\step) -- (-2.75*\step,1.0*\step);
\draw[color=black, thick] (-4.25*\step,2.5*\step) -- (-4.25*\step,1.0*\step);

\draw[color=black, thick] (-4.25*\step,1.0*\step) -- (-2.75*\step,2.5*\step);
\node[text width=3cm] at (-2.8*\step, 0.4*\step) {\scriptsize \ Geometrical \\ non-linearities};
\node[text width=3cm] at (-2.8*\step, 3.1*\step) {\scriptsize \ \ \ \ \ Active \\ non-linearities};
\draw [->,black,thick] (-2.5*\step, 0.45*\step) to [out=30,in=-30] (-2.5*\step, 1.75*\step);
\draw [->,black,thick] (-4.7*\step, 3.05*\step) to [out=-150,in=150] (-4.5*\step, 1.75*\step);

\fill[black] (-5.0*\step,7.0*\step) -- (-4.0*\step,7.0*\step) -- (-4.0*\step,8.0*\step) -- (-5.0*\step,7.0*\step) -- cycle;
\fill[cyan] (-5.0*\step,7.0*\step) -- (-5.0*\step,8.0*\step) -- (-4.0*\step,8.0*\step) -- (-5.0*\step,7.0*\step) -- cycle;
\fill[black] (-4.0*\step,6.0*\step) -- (-3.0*\step,6.0*\step) -- (-3.0*\step,7.0*\step) -- (-4.0*\step,6.0*\step) -- cycle;
\fill[cyan] (-4.0*\step,6.0*\step) -- (-4.0*\step,7.0*\step) -- (-3.0*\step,7.0*\step) -- (-4.0*\step,6.0*\step) -- cycle;
\fill[black] (-3.0*\step,5.0*\step) -- (-2.0*\step,5.0*\step) -- (-2.0*\step,6.0*\step) -- (-3.0*\step,5.0*\step) -- cycle;
\fill[cyan] (-3.0*\step,5.0*\step) -- (-3.0*\step,6.0*\step) -- (-2.0*\step,6.0*\step) -- (-3.0*\step,5.0*\step) -- cycle;
\fill[black] (-2.0*\step,4.0*\step) -- (-1.0*\step,4.0*\step) -- (-1.0*\step,5.0*\step) -- (-2.0*\step,4.0*\step) -- cycle;
\fill[cyan] (-2.0*\step,4.0*\step) -- (-2.0*\step,5.0*\step) -- (-1.0*\step,5.0*\step) -- (-2.0*\step,4.0*\step) -- cycle;
\fill[black] (-2.0*\step,5.0*\step) -- (-1.0*\step,5.0*\step) -- (-1.0*\step,6.0*\step) -- (-2.0*\step,5.0*\step) -- cycle;
\fill[black] (-2.0*\step,6.0*\step) -- (-1.0*\step,6.0*\step) -- (-1.0*\step,7.0*\step) -- (-2.0*\step,6.0*\step) -- cycle;
\fill[black] (-2.0*\step,7.0*\step) -- (-1.0*\step,7.0*\step) -- (-1.0*\step,8.0*\step) -- (-2.0*\step,7.0*\step) -- cycle;
\fill[black] (-3.0*\step,7.0*\step) -- (-2.0*\step,7.0*\step) -- (-2.0*\step,8.0*\step) -- (-3.0*\step,7.0*\step) -- cycle;
\fill[black] (-4.0*\step,7.0*\step) -- (-3.0*\step,7.0*\step) -- (-3.0*\step,8.0*\step) -- (-4.0*\step,7.0*\step) -- cycle;
\draw[help lines, step=\step, thick, color=black] (-6*\step,4*\step) grid (-1*\step,9*\step);
\draw[thick, color=black] (-6*\step,4*\step) -- (-1*\step,4*\step);
\node[] at (-5.5*\step, 8.5*\step) {\scriptsize $\begin{pmatrix}
C_{2} \\
\sigma
\end{pmatrix}$};

\node[] at (-4.5*\step, 8.5*\step) {\tiny $\begin{pmatrix}
-1 \\
-1
\end{pmatrix}$};

\node[] at (-3.5*\step, 8.5*\step) {\tiny $\begin{pmatrix}
-1 \\
1
\end{pmatrix}$};

\node[] at (-2.5*\step, 8.5*\step) {\tiny $\begin{pmatrix}
1 \\
-1
\end{pmatrix}$};

\node[] at (-1.5*\step, 8.5*\step) {\tiny $\begin{pmatrix}
1 \\
1
\end{pmatrix}$};

\node[] at (-5.5*\step, 7.5*\step) {\tiny $\begin{pmatrix}
-1 \\
-1
\end{pmatrix}$};

\node[] at (-5.5*\step, 6.5*\step) {\tiny $\begin{pmatrix}
-1 \\
1
\end{pmatrix}$};

\node[] at (-5.5*\step, 5.5*\step) {\tiny $\begin{pmatrix}
1 \\
-1
\end{pmatrix}$};

\node[] at (-5.5*\step, 4.5*\step) {\tiny $\begin{pmatrix}
1 \\
1
\end{pmatrix}$};

 \end{tikzpicture}
 \caption{\small \textbf{Non-linear couplings between symmetry classes.} (left) Inter-class couplings for the dihedral group $D_{2}$. (right) Inter-class couplings for the dihedral group $D_{6}$. The square $(i,j)$ is colored when energy can flow from the mode of row $i$ to the mode of column $j$. The upper left blue (resp. lower right black) triangles corresponds to the elasto-active feedback non-linearities (resp. geometrical non-linearities) transferring energy from the class of the row $i$ to the class of the column $j$.}
 \label{couplings_table}
\end{figure}

\section{$N$ particles dynamics restricted to two modes} \label{section:dynamics_restricted_to_two_modes}
The strong condensation of the dynamics on a pair of modes cannot be strict, in general, because of the normalization condition of each individual polarity.
This is only possible if the two modes $\boldsymbol{\varphi}_1$ and $\boldsymbol{\varphi}_2$ are fully delocalized  and locally orthogonal : $|\boldsymbol{\varphi}_k^i|=|\boldsymbol{\varphi}_k^j|$ for every sites $i$ and $j$ and $k\in\{1,2\}$, and $\boldsymbol{\varphi}_1^i\perp\boldsymbol{\varphi}_2^i$ for every site $i$. Appart from very specific cases, like the one particle dynamics in the degenerate case studied in the previous section, the pairs of modes of an elastic structure pinned at its boundary do not satisfy such conditions exactly.
However, investigating the dynamics restricted to two modes can still provide us with a condition under which a rotating solution sets in.

The dynamics projected on two modes reads:
\begin{subequations} \label{eq:restricted_to_two_modes}
\renewcommand{\theequation}{\theparentequation.\arabic{equation}}
\begin{align}
 \dot{a}_{1}^{u} &= \pi a_{1}^{n} - \omega_{1}^{2} a_{1}^{u}, \label{eq1:restricted_to_two_modes} \\
  \dot{a}_{2}^{u} &= \pi a_{2}^{n} - \omega_{1}^{2} a_{2}^{u}, \label{eq2:restricted_to_two_modes} \\
  \dot{a}_{1}^{n} &= - \Gamma_{12} \left( \omega_{1}^{2} a_{1}^{u} a_{2}^{n} - \omega_{2}^{2}a_{2}^{u}a_{1}^{n} \right) a_{2}^{n}, \label{eq3:restricted_to_two_modes} \\
  \dot{a}_{2}^{n} &= \Gamma_{12} \left( \omega_{1}^{2} a_{1}^{u} a_{2}^{n} - \omega_{2}^{2}a_{2}^{u}a_{1}^{n} \right) a_{1}^{n}, \label{eq4:restricted_to_two_modes}
\end{align}
\end{subequations}
where there is only one coupling constant :
\begin{equation}
\Gamma_{12} =  \Gamma_{1212} = - \Gamma_{2112} = - \Gamma_{1221} = \Gamma_{2121} = \sum_{i} \left( \boldsymbol{\varphi}_1^i \times \boldsymbol{\varphi}_2^i\right)^{2}
\end{equation}
We note that $a_1^n \dot a_1^n+a_2^n\dot a_2^n=0$, hence that the norm $|a^n|=\left({a_1^n}^2+{a_2^n}^2\right)^{1/2}$ is constant; however, it is not necessarily 1.
Introducing the rescaled quantities $\bar a_k^n= a_k^n/|a^n|$, where $\bar a_k^n$ is now normalized, $\bar a_k^u= \Gamma_{12}|a^n| a_k^u $ and $\bar\omega_k^2=\omega_k^2 / (\Gamma_{12}|a^n|^2)$, the above equations read
\begin{subequations} \label{eq:restricted_to_two_modes_rescaled}
\renewcommand{\theequation}{\theparentequation.\arabic{equation}}
\begin{align}
\left(\Gamma_{12} |a^n|^2 \right)^{-1} \dot{\bar a}_{1}^{u} &= \pi \bar a_{1}^{n} - \bar \omega_{1}^{2} \bar a_{1}^{u}, \\
\left(\Gamma_{12} |a^n|^2 \right)^{-1} \dot{\bar a}_{2}^{u} &= \pi \bar a_{2}^{n} - \bar \omega_{2}^{2} \bar a_{2}^{u}, \\
\left(\Gamma_{12} |a^n|^2 \right)^{-1} \dot{\bar a}_{1}^{n} &= - \left( \bar\omega_{1}^{2} \bar a_{1}^{u} \bar a_{2}^{n} - \bar \omega_{2}^{2}\bar a_{2}^{u}a_{1}^{n} \right) \bar a_{2}^{n},\\
\left(\Gamma_{12} |a^n|^2 \right)^{-1} \dot{\bar a}_{2}^{n} &= \left( \bar\omega_{1}^{2} \bar a_{1}^{u} \bar a_{2}^{n} - \bar \omega_{2}^{2}\bar a_{2}^{u}a_{1}^{n} \right) \bar a_{1}^{n}, \label{eq4:restricted_to_two_modes}
\end{align}
\end{subequations}
Up to a rescaling of the time, these are the equations of motion of a particle in an anisotropic potential (Eqs. (\ref{general_dynamical_system_first})).
In the degenerate case $\omega_1=\omega_2=\omega_0$, rotating solutions exist for
\begin{equation}
\pi>\bar\omega_0^2=\frac{\omega_0^2}{\Gamma_{12}|a^n|^2}.
\end{equation}
When the modes 1 and 2 are fully delocalized and locally orthogonal, the condensation can be strict and the restriction to these modes is exact. In this case $\Gamma_{12}=1/N$ and $|a^n|=\sqrt{N}$ and one recovers the result obtained for one particle in the degenerate case.
When these conditions are not satisfied, $|a^n|<\sqrt{N}$ and more modes are activated, which are selected from symmetry considerations as discussed above, in section \hyperref[subsubsection:symmetry]{6}, when considering the geometries specifically studied experimentally.
In the numerical simulations of the triangular lattice, we find that $|a^{n}_{12}| < 0.93\sqrt{N}$.

Altogether, one notes that the higher the scaled condensation level $|a^{n}_{12}|/\sqrt{N} $ and the stronger the scaled coupling $N \Gamma_{12}$, the lower is the threshold for the existence of a periodic dynamics.

\section{Dynamics of linear structures} \label{section:toy_model}
As we shall see, considering linear structures ensures the existence of pairs of orthogonal modes, hence allowing for further progress in the study of the dynamics.

\subsection{Definition of the 1d chain, eigenmodes}\label{}
We consider a chain with $N$ free particles $1\leq j\leq N$ and pinned edges $j=0$ and $j=N+1$.
The chain is oriented along $\boldsymbol{\hat{x}}$, so that the equilibrium positions are $x_j=\alpha j$, $y_j=0$.
The parameter $\alpha$ is the the ratio between the length of the springs in the equilibrium configuration $l_{e}$ and the natural length of the springs $l_{0}$; the chain thus bears a dimensionless tension $T=\alpha-1$.

The dynamical matrix is minus the discrete Laplacian in both directions, with a factor $A_{\alpha} = 1 - \alpha^{-1}$ in the direction $y$, and reads
\begin{equation}
	\mathbb{M} =
	\begin{pmatrix}
		2 & 0 & -1 & 0 & 0 & 0 & \cdots & 0 \\
		0 & 2A_{\alpha} & 0 & -A_{\alpha} & 0 & 0 & & 0 \\
		-1 & 0 & 2 & 0 & -1 & 0 & & 0 \\
		0 & -A_{\alpha} & 0 & 2A_{\alpha} & 0 & -A_{\alpha} & & 0 \\
		\vdots &  &  &  &  &  & \ddots & \vdots \\
		0 & 0 & 0 & 0 & \cdots & -A_{\alpha} & 0 & 2A_{\alpha} \\
	\end{pmatrix},
	\label{line:dynamical_matrix}
\end{equation}
where odd (resp. even) lines and columns correspond to displacements along $\boldsymbol{\hat{x}}$ (resp. $\boldsymbol{\hat{y}}$).
The directions $x$ and $y$ decouple; as a consequence there are $N$ eigenmodes along $\boldsymbol{\hat{x}}$ (resp. $\boldsymbol{\hat{y}}$), which we denote $\boldsymbol{\varphi}_{x,k}$ (resp. $\boldsymbol{\varphi}_{y,k}$) with eigenfrequencies $\omega_{x,k}$ (resp. $\omega_{y,k}$) :
\begin{subequations}\label{line:modes_spectrum}
\renewcommand{\theequation}{\theparentequation.\arabic{equation}}
\begin{align}
\boldsymbol{\varphi}_{x,k}^j &= \sqrt{\frac{2}{N+1}} \sin\left(\frac{jk\pi}{N+1}\right) \boldsymbol{\hat{x}}; \quad \quad  \omega_{x,k}^{2} = 4\sin\left(\frac{k\pi}{2(N+1)}\right)^2, \\
\boldsymbol{\varphi}_{y,k}^j &= \sqrt{\frac{2}{N+1}} \sin\left(\frac{jk\pi}{N+1}\right) \boldsymbol{\hat{y}}; \quad \quad  \omega_{y,k}^{2} = 4 A_{\alpha} \sin\left(\frac{k\pi}{2(N+1)}\right)^2.
\end{align}
\end{subequations}

\begin{figure}[h!]
\centering
\vspace*{-20mm}
\begin{tikzpicture}

\node[anchor=south west,inner sep=0] at (0.0,3.5)
{\includegraphics[width=\textwidth]{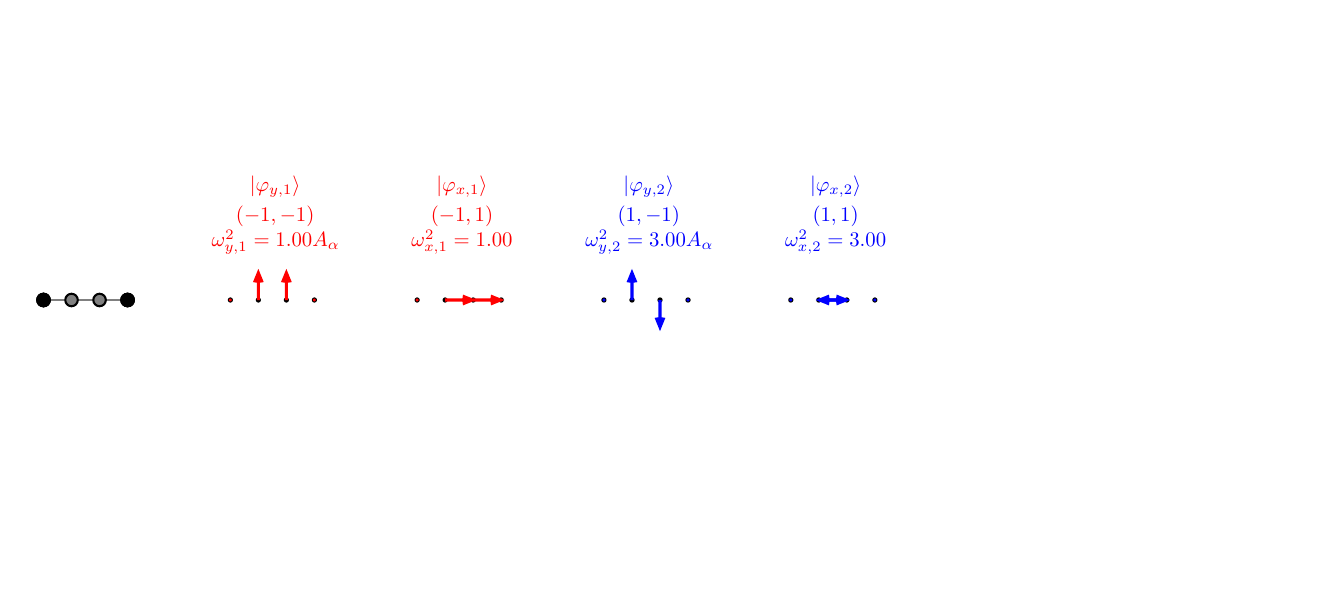}};
\draw[] (0.0, 6.7) -- (\textwidth, 6.7);

\node[anchor=south west,inner sep=0] at (0.0,0.0)
{\includegraphics[width=\textwidth]{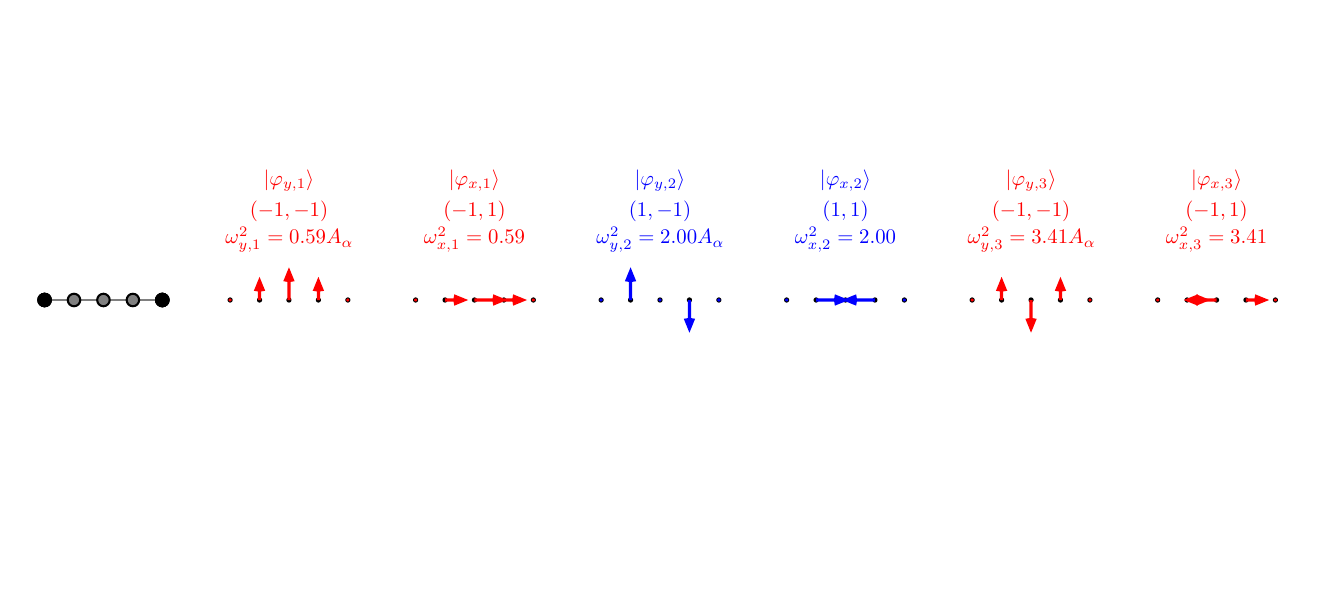}};
\draw[] (0.0, 3.2) -- (\textwidth, 3.2);

\node[anchor=south west,inner sep=0] at (0.0,-3.5)
{\includegraphics[width=\textwidth]{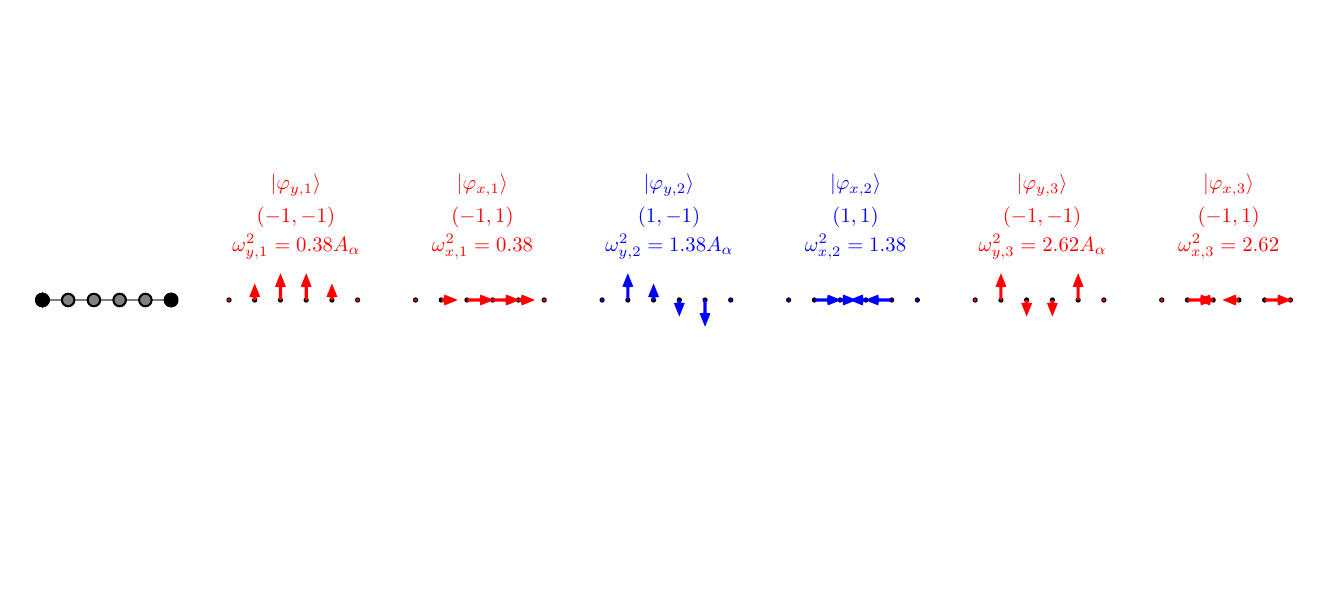}};
\draw[] (0.0, -0.3) -- (\textwidth, -0.3);

\node[anchor=south west,inner sep=0] at (0.0,-7.0)
{\includegraphics[width=\textwidth]{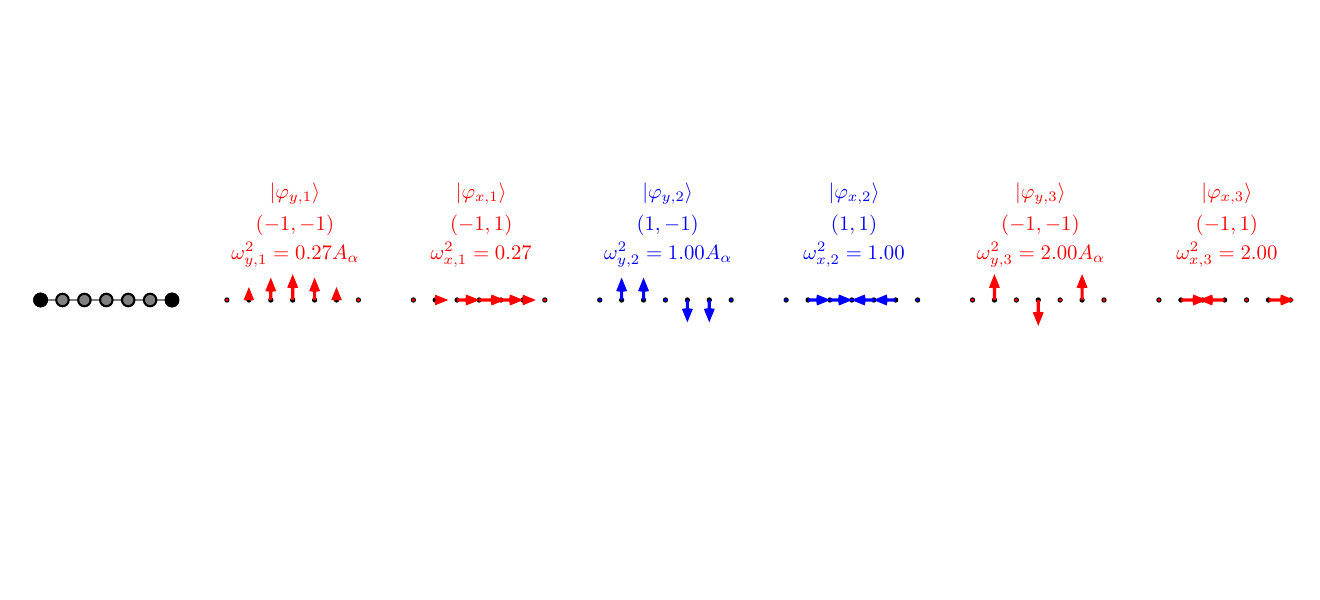}};

\node[] at (0.1,8.5) {\small (a)};
\node[] at (0.1,5.0) {\small (b)};
\node[] at (0.1,1.5) {\small (c)};
\node[] at (0.1,-2.0) {\small (d)};

\end{tikzpicture}
\vspace*{-35mm}
\caption{\small \textbf{Normal modes for linear structures (a) $N = 2$, (b) $N = 3$, (c) $N = 4$, (d) $N = 5$,} sorted by order of growing energies, and colored by their associated eigenvalues with respect to the rotation operation of  the dihedral group $D_{2}$, characterizing the symmetry of the elastic structure. The modes are computed in the limit of infinite tension, and only the six first modes are shown. For every mode, we show the mode's index $k$, the eigenvalues associated with the symmetry operations $(\tau, \sigma)$, and the associated squared eigenfrequency $\omega_{k}^{2}$.}
\label{modes_line}
\vspace{-5mm}
\end{figure}

The eigenmodes and eigenfrequencies for $N=2,3,4$ and $5$ are shown on Fig.~\ref{modes_line}.
The modes in the directions $x$ and $y$ are obviously locally orthogonal.
Moreover, modes with the same index $k$ have the same norm on each site, so that we introduce $\varphi_k^j=\boldsymbol{\varphi}_{x,k}^j\cdot\boldsymbol{\hat{x}}=\boldsymbol{\varphi}_{y,k}^j\cdot\boldsymbol{\hat{y}}$.
Finally, in the limit $\alpha\to\infty$, which corresponds to infinite tension or zero natural length, $A_\alpha\to 1$  and the modes are degenerated: $\omega_{x,k}=\omega_{y,k}=\omega_k$ for $1\leq k\leq N$; we restrict ourselves to this case in the following.

\subsection{General framework for locally orthogonal and degenerated eigenmodes}
\subsubsection{Bounds on the stability thresholds}

The general bounds derived above (Eqs.~(\ref{eq:stability_threshold_lb}) and (\ref{eq:upperbound})) translate into
\begin{equation}
\omega_1^2\leq\pi_c \left(|\boldsymbol{\hat n}^0\rangle \right)\leq 2 \omega_1^2.
\end{equation}
where the upper bound is obtained with the pair of modes $\boldsymbol{\varphi}_{x,1}$ and $\boldsymbol{\varphi}_{y,1}$.

\subsubsection{Single-frequency limit cycles}
We look for a collective actuation pattern where all the particles turn with the same constant angular velocity: the orientation $\theta_j(t)$ of the particle $j$ follows
\begin{equation}
	\theta_j(t) = \Omega t+\phi_j,
\end{equation}
where $\phi_j$ is a constant phase.
Integrating Eq.~(\ref{eq:proj_u}), we deduce that:
\begin{subequations}
\begin{align}
a^u_{x,k}(t) &= a^u_{x,k}(0) e^{- \omega_k^2 t} + \pi \sum_j \int_0^t \varphi^j_k\cos(\theta_j(t-t')) e^{- \omega_k^2 t'} \ \mathrm{d} t' \\
a^u_{y,k}(t) &= a^u_{y,k}(0) e^{- \omega_k^2 t} + \pi \sum_j \int_0^t \varphi^j_k\sin(\theta_j(t-t')) e^{- \omega_k^2 t'} \ \mathrm{d} t'
\end{align}
\end{subequations}
In the long time limit,
\begin{subequations}
\renewcommand{\theequation}{\theparentequation.\arabic{equation}}
\begin{align}
a_{x,k}^u(t) &= \frac{\pi}{\omega_k^4+\Omega^2}\sum_j\varphi_k^jf_k(\Omega t+\phi_j),\\
a_{y,k}^u(t) &= \frac{\pi}{\omega_k^4+\Omega^2}\sum_j\varphi_k^jf_k\left(\Omega t+\phi_j-\frac{\pi}{2}\right),
\end{align}
\end{subequations}
where we have defined
\begin{equation}
f_k(\theta) = \omega_k^2\cos(\theta)+\Omega\sin(\theta).
\end{equation}

Using the above expressions in the equation for $\dot\theta_j=\Omega$ (Eq.~(\ref{eq2:dimensionless_noiseless_braket_theta})), we get
\begin{equation}\label{Omega_self_consistent_equations}
\Omega = \sum_k\omega_k^2\varphi_k^j \left[\sin(\theta_j)a_{x,k}^u-\cos(\theta_j)a_{y,k}^u \right]
= \sum_{k,i} \frac{\pi\omega_k^2}{\omega_k^4+\Omega^2}\varphi_k^i\varphi_k^j f_k\left(\phi_i-\phi_j+\frac{\pi}{2}\right).
\end{equation}

We obtain a set of $N$ equations with $N$ unknowns: $\Omega$ and $N-1$ phases (we can always fix one of them).
$\Omega=0$ is always a solution, corresponding to a fixed point.
Depending on $\pi$, there may be other solutions.

Note that the radii of rotation of the particles can be computed at any time by summing over the modes:
\begin{equation}
R_j=\sqrt{\bsu_j^2}=\sqrt{\sum_{k,l}\varphi_k^j\varphi_l^j(a_{x,k}a_{x,l}+a_{y,k}a_{y,l})}.
\end{equation}

\subsubsection{Stability of the limit cycles}
Each solution may be tested for stability.
To determine the stability of a rotating solution, we use the comoving frame and introduce the coefficients $\beta$ such that:
\begin{subequations}
\renewcommand{\theequation}{\theparentequation.\arabic{equation}}
\begin{align}
	a^u_{x,k}(t) &= \beta_{x,k}(t) \cos(\Omega t) - \beta_{y,k}(t) \sin(\Omega t), \label{eq1:displacement_line_CA} \\
	a^u_{y,k}(t) &= \beta_{x,k}(t) \sin(\Omega t) + \beta_{y,k}(t) \cos(\Omega t). \label{eq2:displacement_line_CA}
\end{align}
\end{subequations}
We now write these coefficients as the rotating solution plus a perturbation:
\begin{subequations}
\renewcommand{\theequation}{\theparentequation.\arabic{equation}}
\begin{align}
	\beta_{x,k}(t) &= \beta_{x,k}^{(0)} +  \beta_{x,k}^{(1)}(t), \\
	\beta_{y,k}(t) &= \beta_{y,k}^{(0)} +  \beta_{y,k}^{(1)}(t), \\
	\theta_j(t) &= \theta_j^{(0)} (t) + \theta_j^{(1)} (t),
\end{align}
\end{subequations}
with
\begin{subequations}
\renewcommand{\theequation}{\theparentequation.\arabic{equation}}
\begin{align}
\beta_{x,k}^{(0)} & = \frac{\pi}{\omega_k^4+\Omega^2}\sum_j\varphi_k^jf_k(\phi_j),\\
\beta_{y,k}^{(0)} & = \frac{\pi}{\omega_k^4+\Omega^2}\sum_j\varphi_k^jf_k\left(\phi_j-\frac{\pi}{2}\right),\\
\theta_j^{(0)} & = \Omega t+\phi_j.
\end{align}
\end{subequations}
The dynamical equations for these perturbations are derived from (\ref{eq:proj_u}) and (\ref{eq2:dimensionless_noiseless_braket_theta}):
\begin{subequations}
\renewcommand{\theequation}{\theparentequation.\arabic{equation}}
\begin{align}
\dot\beta_{x,k}^{(1)} &= -\omega_k^2\beta_{x,k}^{(1)}+\Omega\beta_{y,k}^{(1)}-\pi\sum_i\varphi_k^i\sin(\phi_i) \theta_i^{(1)},\\
\dot\beta_{y,k}^{(1)} &= -\omega_k^2\beta_{y,k}^{(1)}-\Omega\beta_{x,k}^{(1)}+\pi\sum_i\varphi_k^i\cos(\phi_i) \theta_i^{(1)},\\
\dot\theta_j^{(1)} & = \sum_k\omega_k^2\varphi_k^j \left[\sin(\phi_j)\beta_{x,k}^{(1)}-\cos(\phi_j)\beta_{y,k}^{(1)} \right]
+\sum_{k,i} \frac{\pi\omega_k^2}{\omega_k^4+\Omega^2}\varphi_k^i\varphi_k^jf_k(\phi_i-\phi_j)\theta_j^{(1)}.
\end{align}
\end{subequations}

\subsubsection{Geometrical restriction on the existence of rotating solutions}

A simple condition can be derived to determine the stability of the rotating solution found above. Let's remind the equations for a single particle:

\begin{subequations}
\renewcommand{\theequation}{\theparentequation.\arabic{equation}}
\begin{align}
	\dot{\boldsymbol{x}}_{i} &= \pi \boldsymbol{\hat{n}}_{i} + \boldsymbol{F}_{i}^\textrm{el} \label{polarity_dynamics_single_particle_eq1} \\
	\dot{\theta}_{i} &= \boldsymbol{F}_{i} \cdot \boldsymbol{\hat{n}}_{i}^{\perp} \label{polarity_dynamics_single_particle_eq2}
\end{align}
\end{subequations}

where we may express Eq.~(\ref{polarity_dynamics_single_particle_eq2}) as:

\begin{equation} \label{polarity_dynamics_single_particle_theta}
 \dot{\theta}_{i} = \dot{\boldsymbol{x}}_{i} \cdot \boldsymbol{\hat{n}}_{i}^{\perp}
\end{equation}

Consider an active particle in the condensed state of the linear chain (i.e. circular motion). From the dynamics periodicity, the angular speed of the position vector and of the polarity vector are the same. Thus:

\begin{subequations}
\renewcommand{\theequation}{\theparentequation.\arabic{equation}}
\begin{align}
	\dot{\boldsymbol{x}}_{i} &= \Omega R_{i} \boldsymbol{\hat{r}}_{i}^{\perp} \\
	\dot{\theta}_{i} &= \Omega
\end{align}
\end{subequations}

Where $R_{i}$ is the radius of particle $i$'s trajectory. Then, replacing in Eq~(\ref{polarity_dynamics_single_particle_theta}), and discarding the $\Omega = 0$ case:

\begin{equation}
 \frac{1}{R_{i}} = \boldsymbol{\hat{r}}_{i}^{\perp} \cdot \boldsymbol{\hat{n}}_{i}^{\perp} = \boldsymbol{\hat{r}}_{i} \cdot \boldsymbol{\hat{n}}_{i} \leq 1
\end{equation}

This means that rotating solutions exist only when the trajectory radius is greater than 1. As $\pi$ decreases, the radii of rotation of the active units in the rotating solution decrease until the outer-most particles cross the threshold and the rotating solution doesn't exist anymore. In this state, the polarity vector ``drifts'' with respect to the position of the particle.

\subsection{Application to the chain}\label{}

\subsubsection{$N=2$ chain}

The eigenfrequencies are $\omega_1^2=1$ and $\omega_2^2=3$.

\begin{figure}[h!]
\centering
\vspace*{-3mm}
\begin{tikzpicture}

\node[] at (0.0,0.0)
{\includegraphics[width=0.45\textwidth]{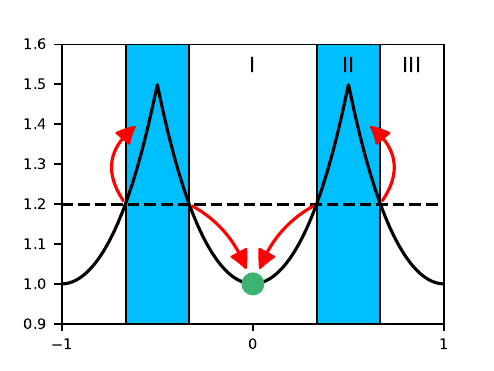}};

\node[rotate=90] at (-3.8,0.0) {\small $\pi_{c}$};

\node[] at (0.1,-2.85) {\small $\Delta \theta$};

\node[] at (-2.7,-0.05) {\small $\pi$};

\end{tikzpicture}
\vspace*{-2mm}
\caption{{\textbf{$N=2$ chain: Stability threshold of the fixed points.} The value of $\pi$ determines which configurations $\Delta \theta$ are stable ($\pi < \pi_{c}(\Delta \theta)$) and which are not. zone I and III: unstable configurations; zone II: stable configurations. The final configuration reached depends on the initial configuration $\Delta \theta$. All configurations starting in zone I are unstable, and move in phase space up to meet the limit cycle (green dot), corresponding to a condensation on modes $| \boldsymbol{\varphi}_{y,1} \rangle$ and $| \boldsymbol{\varphi}_{x,1} \rangle$. All configurations starting in zone II are stable and thus stay immobile. The red arrows indicate where the corresponding fixed point goes once it destabilizes.}}
\label{fig:chain_N2_stab_fp}
\end{figure}

We start by calculating the stability threshold of the fixed points.
A fixed point is defined by the orientations $\theta_1$ and $\theta_2$ of the two particles; however, from rotational invariance all the fixed points with the same value of $\Delta\theta=\theta_1-\theta_2$ are equivalent.
Using Eq.~(\ref{eq:stability_threshold_n}), we find their stability threshold, illustrated on Fig.~\ref{fig:chain_N2_stab_fp}:
\begin{equation}
\pi_c(\Delta\theta)= \frac{3}{2+|\cos(\Delta\theta)|}.
\end{equation}

We now look for a rotating solution.
Setting $\phi_1=0$, the equations (\ref{Omega_self_consistent_equations}) for $\Omega$ and $\phi_2$ read
\begin{subequations}
\renewcommand{\theequation}{\theparentequation.\arabic{equation}}
\begin{align}
\Omega &= \pi \left[ \frac{\Omega(1+ \cos(\phi_2)) -\sin(\phi_2)}{2(1 + \Omega^2)} + 3 \frac{\Omega(1- \cos(\phi_2)) + 3 \sin(\phi_2)}{2(9 + \Omega^2)} \right] \\
 &= \pi \left[ \frac{\Omega(1+ \cos(\phi_2)) + \sin(\phi_2)}{2(1 + \Omega^2)} + 3 \frac{\Omega(1- \cos(\phi_2)) - 3 \sin(\phi_2)}{2(9 + \Omega^2)} \right]
\end{align}
\end{subequations}
We see that $\phi_2=0$ or $\phi_2=\pi$.
For $\phi_2=0$ the angular velocity is given by
\begin{equation}
\Omega=\sqrt{\pi-1},
\end{equation}
which is a valid solution as long as $\pi\geq 1$.
To determine the stability of this solution, we need to study the spectrum of the matrix
\begin{equation}
	\mathbb{C}_2 = \begin{pmatrix}
		- 1 & \Omega & 0 & 0 & 0 & 0 \\
		- \Omega & - 1 & 0 & 0 & \frac{\pi}{\sqrt{2}} & \frac{\pi}{\sqrt{2}} \\
		0 & 0 & - 3 & \Omega & 0 & 0 \\
		0 & 0 & - \Omega & - 3 & \frac{\pi}{\sqrt{2}} & - \frac{\pi}{\sqrt{2}} \\
		0 & - \frac{1}{\sqrt{2}} & 0 & - \frac{3}{\sqrt{2}} & 1 & 0 \\
		0 & - \frac{1}{\sqrt{2}} & 0 & \frac{3}{\sqrt{2}} & 0 & 1
	\end{pmatrix}
\end{equation}
We are interested in particular in the zero eigenvalues.
This matrix always admits one that corresponds to global rotations; it does not preclude the stability of the solution.
There is another null eigenvalue when $\pi=1$, meaning that the rotating solution is stable on its whole range of existence.
We can also compute the radius of rotation and obtain $R=\sqrt{\pi}$.
This rotating solution is thus very similar to the one of the single particle.
Finally, one can show that the rotating solution corresponding to $\phi_2=\pi$ is unstable.

As a conclusion, for $\pi<1$ all the fixed points are stable and no rotating solution exists.
A rotating solution exists and is stable for $\pi>1$.
For $1<\pi<3/2$, the rotating solution coexists with stable fixed points.
In this range, starting close to an unstable fixed point, the system either evolves to the limit cycle (for fixed points close to $\Delta\theta=0$) or to a stable fixed point (for fixed points close to $\Delta\theta=\pi$).

\subsubsection{$N=3$ chain}
We restrict our analytical calculations to $\theta_1=\theta_3=\theta_2+\Delta\theta$ and find the stability threshold for the fixed points:
\begin{equation}
\pi_c(\theta_1,\theta_1-\Delta\theta,\theta_1)=\frac{2}{2+\sqrt{2}|\cos(\Delta\theta)|}.
\end{equation}
The stability threshold ranges from $\pi_c(\Delta\theta=0[\pi])=2-\sqrt{2}$ to $\pi_c(\Delta\theta=\pm\pi/2)=1$.
We note that $\pi_c(\Delta\theta=0[\pi])=\pi_{c}^{\text{min}}=\omega_\mathrm{min}^2$, confirming that these are the most unstable fixed points.
On the other hand, we confirm numerically that $\Delta\theta=\pm\pi/2$ correspond to the most stable fixed points.

We turn to the rotating solution.
For symmetry reasons, we assume that $\phi_1 = \phi_3$, and we set these phases to 0.
From Eq.~(\ref{Omega_self_consistent_equations}), we can establish the relation between $\Omega$ and $\pi$:

\begin{equation}
	\frac{4 + 12 \Omega^2 + \Omega^4}{2\pi} = 2 + \Omega^2 \pm  \frac{8 \Omega(\Omega^2 - 2)}{\sqrt{4 + 140 \Omega^2 + \Omega^4}}.
\end{equation}

\begin{figure}[h!]
\centering
\vspace*{-1mm}
\begin{tikzpicture}

\node[] at (0.0,0.0) {\includegraphics[width=5.0cm]{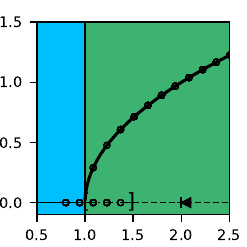}};
\node[] at (5.3,0.0) {\includegraphics[width=5.0cm]{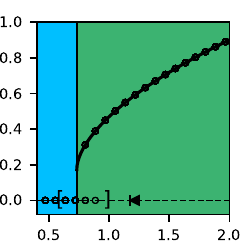}};
\node[] at (10.6,0.0) {\includegraphics[width=5.0cm]{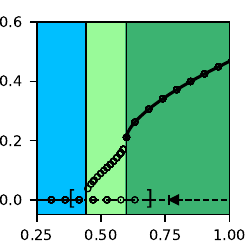}};

\node[] at (0.15,2.5) {\includegraphics[height=0.36cm]{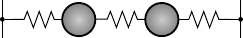}};
\node[] at (5.45,2.5) {\includegraphics[height=0.36cm]{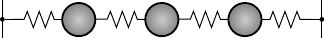}};
\node[] at (10.75,2.5) {\includegraphics[height=0.36cm]{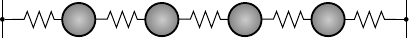}};

\node[] at (0.0,-4.7) {\includegraphics[width=4.5cm]{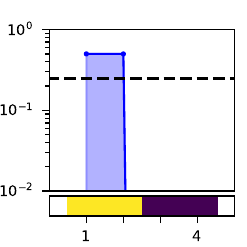}};
\node[] at (5.3,-4.7) {\includegraphics[width=4.5cm]{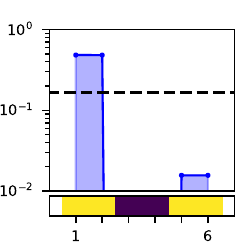}};
\node[] at (10.6,-4.7) {\includegraphics[width=4.5cm]{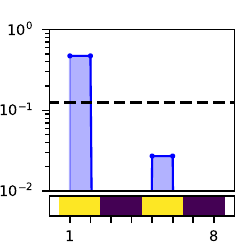}};

\node[] at (0.3,-6.9) {\small $k$ index};
\node[] at (5.6,-6.9) {\small $k$ index};
\node[] at (10.9,-6.9) {\small $k$ index};

\node[anchor=north, rotate=90] at (-2.85,-4.5) {\small $\langle \langle \boldsymbol{\varphi_{k}} | \boldsymbol{\hat{n}} \rangle^{2} \rangle_{t}$};
\node[anchor=north, rotate=90] at (2.45,-4.5) {\small $\langle \langle \boldsymbol{\varphi_{k}} | \boldsymbol{\hat{n}} \rangle^{2} \rangle_{t}$};
\node[anchor=north, rotate=90] at (7.75,-4.5) {\small $\langle \langle \boldsymbol{\varphi_{k}} | \boldsymbol{\hat{n}} \rangle^{2} \rangle_{t}$};


\node[] at (0.1,-2.6) {\small $\pi$};
\node[] at (5.4,-2.6) {\small $\pi$};
\node[] at (10.7,-2.6) {\small $\pi$};

\node[rotate=90] at (-2.7,0.2) {\small $\Omega$};
\node[rotate=90] at (2.6,0.2) {\small $\Omega$};
\node[rotate=90] at (7.9,0.2) {\small $\Omega$};

\node[rotate=90, anchor=east] at (9.05,2.05) {\footnotesize Frozen-Disordered};
\node[rotate=90, anchor=east] at (10.05,2.05) {\footnotesize Heterogeneous};
\node[rotate=90, anchor=west] at (12.52,-1.55) {\footnotesize Actuation};
\node[rotate=90, anchor=west] at (12.22,-1.55) {\footnotesize Collective};

\node[rotate=90, anchor=east] at (3.75,2.05) {\footnotesize Frozen-Disordered};
\node[rotate=90, anchor=west] at (7.22,-1.55) {\footnotesize Actuation};
\node[rotate=90, anchor=west] at (6.92,-1.55) {\footnotesize Collective};

\node[rotate=90, anchor=east] at (-1.55,2.05) {\footnotesize Frozen-Disordered};
\node[rotate=90, anchor=west] at (1.92,-1.55) {\footnotesize Actuation};
\node[rotate=90, anchor=west] at (1.62,-1.55) {\footnotesize Collective};

\node[rotate=90] at (-1.02,-0.96) {\small $\pi_{c}^{\text{min}}$};
\node[rotate=90] at (0.15,-0.96) {\small $\pi_{c}^{\text{max}}$};
\node[rotate=90] at (1.15,-1.07) {\small $\pi_{c}^{\text{upp}}$};

\node[rotate=90] at (3.95,-0.92) {\small $\pi_{c}^{\text{min}}$};
\node[rotate=90] at (4.96,-0.92) {\small $\pi_{c}^{\text{max}}$};
\node[rotate=90] at (5.45,-1.04) {\small $\pi_{c}^{\text{upp}}$};

\node[rotate=90] at (9.5,-0.91) {\small $\pi_{c}^{\text{min}}$};
\node[rotate=90] at (11.1,-0.91) {\small $\pi_{c}^{\text{max}}$};
\node[rotate=90] at (11.55,-1.0) {\small $\pi_{c}^{\text{upp}}$};

\node[] at (0.15,3.1) {\small (a)};
\node[] at (5.45,3.1) {\small (b)};
\node[] at (10.75,3.1) {\small (c)};

\end{tikzpicture}
\vspace*{-2mm}
\caption{\small{\textbf{Bifurcation of the rotating solution in linear chains.} (a) $N=2$. (b) $N=3$. (c) $N = 4$. (top) Bifurcation diagrams of stationary solutions corresponding to single frequency limit cycles. Solid lines represent stable solutions and dashed lines unstable solutions. In (a), $\pi_{c}^{\text{min}} = 1.0$, $\pi_{c}^{\text{max}} = 1.5$, $\pi_{c}^{\text{upp}} = 2.0$, $\pi_{FD} = \pi_{CA} = 1$. In (b), $\pi_{c}^{\text{min}} = 0.586$, $\pi_{c}^{\text{max}} = 1.0$, $\pi_{c}^{\text{upp}} = 1.17$, $\pi_{FD} = \pi_{CA} = 0.731$. In (c), $\pi_{c}^{\text{min}} = 0.382$, $\pi_{c}^{\text{max}} = 0.691$, $\pi_{c}^{\text{upp}} = 0.764$, $\pi_{FD} = 0.440$, $\pi_{CA} = 0.599$.} (bottom) Spectral decomposition of simulated collective actuation dynamics (for $\pi = 2.0$) on the normal modes of the different linear structures, sorted by order of growing energies. The horizontal dashed lines indicate equipartition. The bottom color bars codes for the symmetry class of the modes (Supplementary Information section 6).}
\label{fig:chain_N3}
\vspace{-5mm}
\end{figure}

Plotting $\Omega$ versus $\pi$ reveals a bifurcation that occurs for $\pi_{CA} \simeq 0.7310$, $\Omega^{CA} \simeq 0.1733$ (Fig.~\ref{fig:chain_N3}): there is no rotating solution for $\pi<\pi_{CA}$, while there are two solutions for $\pi>\pi_{CA}$, a stable one and an unstable one.
Contrary to the $N=2$ chain, the rotation starts with a finite angular velocity $\Omega_{CA}$.

\subsubsection{$N=4$ chain}
The most unstable fixed points destabilize at $\pi_c^{\text{min}}=\omega_\mathrm{min}^2=4\sin(\pi/10)^2\simeq 0.3820$.
The most stable fixed points are obtained numerically; they destabilize at $\pi_c^{\text{max}} \simeq 0.6908$.

Assuming that $\phi_1=\phi_4$ and $\phi_2=\phi_3$, we derive from Eq.~(\ref{Omega_self_consistent_equations}) the equation for the rotating solution:
\begin{equation}
	\frac{1 + 7 \Omega^2 + \Omega^4}{\pi}
	=
	\frac{3}{2} (1 + \Omega^2) \pm (1 - \Omega^2) \sqrt{1 - \left(\frac{1-\Omega^2}{6\Omega}\right)^2}.
\end{equation}

The solution presents a bifurcation $\pi_{CA} \simeq 0.5987$, $\Omega_{CA} \simeq 0.2082$.

\newpage
\section{Dynamics of the experimental structures}

\subsection{Normal modes}
\begin{figure}[h!]
\centering
\vspace*{-20mm}
\begin{tikzpicture}

\node[anchor=south west,inner sep=0] at (0.0,0.0)
{\includegraphics[width=\textwidth]{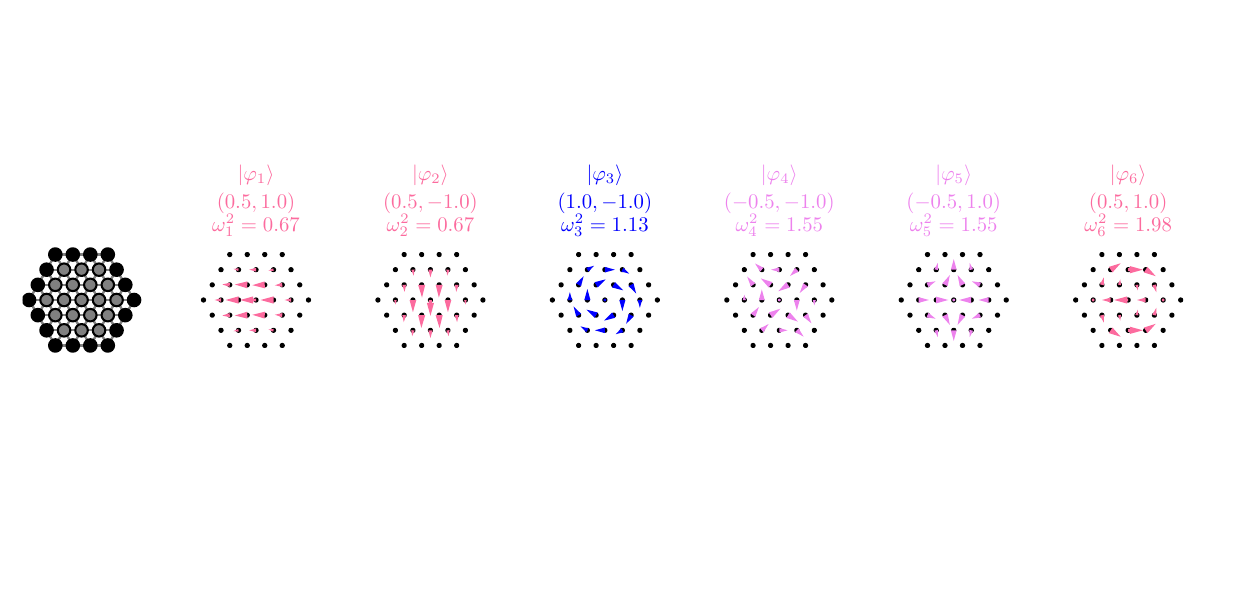}};
\node[anchor=south west,inner sep=0] at (0.0,-3.0)
{\includegraphics[width=\textwidth]{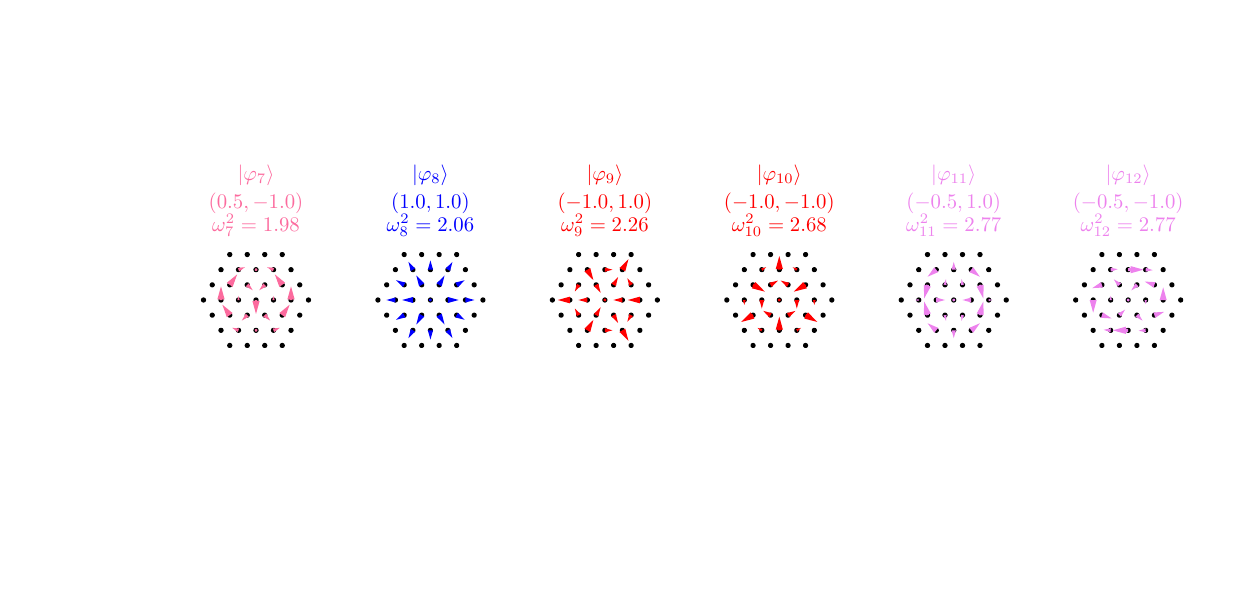}};

\draw[] (0.0, -0.1) -- (\textwidth, -0.1);

\node[anchor=south west,inner sep=0] at (0.0,-6.8)
{\includegraphics[width=\textwidth]{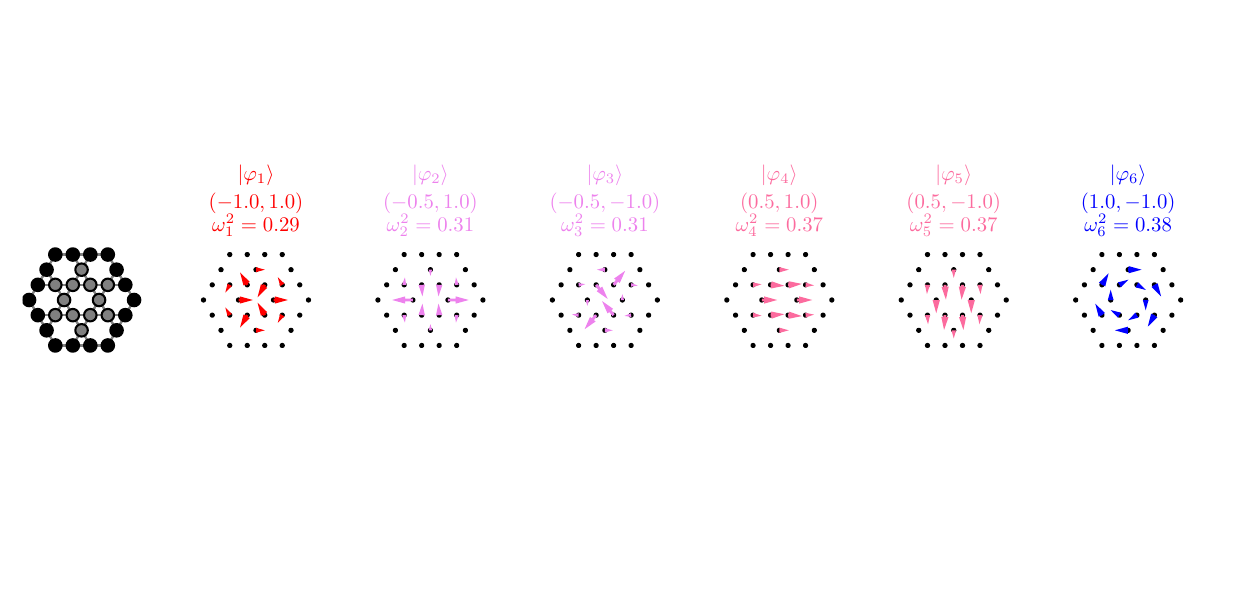}};
\node[anchor=south west,inner sep=0] at (0.0,-9.8)
{\includegraphics[width=\textwidth]{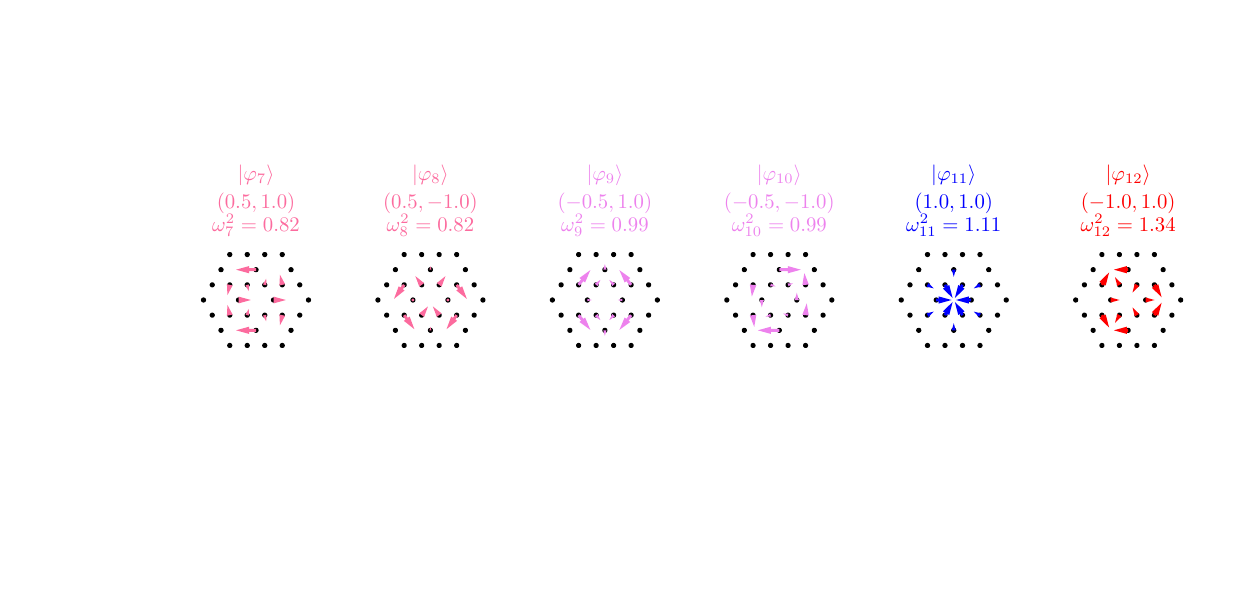}};

\node[] at (0.1,5.0) {\small (a)};
\node[] at (0.1,-1.8) {\small (b)};

\end{tikzpicture}
\vspace*{-35mm}
\caption{\small \textbf{Normal modes of the structures studied experimentally: (a) the triangular lattice pinned at the edges ($\alpha = 1.27$), (b) the kagome lattice pinned at the edges ($\alpha = 1.02$).} The modes are sorted by order of growing energies, and colored by their associated eigenvalues with respect to the rotation operation of the dihedral group of symmetry $D_{6}$ of the structure. The modes are computed for the experimental values of the tension, and only the twelves first modes are shown. For every mode, the figure highlights the mode's index $k$, the eigenvalues associated with the symmetry operations $(\tau, \sigma)$, and the associated squared eigenfrequency $\omega_{k}^{2}$.}
\label{modes_hex}
\end{figure}

Fig.~\ref{modes_hex} provides the twelve first modes, sorted by class of symmetry, of both experimental structures.

\subsection{Bifurcation scenarii}
From Fig.~\ref{modes_hex}, one sees that the modes concerned by the condensation ($|\boldsymbol{\varphi}_{1}\rangle$ and $|\boldsymbol{\varphi}_{2}\rangle$ for the triangular lattice; $|\boldsymbol{\varphi}_{4}\rangle$ and $|\boldsymbol{\varphi}_{5}\rangle$ for the kagome lattice) are neither fully delocalized nor locally orthogonal. The $D_{6}$ symmetry and the pinning of the nodes at the edges indeed forbid the modes to have such properties. Typically the modes have a larger polarization away from the edges. They are also not strictly locally orthogonal one with another.

The modes $|\boldsymbol{\varphi}_{1}\rangle$ and $|\boldsymbol{\varphi}_{2}\rangle$ of the triangular lattice are composed of geometrical domains of nodes, which are equidistant to the central node: (i) the central node (ii) a ring of first neighbors (iii) a ring of second neighbors closest to the center (iv) a ring of second neighbors further away from the center. Similarly, the modes $|\boldsymbol{\varphi}_{4}\rangle$ and $|\boldsymbol{\varphi}_{5}\rangle$ of the kagome lattice are composed of domains of nodes equidistant from the center (i) a ring of first neighbors (ii) and a ring of second neighbors.

\begin{figure}[h!]
\centering
\captionsetup{format=plain}
 \begin{tikzpicture}

\node[] at (0.0,5.4)
{\includegraphics[height=6.0cm]{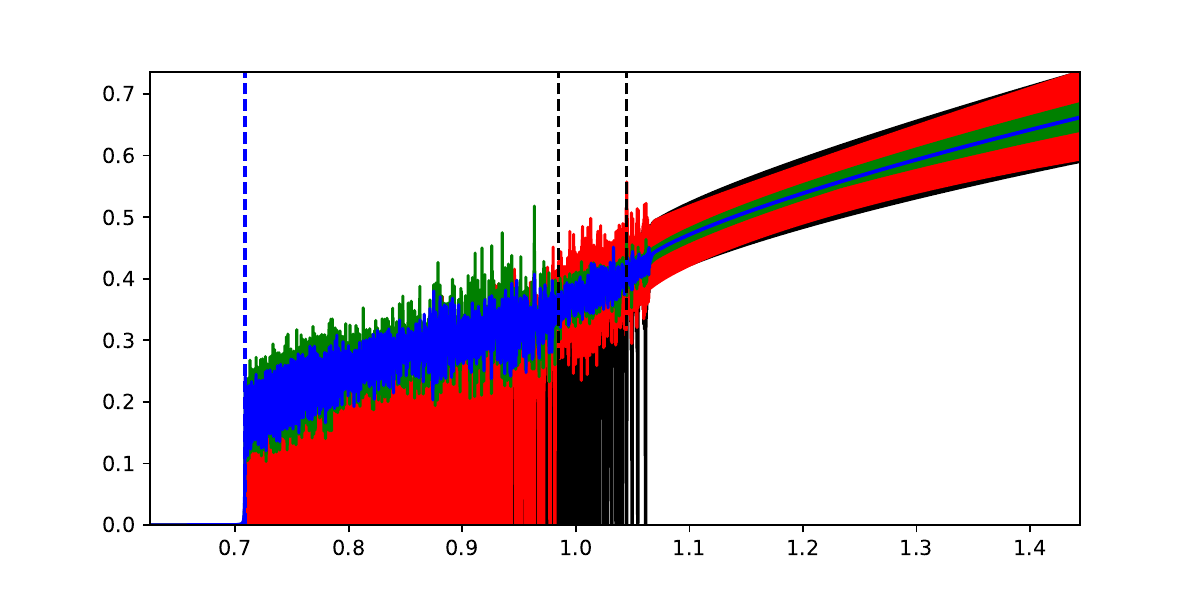}};
\node[] at (0.0,10.5)
{\includegraphics[height=6.0cm]{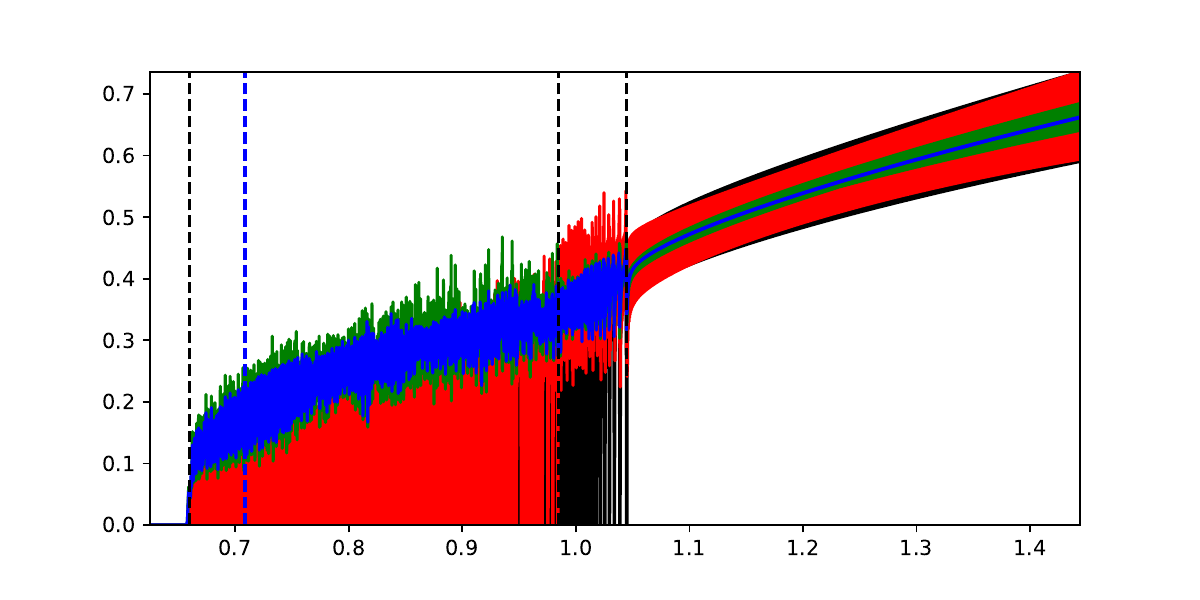}};

\node[] at (2.7,4.7)
{\includegraphics[height=2.5cm]{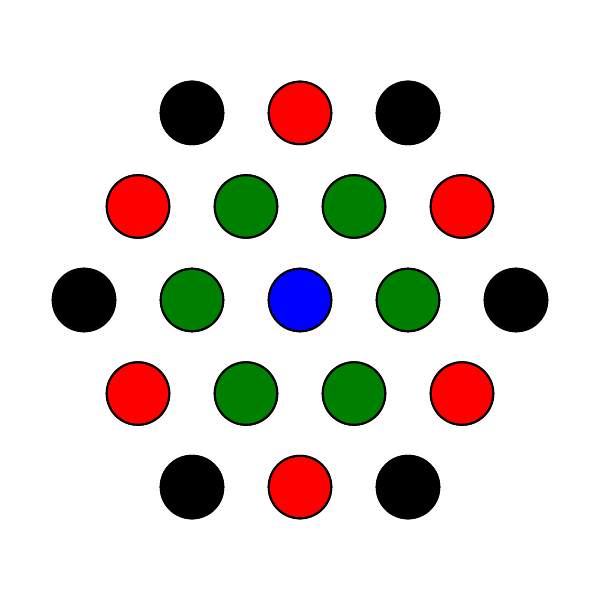}};

\node[] at (2.7,9.7)
{\includegraphics[height=2.5cm]{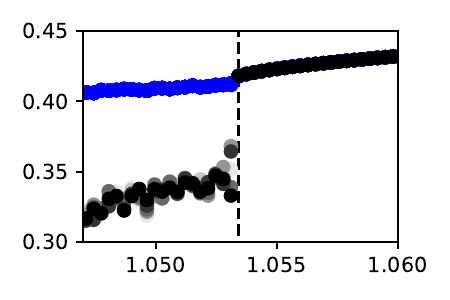}};
\node[] at (0.0,0.0)
{\includegraphics[height=5.5cm]{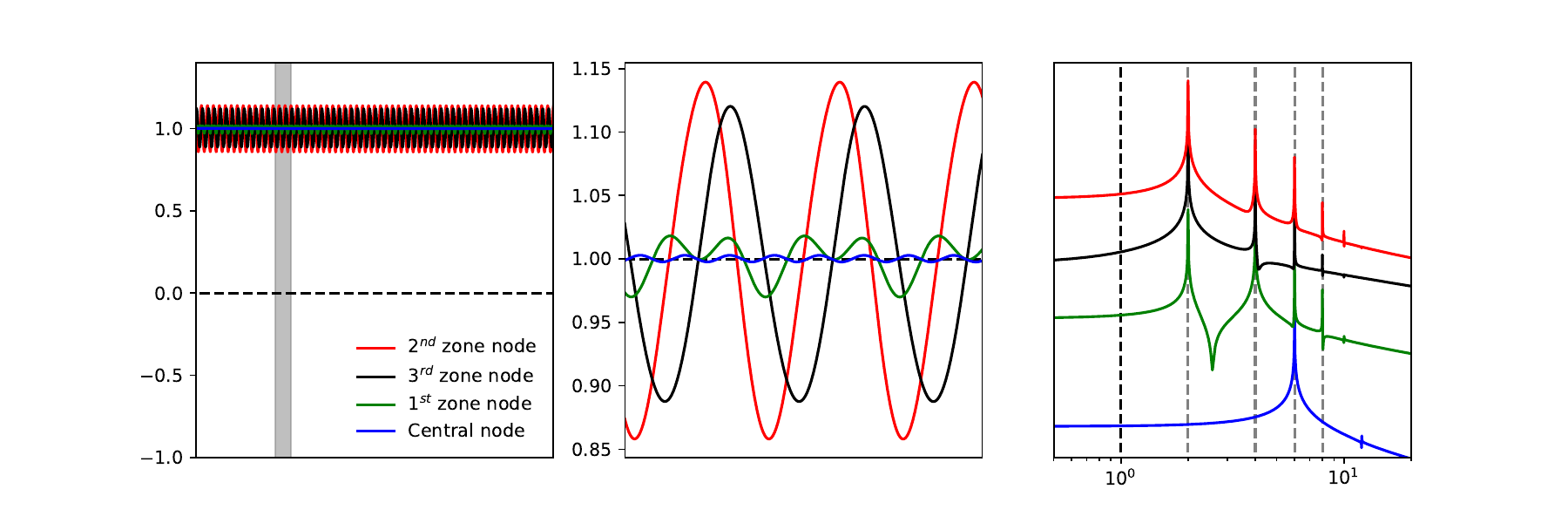}};
\node[] at (0.0,-4.5)
{\includegraphics[height=5.5cm]{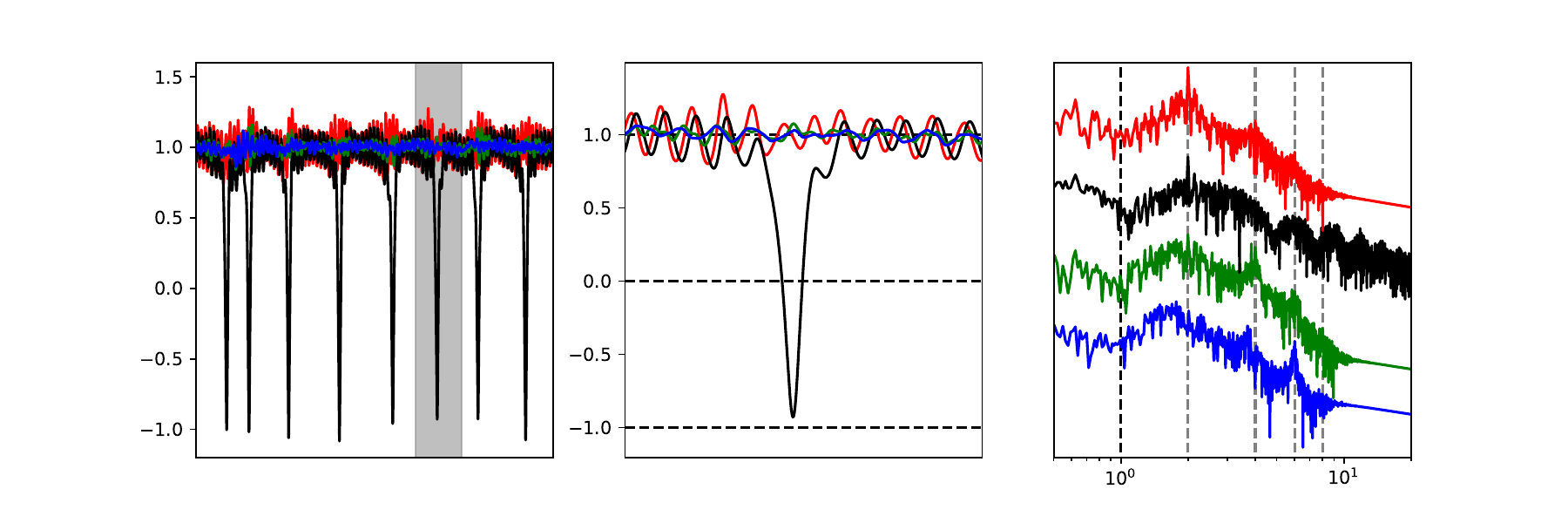}};


\node[rotate=90] at (-5.4,10.5) {\small $| \omega_{i} |$};
\node[rotate=90] at (-5.4,5.4) {\small $| \omega_{i} |$};
\node[rotate=90] at (-7.0,-0.2) {\small $ \omega_{i}/\langle \omega \rangle$};
\node[rotate=90] at (-7.0,-4.5) {\small $ \omega_{i}/\langle \omega \rangle$};

\node[] at (-4.1,-6.85) {\small $t$};
\node[] at (0.3,-6.85) {\small $t$};
\node[] at (4.8,-7.0) {\small $ \omega_{i}/\langle \omega \rangle$};

\node[] at (3.9,9.15) {\small $\pi$};
\node[] at (3.0,11.0) {\small $\pi_{CA}$};
\node[rotate=90] at (0.7,9.88) {\small $\langle \omega_{i} \rangle_{t}$};

\node[rotate=90] at (2.55, 0.1) {\small Power spectrum (a.u.)};
\node[rotate=90] at (2.55, -4.4) {\small Power spectrum (a.u.)};

\node[] at (0.0, 2.6) {\small $\pi$};

\draw[->, thick] (-0.9,11.5) -- (-2.8,10.7);

\draw[->, thick] (-2.8,5.7) -- (-0.9,6.5);

\node[] at (0.75,12.5) {\small $\pi_{CA}$};

\node[rotate=90] at (6.9, -0.0) {\small $\pi > \pi_{CA}$};
\node[rotate=90] at (6.9, -4.5) {\small $\pi < \pi_{CA}$};

\node[] at (2.8,12.5) {\small (a)};
\node[] at (4.4, 3.5) {\small (b)};
\node[] at (-5.9, -1.7) {\small (c)};
\node[] at (-1.35, 1.8) {\small (d)};
\node[] at (3.15, 1.8) {\small (e)};
\node[] at (-2.7, -2.7) {\small (f)};
\node[] at (1.8, -2.7) {\small (g)};
\node[] at (6.3, -2.7) {\small (h)};

 \end{tikzpicture}
 \caption{\small \textbf{Simulations of the bifurcation to the CA condensation in the case of the triangular lattice.} (a) Absolute value of the instantaneous angular velocities $|\omega_{i}|$ during an annealing experiment with decreasing $\pi$. Only four polarities are shown, each representative of a given zone. The black vertical bars show the value of $\pi$ at which each zone desynchronize. The blue vertical bar highlights the value of $\pi$ at which the fixed point reached at the end of the process should destabilize (from Eq. (\ref{eq:stability_threshold_n})).
 Inser: time averaged angular velocity for all polarities zoomed on the first desynchronization. The gray level curves represent the $3^{rd}$ zone polarities. The rest is shown in blue. (b) Inverse annealing process. Inser: definition of the zones. (c/d/e) Dynamics in the physical space just before the first desynchronization; (d) zoom on the gray zone of (c); (e) power spectrums shifted vertically for clarity reasons. (f/g/h) Dynamics in the physical space just after the first desynchronization; (d) zoom on the gray zone of (f); (h) power spectrums shifted vertically for clarity reasons.}
 \label{bifurcation_simu}
\end{figure}

Fig.~\ref{bifurcation_simu} describes the details of the noiseless transition from the collective actuation to the frozen regime in the case of the triangular lattice. In the collective actuation regime, the polarities rotate at a given mean frequency, dressed with periodic modulations, (Figs.~\ref{bifurcation_simu}-c,d). Indeed, as the modes concerned by the condensation are not strictly locally orthogonal, the oscillation cannot take place at a single-frequency, and is modulated by even multiples of the mean rotation rate (Fig.~ \ref{bifurcation_simu}-e). As $\pi$ decreases below $\pi_{CA}$, the periodic collective actuation regime turns unstable, and the outer domain desynchronize from the mean oscillation. This yields a discontinuous jump in the collective oscillation frequency $\Omega$ (Inset of Fig.~\ref{bifurcation_simu}a), and aperiodic turnarounds of the outer domain polarities (Figs.~\ref{bifurcation_simu}f and g). The system has entered into the heterogeneous regime, where the collective actuation and the frozen regimes coexist spatially.

\vspace{0.2cm}
\noindent\makebox[\linewidth]{\color{black} \rule{10.0cm}{1.0pt}}
\vspace{0.1cm}

\end{document}